\documentclass[12pt,preprint]{aastex}
\usepackage{amssymb, amsmath, apjfonts, emulateapj5, natbib, color}
\input psfig.tex

\def\degree{{\circ}}
\newdimen\digitwidth      
\setbox1=\hbox{0}       
\digitwidth=\wd1        
\catcode`"=\active      
\def"{\kern\digitwidth}

\def\aa{{A\&A}}

\def\aj{{AJ}}

\def\amin{$^\prime$}
\def\annrev{{ARA\&A}}
\def\apj{{ApJ}}
\def\apjs{{ApJS}}
\def\asec{$^{\prime\prime}$}

\def\deg{$^{\circ}$}

\def\lax{{$\mathrel{\hbox{\rlap{\hbox{\lower4pt\hbox{$\sim$}}}\hbox{$<$}}}$}}
\def\gax{{$\mathrel{\hbox{\rlap{\hbox{\lower4pt\hbox{$\sim$}}}\hbox{$>$}}}$}}
\def\simlt{\lower.5ex\hbox{$\; \buildrel < \over \sim \;$}}
\def\simgt{\lower.5ex\hbox{$\; \buildrel > \over \sim \;$}}

\def\mnras{{MNRAS}}

\def\percm2{cm$^{-2}$}

\slugcomment{To appear in {\it The Astrophysical Journal Supplement}.}
\lefthead{Li et al.}
\righthead{The Carnegie-Irvine Galaxy Survey. II.}

\begin{document}

\title{THE CARNEGIE-IRVINE GALAXY SURVEY. II. ISOPHOTAL ANALYSIS}

\author{ Zhao-Yu Li\altaffilmark{1, 2}, Luis C. Ho\altaffilmark{2}, Aaron J. 
Barth\altaffilmark{3}, and Chien Y. Peng\altaffilmark{4}}

\altaffiltext{1}{Department of Astronomy, School of Physics, Peking 
University, Beijing 100871, China}

\altaffiltext{2}{The Observatories of the Carnegie Institution for Science, 
813 Santa Barbara Street, Pasadena, CA 91101, USA}

\altaffiltext{3}{Department of Physics and Astronomy, 4129 Frederick Reines 
Hall, University of California, Irvine, CA 92697-4575, USA}

\altaffiltext{4}{NRC Herzberg Institute of Astrophysics, 5071 West Saanich 
Road, Victoria, British Columbia, V9E 2E7, Canada}

\begin{abstract}
The Carnegie-Irvine Galaxy Survey (CGS) is a comprehensive investigation of 
the physical properties of a complete, representative sample of 605 bright 
($B_T \leq 12.9$ mag) galaxies in the southern hemisphere.  This contribution
describes the isophotal analysis of the broadband (\emph{BVRI}) optical 
imaging component of the project.  We pay close attention to sky subtraction, 
which is particularly challenging for some of the large galaxies in our 
sample.  Extensive crosschecks with internal and external data confirm that 
our calibration and sky subtraction techniques are robust with respect to the 
quoted measurement uncertainties.  We present a uniform catalog of 
one-dimensional radial profiles of surface brightness and geometric 
parameters, as well as integrated colors and color gradients.  Composite 
profiles highlight the tremendous diversity of brightness distributions found 
in disk galaxies and their dependence on Hubble type.  A significant fraction 
of S0 and spiral galaxies exhibit non-exponential profiles in their outer 
regions.  We perform Fourier decomposition of the isophotes to quantify 
non-axisymmetric deviations in the light distribution.  We use 
the geometric parameters, in conjunction with the amplitude and phase of the 
$m=2$ Fourier mode, to identify bars and quantify their size and strength.  
Spiral arm strengths are characterized using the $m=2$ Fourier profiles and 
structure maps.  Finally, we utilize the information encoded in the $m=1$ 
Fourier profiles to measure disk lopsidedness.  The databases assembled here 
and in Paper~I lay the foundation for forthcoming scientific 
applications of CGS.
\end{abstract}

\keywords{atlases --- galaxies: fundamental parameters --- galaxies: general 
--- galaxies: photometry --- galaxies: structure --- surveys}

\section{Introduction}

This paper, the second in a series, presents the isophotal analysis for the 
optical images of the Carnegie-Irvine Galaxy Survey (CGS), a detailed study 
of a statistically complete sample of nearby, bright galaxies in the southern 
sky (Ho et al. 2011, hereinafter Paper~I).  The immediate aim of this paper
is to reduce our extensive set of images to a uniform database of 
one-dimensional (1-D) radial profiles of surface brightness and geometric 
parameters, on which much of our subsequent scientific analysis will depend.
Although we intend to apply more sophisticated methods of analysis to the 
images (Peng et al. 2010; S. Huang et al. 2011, in preparation), the 1-D analysis
already contains a wealth of useful information that can be exploited for 
science.  Moreover, 1-D analysis has the virtue of simplicity.  It can be 
efficiently applied to a large sample of objects, allowing a quick overview of 
the global properties of the survey.

The brightness profiles of galaxies have long helped to guide our 
understanding of their physical nature.  Despite the visual complexity of 
their images, the 1-D radial brightness profiles of galaxies in the nearby 
universe actually show a surprising degree of order.  De~Vaucouleurs (1948) 
first noticed that the light distributions of elliptical galaxies generally 
follow a $r^{1/4}$ profile, which has been interpreted as a signature of 
dissipationless formation processes \citep{vanalbada82, katz91}.
Later studies, beginning with \citet{caon93}, increasingly recognized
that many ellipticals, in fact, do not strictly follow the $r^{1/4}$ law, but
instead are better described by the more general $r^{1/n}$ function of
\citet{sers68}, of which de~Vaucouleurs' law is a special case ($n=4$).
Indeed, the S\'{e}rsic function has since been generally adopted as the
standard formula for fitting the brightness profiles of ellipticals
\citep[e.g.,][]{grah96, truj01, korm09}.  
Our modern view of bulges has also grown steadily more complex over time.
Once thought to be "mini-ellipticals" with $r^{1/4}$ profiles, bulges, too,
are now known to be better described by a S\'ersic $r^{1/n}$ function
\citep{ansa94, andr95, dejo96, cour96, maca03}.  The S\'ersic indices of 
bulges have a broad distribution of observed values, from $n < 1$ to $n > 4$
\citep[e.g.,][]{maca03, fidr08, gadotti08}, and it is argued that they reflect
different formation physics.  Spheroids with $n$ \lax\ 2 are regarded as
pseudobulges \citep{fidr08}, which formed through internal, secular
processes, while those with $n$ \gax\ 2 are classical bulges, which, like the
ellipticals, were assembled more rapidly, most likely with the assistance of
mergers \citep{koke04}.

The brightness profiles of the disks of S0 and spiral galaxies have been 
traditionally described by a single exponential function \citep{deva59, 
free70}, which arises as a natural consequence of viscous transport in a disk
\citep{yoso89, zhwy00, fecl01, slyz02}, perhaps mediated by star formation and 
feedback processes \citep{robe04, gove07}. In actuality, very few disks are so 
simple.  Many possess breaks and inflections in their outer radial profile 
\citep{vand79, vase81, maga97, pohl00, degr01, erwi05, erbp05, potr06}.  No 
general consensus yet exists as to their cause, but they offer 

\begin{figure*}[t]
\centerline{\psfig{file=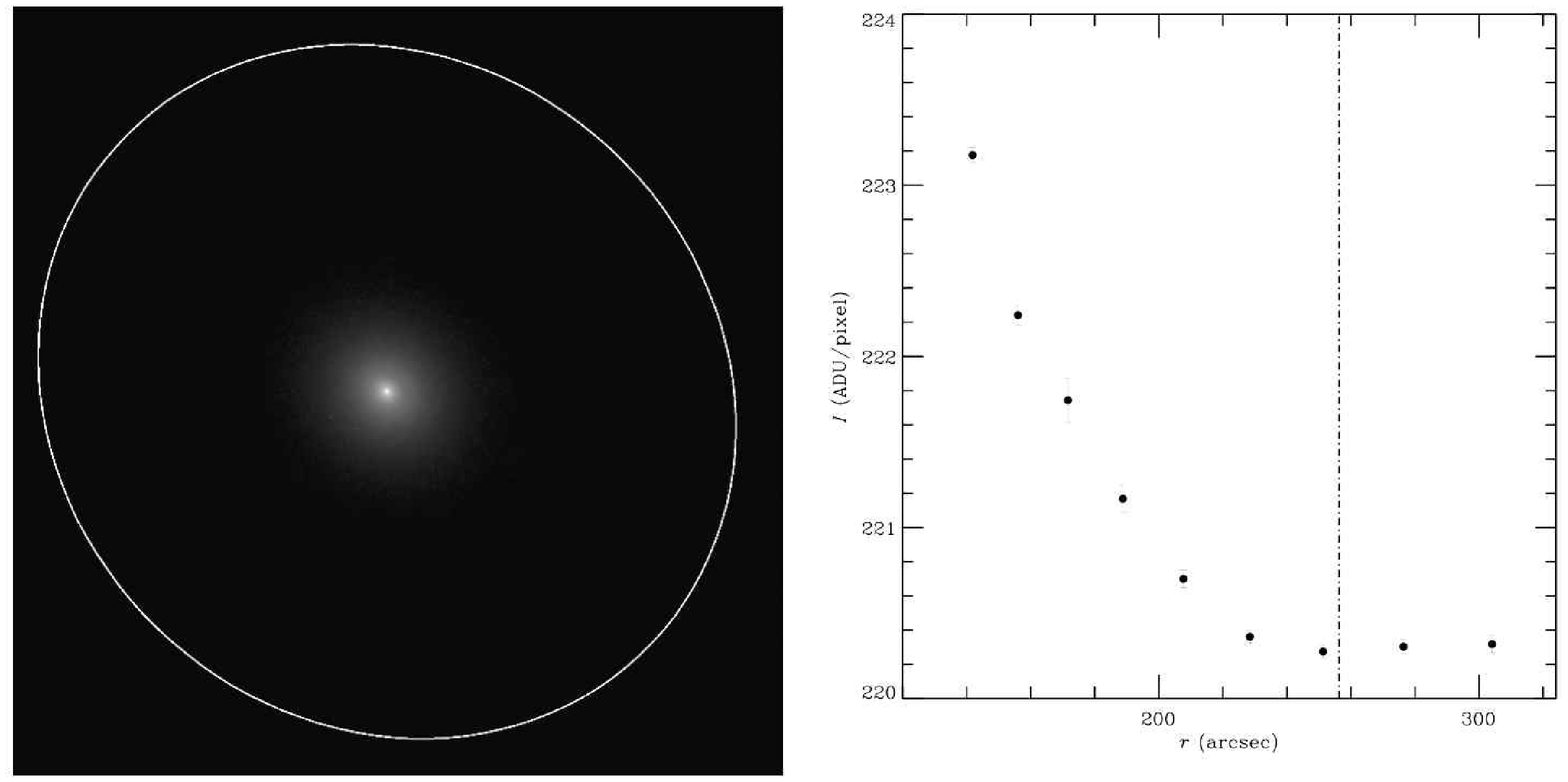,width=17.5cm,angle=0}}
\figcaption[fig1.ps]{
Illustration of how we determine the sky radius for the $B$-band image of
NGC~1400. Left: star-cleaned image, showing the full field-of-view of
8\farcm9$\times$8\farcm9.  Right: radial profile of the isophotal
intensity, in units of ADU~pixel$^{-1}$.  The vertical dash-dotted line marks
the radius where the isophotal intensities start to oscillate rather than
continue to decrease; this radius, designated by the keyword {\tt SKY\_RAD} in
the image header, denotes the region outside of which the sky dominates. The
corresponding isophotal ellipse is overplotted in the left-hand panel.
\label{figure:skyrad}}
\end{figure*}

\noindent
important clues 
to a host of physical processes pertinent to galaxy formation 
\citep{vanderkruitfreeman11}.  

Apart from intensity profiles, isophotal analysis of galaxy images yields 
other useful diagnostics.  The radial variation of the ellipticity and 
position angle, for example, provides an efficient means to identify bars and 
to quantify their length and strength \citep[e.g.,][]{lain02, erwi05, mene07}.
Fourier decomposition of the isophotes provides yet another method to probe 
non-axisymmetric perturbations in the light distribution.
The relative amplitude of the $m=2$ mode, in combination with its phase angle, 
has been shown to be effective in isolating bars \citep{elmegreen85, buta86, 
ohta90} and spirals \citep{elmegreen89, rixzar95, odewahn02}.  Both of these 
features are common constituents in disk galaxies, and both are thought to 
play a dynamical role in facilitating angular momentum transport and driving 
secular evolution.  Likewise, a significant fraction of disk galaxies exhibits 
global lopsidedness in their stellar light distribution \citep{rixzar95, 
zari97, bour05, reic08}, whose main culprit remains in dispute 
\citep{joco08}.  As shown by \citet{rixzar95}, this type of non-axisymmetric 
perturbation is again conveniently revealed through Fourier analysis of the 
isophotes, in this case through the $m=1$ mode.

This paper is organized as follows.  Section~2 gives a brief overview of the 
CGS sample, the observations, and some basic characteristics of the images.  
Section~3 describes our method of sky subtraction.  The procedural details of 
isophotal analysis are presented in Section~4, including our method for 
extracting geometric parameters, surface brightness profiles, and Fourier 
components.  We generate composite light distributions (Section~5) to identify 
statistical trends in disk profiles, and assemble integrated colors and 
color gradients (Section~6).  The products from the isophotal and Fourier 
analysis are used to quantify the strengths of bars (Section~7), spiral arms 
(Section~8), and lopsidedness (Section~9).  Section~10 assesses the reliability
of our measurements using internal and external tests.  Section~11 gives a 
brief summary and an outline of future plans.  The database of isophotal 
parameters is described in the Appendix.

\section{Sample Properties}

The CGS covers a statistically complete sample of 605 bright, nearby galaxies 
of all morphological types in the southern hemisphere, with $B$-band total 
magnitude $B_{T} \leq 12.9$ and $\delta < 0^\degree$. These very general 
selection criteria enable us to probe galaxies with a broad range of physical 
properties and morphologies.  The primary parent sample\footnote{As described
in Paper~I, we observed an additional 11 galaxies that do not formally meet
the selection criteria of CGS.  We still analyze them here but will not use 
their results to draw statistical inferences on the sample.} comprises 17\% 
ellipticals, 18\% S0 and S0/a, 64\% spirals, and 1\% irregulars.  The bulk of
the sample is relatively nearby (median $D_L$ = 24.9 Mpc), luminous (median 
$M_{B_T} = -20.2$ mag), and well resolved.  The typical seeing of CGS is 
$\sim 1$\asec, and the sample has a median isophotal angular diameter of 
$D_{25}$ = 3\farcm3 at a surface brightness level of $\mu_B=25$ mag 
arcsec$^{-2}$.

Paper~I describes the observing strategy, data reductions, and photometric 
calibration of the optical imaging component of the project.  We only repeat 
a few essential details here.  The broadband \emph{BVRI} images have a 
field-of-view of 8\farcm9$\times$8\farcm9 and a pixel scale of 0\farcs259, 
which is well matched to the good seeing typically achieved with the 
du~Pont 2.5-m telescope at Las Campanas Observatory.  The median seeing 
of the survey, as determined from over 6000 science images, is 1\farcs17, 
1\farcs11, 1\farcs01, and 0\farcs96 in the $B$, $V$, $R$, and $I$ band, 
respectively.  A little more than half of the galaxies were observed under 
photometric conditions, with median photometric errors of 0.08, 0.04, 0.03, 
and 

\begin{figure*}[t]
\centerline{\psfig{file=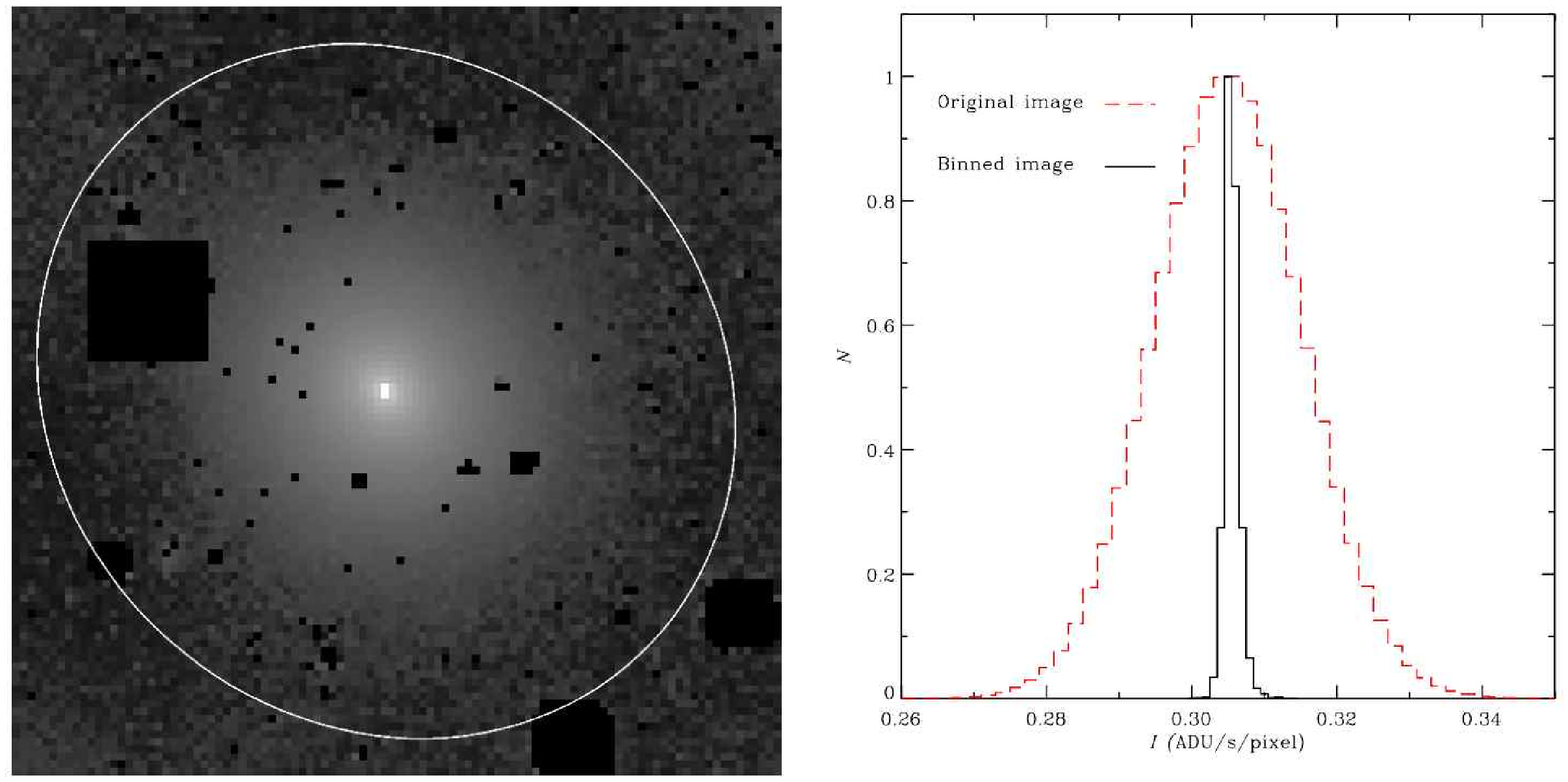,width=17.5cm,angle=0}}
\figcaption[fig2.ps]{
Left: $B$-band image of NGC~1400, binned by $20\times 20$,
with the ellipse in Figure~\ref{figure:skyrad} overplotted.  We show the full
field-of-view of 8\farcm9$\times$8\farcm9.  The sky value and its uncertainty
are simply the mean and standard deviation of the pixel values outside of this
ellipse, after excluding the masked objects, which are shown as black regions.
Right: normalized histograms of the background pixels of the original
and the binned images; their peak positions are $0.3060\pm0.0102$ and
$0.3059\pm0.0011$ ADU~s$^{-1}$~pixel$^{-1}$, respectively.
(A color version of this figure is available in the online journal.)
\label{figure:skybin}}
\end{figure*}

\noindent
0.04 mag for the \emph{B, V, R} and $I$ filters, respectively.  We devised 
a calibration strategy to establish an approximate photometric zero point for 
the non-photometric observations, for which the corresponding photometric 
errors are 0.21, 0.11, 0.10, and 0.09 mag.  After correcting for large-scale 
gradients in the background, the flatness of the final images is about 0.6\%, 
and the typical depth of the surface brightness, defined as $1\, \sigma$ 
above the background, has a median value of $\mu \approx 27.5, 26.9, 26.4,$ 
and 25.3 $\rm mag\  arcsec^{-2}$ in the $B, V, R,$ and $I$ bands, 
respectively. 

We derived a number of data products from the reduced, calibrated images.
These include red--green--blue color composites generated from the $B$, $V$, 
and $I$ bands, images cleaned of foreground stars and background galaxies, 
a stacked image from a weighted combination of the four filters optimized
to enhance regions of low surface brightness, structure maps designed to 
accentuate high-spatial frequency features, and color index maps from 
different combinations of the filters.

\section{Sky Determination}

Sky determination is a crucial, fundamental step in the data analysis.  Many 
of the basic galaxy parameters we are interested in measuring (magnitudes, 
colors, characteristic size and brightness level, etc.) are predicated
on having the sky level properly subtracted.  Importantly, under-subtraction 
or over-subtraction of the sky value can introduce spurious curvature into 
the brightness profile, especially in the faint, outer regions of the 
galaxy \citep[e.g.,][]{erwi08}.  \citet{maca03} studied the influence of the 
sky value on the  bulge and disk parameters for a sample of spirals, and 
concluded that the disk, but to a lesser extent even the bulge, parameters 
are sensitive to the sky value.

There are a variety of ways to measure the sky value of a CCD image.  Science 
data such as ours, however, wherein an extended object fills a substantial 
portion of the chip, pose unique challenges.  This is especially so because
the background of our images is not always entirely uniform (Paper~I).  We 
adopt a two-step approach.  As in \citet{noovan07}, we generate the isophotal 
intensity radial profile of the galaxy to the edge of the field to determine 
the radius beyond which the sky background dominates the signal.  As 
illustrated in Figure~\ref{figure:skyrad}, the transition from the galaxy's 
outer boundary to the sky-dominated region manifests itself as flattening of 
the radial profile, beyond which it oscillates about a constant intensity 
level.  To be specific, we define the outer radius to be the first data point 
where the measured isophotal intensity rises instead of decreases 
monotonically. Typically the outer radius is large enough to avoid the spiral 
arms or other features that may cause a real rise in the outer brightness 
profile.  In the standard procedure of \citet{noovan07}, the average value of 
this isophotal intensity outside the outer radius and the associated standard 
deviation gives estimates of the sky level and 
its uncertainty.  However, this technique is reliable only if the background 
is uniform and well measured \citep{erwi08}.  The field-of-view of our images 
is typically only a factor of $\sim$2 larger than the galaxies, generally too
marginal to provide enough data points in the sky-dominated region to yield a 
statistically robust measurement of the background and its error.  The 
situation is further exacerbated by the occasional presence of residual 
large-scale non-uniformities in the background.  In view of these 
complications, we use Noordermeer \& van~der~Hulst's method only to determine 
the radius of the sky-dominated region (Figure~\ref{figure:skyrad}), which we 
record in the image header under the keyword {\tt SKY\_RAD}.  

To estimate the actual sky value and its associated error, we follow a method 
similar to that used by \citet{erwi08}.  We first smooth the original 
$2042\times2042$ pixel image by binning it down to a $102\times102$ pixel 
image.  This highlights underlying large-scale, systematic fluctuations in the 
background, which is the main factor that ultimately limits the accuracy with 
which 

\vskip 0.3cm
\psfig{file=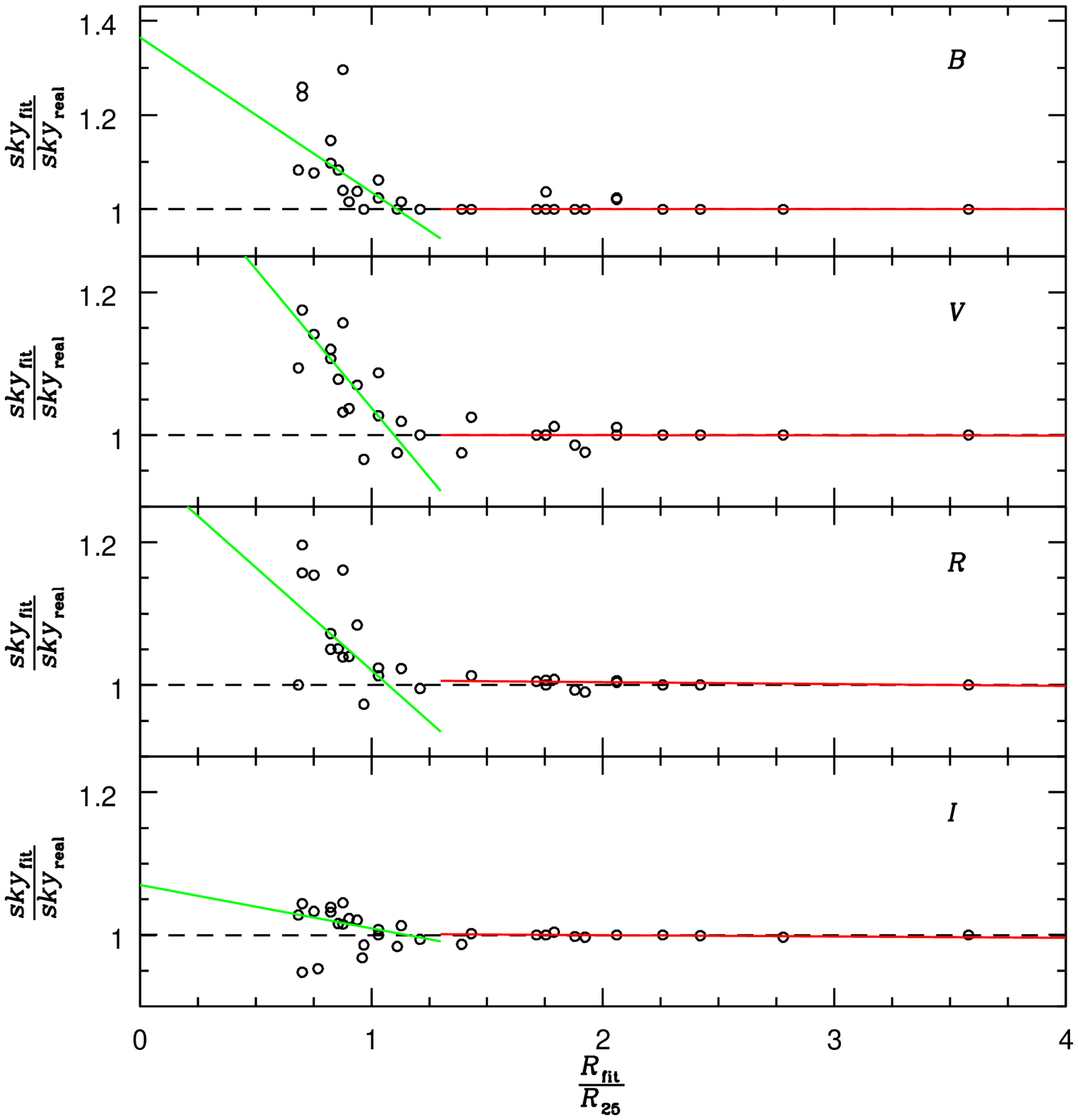,width=8.75cm,angle=0}
\figcaption[fig3.eps]{
Empirical relationship between sky$_{\rm fit}$/sky$_{\rm real}$ and
$R_{\rm fit} / R_{25}$, for each of the four filters.  For $R_{\rm fit}
/ R_{\rm 25}$ \gax\ 1.3, ${\rm sky}_{\rm fit} \approx {\rm sky}_{\rm real}$ as indicated
by the best-fitted red solid lines with slopes close to 0, but when
$R_{\rm fit} / R_{\rm 25}$ \lax\ 1.3, ${\rm sky}_{\rm fit}$ is systematically larger
than ${\rm sky}_{\rm real}$.  The green solid lines represent the best-fit linear
relations given by Equations (1)--(4).
(A color version of this figure is available in the online journal.)
\label{figure:sky_fit_empeqn}}
\vskip 0.3cm

\noindent
we can determine the sky level, and hence the final sensitivity of the 
surface brightness of our images.  The value of each binned pixel is the mean 
value of all the pixels inside a $20\times20$ pixel box, after excluding field 
stars and background galaxies identified in the image's object mask (Paper~I).
The background pixels, then, are defined to be all the pixels in the binned 
image outside of the isophotal ellipse marked by {\tt SKY\_RAD}. The sky level 
is simply the mean of the background pixel values, and its uncertainty is the 
standard deviation of the pixel flux distribution; these values are stored in 
the image header under the keywords {\tt SKY\_VAL} and {\tt SKY\_ERR}.  (The 
standard deviation of the mean, useful for other applications, is stored 
separately under the keyword {\tt SKY\_SIG}.) We have opted to compute the 
mean of the pixel flux distribution, rather than its median or mode, but in 
practice this makes little difference (they generally agree to $\sim$0.1\%) 
because the shape of the distribution is highly symmetrical.  This reflects 
the robustness of our estimate of the sky-dominated region and the 
effectiveness of our object masks in rejecting faint halos around foreground 
stars and background galaxies.  

We have tested the effect of choosing different scales for the smoothing, 
varying the binning box sizes from 5 to 50 pixels. While the average sky 
value remains stable, the width of the pixel distribution decreases with 
increasing smoothing length, leveling off to a near-constant value for 
box sizes \gax 20 pixels ($\sim$5\asec). We interpret this to represent the 
typical scale of large-scale systematic fluctuations in the sky background.
We choose a box size of $20\times20$ pixel as a reasonable compromise in order 
to retain a statistically significant number of data points to compute their 
average and standard deviation.  

Figure~\ref{figure:skybin} illustrates our method of sky estimation, using 
a $B$-band image of NGC~1400.  The image has been binned $20\times 20$, and 
the foreground stars and background galaxies have been masked out.  The sky 
value is the mean of pixel values outside of the ellipse, after excluding the 
masked objects, and the error is simply the standard deviation of the sky 
pixels in the binned image.  The right-hand panel shows normalized histograms 
of the sky pixels of the original and binned image.  Clearly they peak at 
nearly identical locations (the peaks of the two histograms differ by 0.0001 
${\rm ADU\ s^{-1}\ pixel^{-1}}$), but the distribution for the binned image is 
narrower than that of the original image by a factor of 10.

The above-described strategy for sky determination can only be applied to 
galaxies with angular diameters $D_{25}$ \lax\ 5\amin--6\amin.  For the 
$\sim$15\% of the sample more extended than this, it is difficult to 
impossible to determine the radius of the sky-dominated region and obtain 
robust statistics for the sky pixels, and we must resort to a more indirect 
approach.  The signal in the outer regions of the CCD frame consists of galaxy 
light plus a constant sky background.  Assuming that the galaxy component can 
be modeled by a single S\'ersic function, we can fit the observed light profile 
to solve for the underlying sky value.  We perform the fitting on the 1-D 
surface brightness profile (Section~4), after excluding the central regions of 
the galaxy and other features such as the bulge, the bar, or strong spiral 
arms, if present.  Simple experimentation shows that the best-fit sky value 
depends on the fitting radius relative to the size of the galaxy.  Clearly, if 
the fitting radius is large compared to the outer edge of the galaxy, the sky 
will be well determined; however, if the fitting radius lies substantially 
interior to the main body of the galaxy, the inferred sky value will depend 
critically on how well the S\'ersic model represents the intrinsic light 
profile of the galaxy.

We devise an empirical correction, as follows.  We select several galaxies 
that (1) have relatively simple structures, (2) are small compared 
to the CCD's field-of-view, and (3) have well-determined sky values.  Then, 
we fit their surface brightness profiles with different fitting radii 
($R_{\rm fit}$), to mimic the actual situation in galaxies that are too 
angularly extended to have a reliable sky determination.  The resulting 
fitted sky value, ${\rm sky}_{\rm fit}$, is then compared with the independently 
known, correct value ${\rm sky}_{\rm real}$.  Figure~\ref{figure:sky_fit_empeqn} 
shows ${\rm sky}_{\rm fit} / {\rm sky}_{\rm real}$ vs. $R_{\rm fit} / R_{25}$, where 
$R_{25} = 0.5\,D_{25}$.  We can see that ${\rm sky}_{\rm fit} / {\rm sky}_{\rm real} 
\approx 1$ when $R_{\rm fit} / R_{25}$ \gax\ 1.3.  When $R_{\rm fit} / R_{25} \leq 
1.3$, ${\rm sky}_{\rm fit}$ overestimates ${\rm sky}_{\rm real}$, but it does so 
systematically, in such a way that we can apply an approximate empirical 
correction to recover the true sky value.  The best-fitting linear 
relations (and their associated rms scatter), are s follows.

\begin{itemize}
\item {\it B}\ band
\begin{align}
\frac{{\rm sky}_{\rm fit}}{{\rm sky}_{\rm real}} &= 1.364 - 0.327 \times 
\frac{R_{\rm fit}}{R_{25}}, &\sigma = 0.05,
\end{align}

\item {\it V}\ band
\begin{align}
\frac{{\rm sky}_{\rm fit}}{{\rm sky}_{\rm real}} &= 1.427 - 0.389 \times 
\frac{R_{\rm fit}}{R_{25}}, &\sigma = 0.033,
\end{align}

\item {\it R}\ band
\begin{align}
\frac{{\rm sky}_{\rm fit}}{{\rm sky}_{\rm real}} &= 1.309 - 0.287 \times 
\frac{R_{\rm fit}}{R_{25}}, &\sigma = 0.042,
\end{align}

\item {\it I}\ band
\begin{align}
\frac{{\rm sky}_{\rm fit}}{{\rm sky}_{\rm real}} &= 1.070 - 0.060 \times 
\frac{R_{\rm fit}}{R_{25}}, &\sigma = 0.020.
\end{align}
\end{itemize}

\noindent  
The best-fit sky value, ${\rm sky}_{\rm fit}$, and its associated statistical 
error, $\sigma_{\rm fit}$, are recorded under the header keywords {\tt SKY\_VAL}
and {\tt SKY\_ERR} with the comment ``fitted sky value.''  If the above empirical 
correction to ${\rm sky}_{\rm fit}$ is necessary, we fold the scatter of the 
correction relation into the error budget.

\section{Isophotal Analysis}

\subsection{Geometric Parameters and Surface Brightness Profile}

The IRAF\footnote{IRAF is distributed by the National Optical Astronomy 
Observatory, which is operated by the Association of Universities for Research 
in Astronomy, Inc., under cooperative agreement with the National Science 
Foundation.} task {\em ellipse} is commonly used to measure the surface 
brightness profiles of galaxies \citep[e.g.,][]{silels94, miljog99, 
lain02, jogee04, agu05, marjog07, noovan07, bara08}. Following the iterative
method of \citet{jedr87}, we fit the isophotes of the
galaxy with a set of ellipses. This is motivated by the fact that the 
isophotes of most galaxies, especially early-type systems such as ellipticals 
and lenticulars, are quite close to ellipses\footnote{An important exception 
is edge-on galaxies, which are not well suited to ellipse fits.  As the
bulge and disk components have different profiles and ellipticities, their 
relative contributions change with radius and azimuth (i.e. along major or 
minor axes).  A dust lane, if present, also can strongly affect the averaged 
isophotal intensity.  For edge-on galaxies it is preferable to extract the 
isophotal intensities along cuts in the major and minor axis directions 
\citep[e.g.,][]{degr98, fry99, wu02}.}.  In our implementation, the ellipses 
are sampled along the semi-major axis of the galaxy in logarithmic intervals, 
starting from $r \approx $ 0\farcs3 and increasing the radius of each 
successive ellipse by a factor of 1.1.  After reaching the outermost ellipse, 
the fitting reverses direction and moves toward the galaxy center from 
$r \approx $ 0\farcs3, with each subsequent radius decreasing by a factor of 
1.1. 

As in \citet{noovan07}, we determine the isophotal geometric parameters of 
the galaxy in two steps.  In the first step, we estimate the center of the 
galaxy. Ellipses are fitted to the $I$-band image with the center, position 
angle (PA), and ellipticity ($e$) set as free parameters. We use the $I$-band 
image as the fiducial reference because of its relative insensitivity to 
dust extinction and young stars, and because it generally has the best seeing.
The center of the galaxy is the average central position of the ellipses 
inside $\sim$5\asec--7\asec.  In images of regular galaxies the center of the 
best-fit ellipses often converge to a well-defined value, with rms
$\approx$ 0\farcs015. However, in galaxies with dusty nuclear regions, 
the best-fitting central isophotes may give a poor measure of the true 
center.  In such cases, a better estimate of the true center comes from 
isophotes at intermediate radii, $\sim$10\asec--30\asec, far enough to be 
undisturbed by central dust but yet sufficiently close to the nucleus to give 
a faithful measure of its position.  The typical uncertainty of the 
central positions estimated in this way is $\sim$0\farcs02. 

Next, we fix the center just determined and run {\em ellipse} again,
while still setting $e$ and PA free. Our goal is to determine the 
characteristic $e$ and PA of the galaxy based on its best-fit isophotes. 
Typically we take the average value of these parameters in the outer regions 
of the galaxy, where the intensity is about $1 \,\sigma$ above the sky, as 
their characteristic values and use their standard deviations over that 
region as the uncertainties. These parameters usually converge to a constant 
value within that region, with variations of $\sim$0.04 for $e$ and 
$\sim$2\deg\ for PA.  However, the intrinsic geometric parameters of some 
galaxies can be distorted by mergers or interactions, causing $e$ and PA 
to diverge at large radii.  In these cases we simply estimate their values 
manually from the visually best-fitting isophotes near the edge of the galaxy.
The $B$, $V$, $R$, and $I$ images have their center, $e$, and PA values 
determined independently, and they are stored in their corresponding image 
headers.

During this second step, we also record the deviations of the isophotes from 
perfect ellipses, which, as described in \citet{jedr87}, are parameterized by 
the third ($A_3, B_3$) and fourth ($A_4, B_4$) harmonics of the 
intensity distribution.  The $A_3$ and $B_3$ parameters give
``egg-shaped'' or ``heart-shaped'' isophotes \citep{cart78, jedr87}.
\citet{pele90} point out that $A_3$ and $B_3$ appear to be sensitive
diagnostics of dust features in elliptical galaxies. The most interesting
parameter among them is $B_4$: if it is positive, the underlying isophote is
disky with respect to a perfect ellipse; a negative $B_4$ corresponds to a
boxy isophote. Figure~1 of \citet{pele90} gives examples of the different
isophotal shapes for different values of $B_3, A_4$, and $B_4$.

We run {\em ellipse} for a third and final time to extract the average 
intensity of the isophotes, fixing the geometric parameters to the values 
determined above \citep[e.g.,][]{potr06, noovan07}.  For this step, we do not 
allow the geometric parameters to vary in order to reduce the influence of 
bars and other non-axisymmetric features on the average intensity profile, as 
well as to reach convergence in regions where the signal-to-noise ratio (S/N) 
is marginal \citep {erwi08}.  For consistency, we apply this isophote 
measurement to all the galaxies in our sample.  For the lopsided galaxies, we 
also experimented with allowing the isophotal centers to be left as free 
parameters.  We find that the typical difference in the brightness profile, 
compared with the fits based on fixed isophotal centers, is $\sim$ 0.2 $\rm 
mag\ arcsec^{-2}$. Moreover, our tests show that the relative amplitudes of 
the Fourier terms (Section 4.2) decrease significantly when the isophotal
centers are allowed to be free, due to the fact that the free-fitting ellipses 
can better trace the distorted disk in the outer part of the galaxy to produce 
very small fluctuations along each isophote.  This will make the Fourier 
analysis less effective for detecting and quantifying the properties of bars 
or lopsided structures.

After subtracting the sky background from the image, the surface brightness 
is calculated from

\begin{eqnarray}
\mu = -2.5 \log \left( \frac{I_{\rm iso}}{t_{\rm exp} \times A} 
\right) + {\rm zpt},
\end{eqnarray}

\noindent
where $I_{\rm iso}$ is the isophotal intensity after subtracting the sky, 
$t_{\rm exp}$ is the exposure time of the image in units of seconds, $A$ is 
the pixel area in units of $\rm arcsec^2$, and ${\rm zpt}$ is the photometric 
zero point of the image in units of magnitudes.  We calculate the surface 
brightness only from those isophotes whose intensities are larger than 
$I_{\rm sky} + \sigma_{\rm sky}$.   We propagate the errors on $I_{\rm iso}$ 
into errors on $\mu$ in magnitude units.  The surface brightness profiles in 
the $B, V,$ and $R$ bands are constrained to have the same geometric 
parameters as determined in the $I$ band.  

To construct 1-D color profiles, we blur all the images to a common seeing. 
This is done by convolving the image having the better seeing with a 
two-dimensional (2-D) Gaussian function whose full width at half maximum 
(FWHM) is the quadrature difference between the two seeing values.  The 
measured isophotal ellipses of the unblurred $I$-band image are used to 
directly calculate the isophotal intensity of the blurred images in all the 
filters.  Color profiles follow from straightforward differencing of one band 
from another.

\begin{figure*}[t]
\centerline{\psfig{file=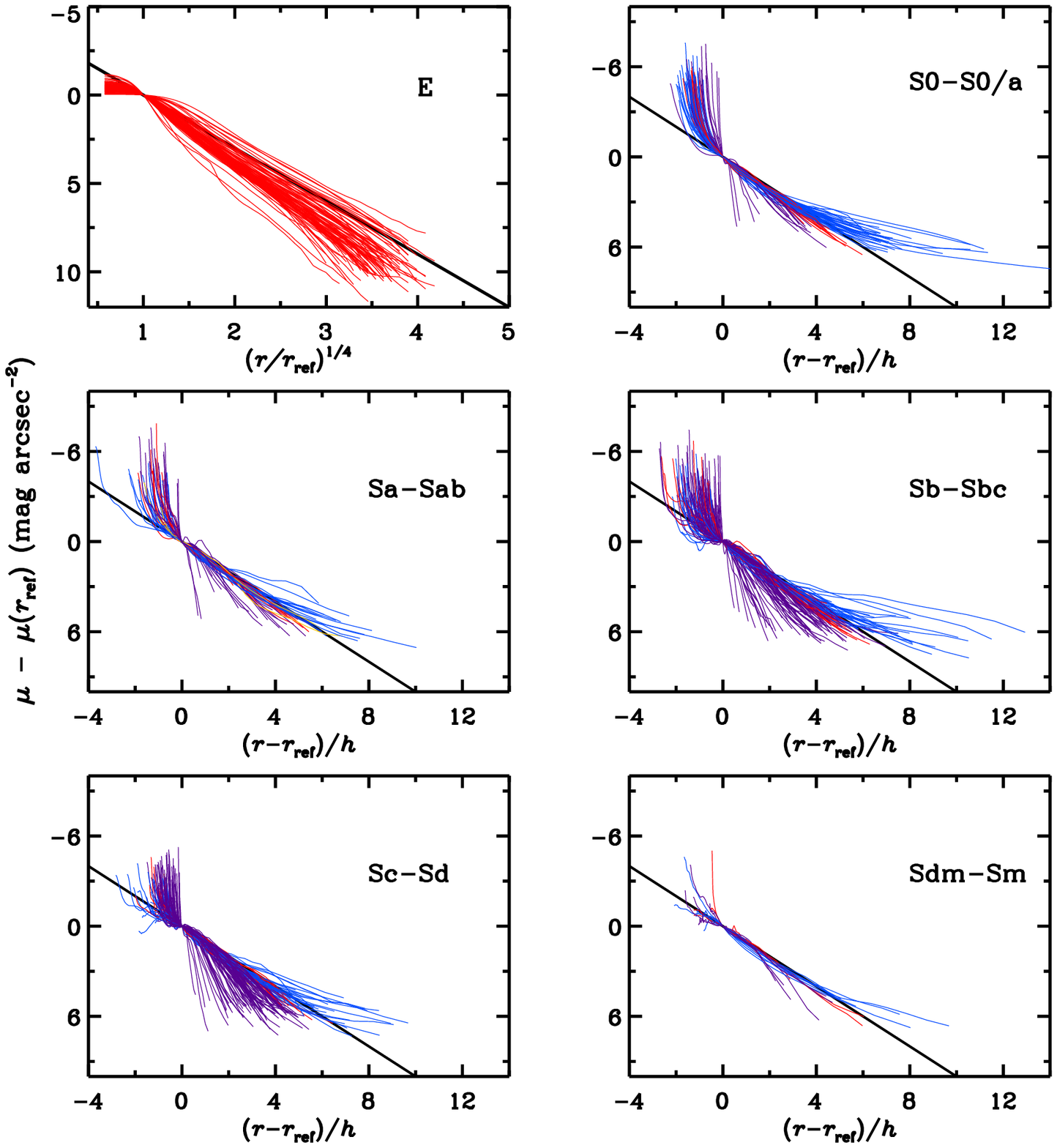,width=17.0cm,angle=0}}
\figcaption[fig4.eps]{
$B$-band composite profiles of our sample, divided by morphological
type. The elliptical galaxies are normalized at a reference radius of
$r_{\rm ref}$ = $R_{20}$, and the profiles are plotted vs.
$(r/r_{\rm ref})^{1/4}$.  The thick black line corresponds to a de~Vaucouleurs
$r^{1/4}$ law.  The profiles for the disk galaxies are scaled according to the
scale length $h$ of the disk outside $r_{\rm ref}$.  Profiles of Type I, II,
and III are marked in red, purple, and blue, respectively.  The thick black
line corresponds to a pure exponential function. 
(A color version of this figure is available in the online journal.)
\label{figure:compprofb}}
\end{figure*}

\begin{figure*}[t]
\centerline{\psfig{file=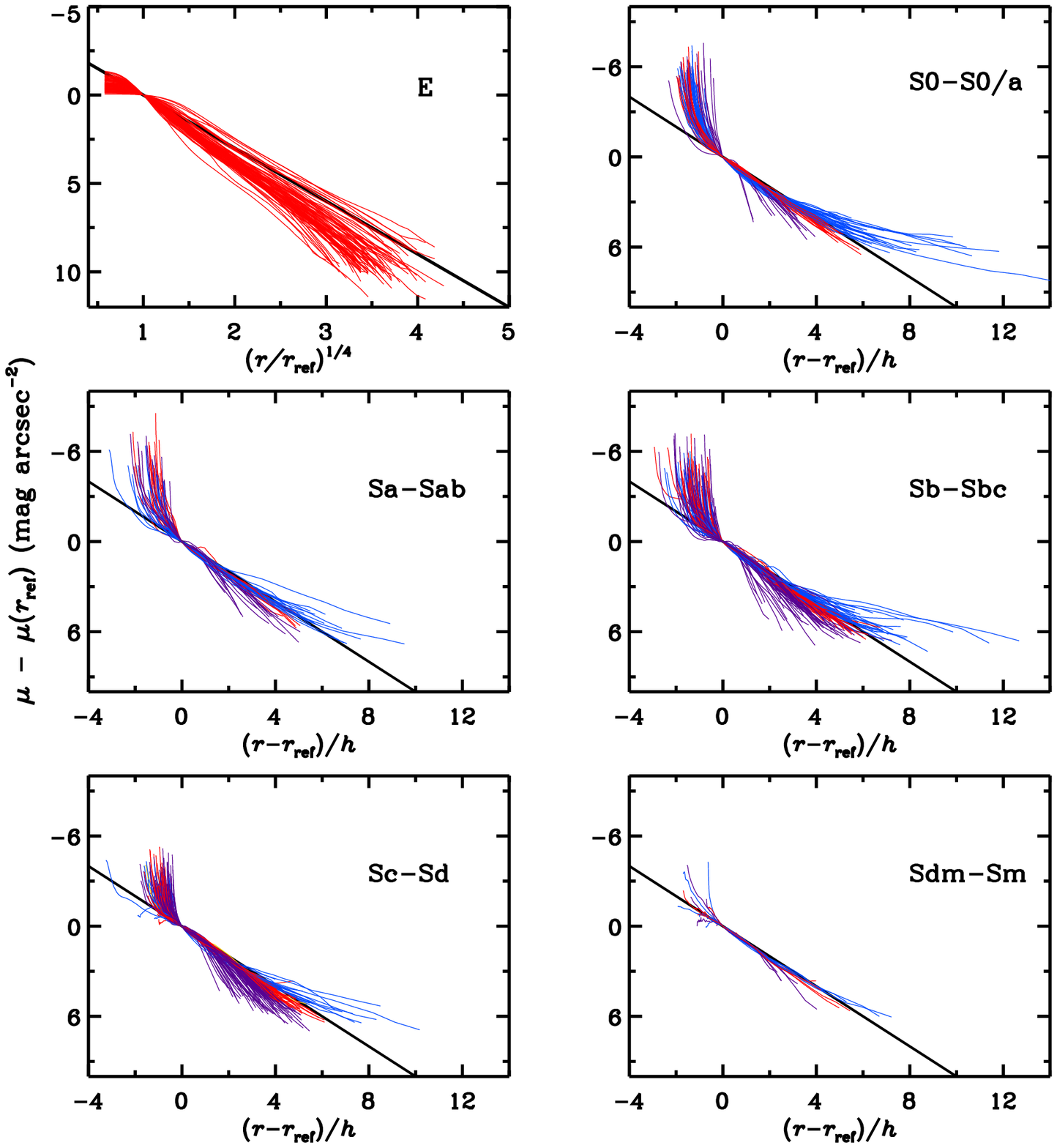,width=17.0cm,angle=0}}
\figcaption[fig5.eps]{
$I$-band composite profiles of our sample. See Figure~\ref{figure:compprofb}
for details. 
(A color version of this figure is available in the online journal.)
\label{figure:compprofi}}
\end{figure*}


\subsection{Fourier Analysis}

We study the harmonic components of the intensity distribution of the
isophotes.  In our work, we decompose the intensity distribution along each
ellipse into a Fourier series of the form

\begin{eqnarray}
I(\theta) = I_{0} + \sum_{j=1}^{\infty} I_{j} \cos j(\theta + \phi_{j}),
\end{eqnarray}

\noindent
where $I$ is the intensity (in units of ADU~s$^{-1}$~pixel$^{-1}$) on the 
ellipse in the direction $\theta$, $I_{0}$ is the average intensity of the 
ellipse, $I_{j}$ measures the strength of the $j$th mode in the series, and 
$\phi_{j}$ is the corresponding phase angle of that mode. The angle $\theta$ 
is defined to be 0\deg\ along the positive $y$-axis and increases 
counterclockwise; $\phi_{j} = 0$\deg\ along the positive $y$-axis and 
increases clockwise.  A high S/N is required to derive a statistically 
significant measurement of the high-order Fourier terms \citep{noovan07}; 
thus, we only perform this decomposition inside the radius where the average 
intensity of the isophote is $3 \,\sigma_{\rm sky}$ larger than the determined 
sky value.  The relative amplitude of the $m$ = 1 ($I_{1}/I_{0}$) and $m = 2$ 
($I_{2}/I_{0}$) mode will be especially useful in our analysis, as they 
reflect the lopsidedness of the galaxy \citep{rixzar95} and the strength of 
the bar or spiral arms \citep{buta86}, respectively.

\citet{buta86} performed a Fourier analysis to study the azimuthal variations 
of the light distribution of NGC~1433, which is also included in the CGS 
sample.  We found good agreement between Buta's values of the relative 
amplitudes of the $m$ = 1 and 2 modes and those calculated by us.  This helps
to confirm the robustness of our method.

\subsection{Database of 1-D Profiles}

The full database of isophotal parameters for the 605 galaxies in CGS
(including the 11 extras not formally part of the survey) is given in the
Appendix, as Figures 19.1-19.616, as well as on the project  Web site
{\tt http://cgs.obs.carnegiescience.edu}.

\section{Composite Profiles}

Composite profiles can help to highlight characteristic statistical trends, 
as well as to isolate interesting outliers, in a class of objects.  We 
normalize the surface brightness profiles to a common reference radius.  For 
the elliptical galaxies in our sample, we set $r_{\rm ref}$ to $R_{20}$, 
the radius wherein 20\% of the total flux is enclosed, and plot the profiles 
as a function of $(r/r_{\rm ref})^{1/4}$.  In this reference frame, a 
classical de~Vaucouleurs $r^{1/4}$ profile traces a 

\begin{figure*}[t]
\centerline{\psfig{file=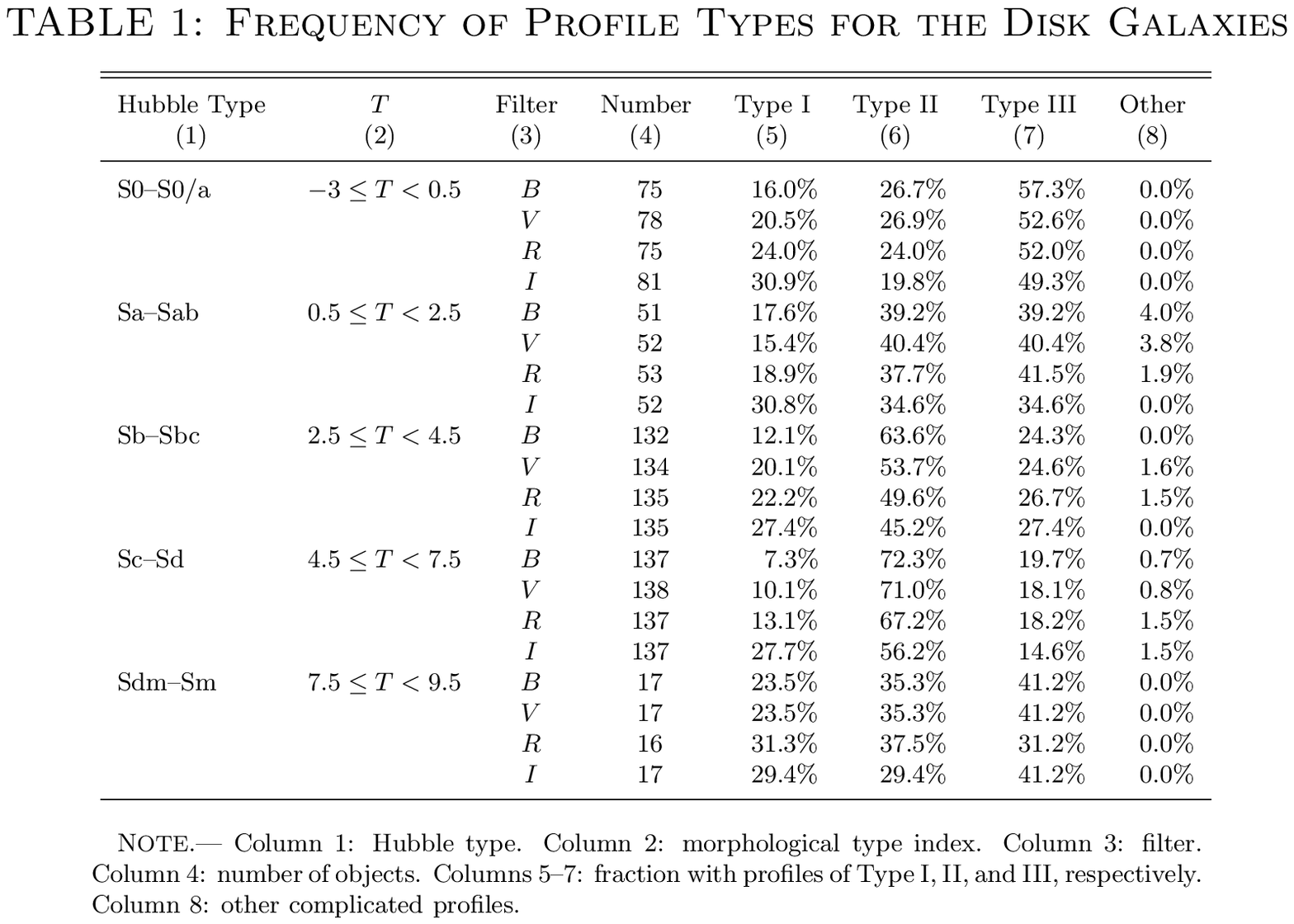,width=16.5cm,angle=0}}
\end{figure*}

\noindent
straight line.
Figure~\ref{figure:compprofb} (top left panel) illustrates 
the now-known fact that not all ellipticals obey the $r^{1/4}$ law, but rather
are better described by a more general S\'ersic function with $r^{1/n}$.  

For the disk (S0 and spirals) galaxies, we set $r_{\rm ref}$ to be roughly the 
boundary between the bulge and disk, and we plot the brightness profiles as a 
function of  $(r - r_{\rm ref})/h$, where $h$ is the scale length of the disk, 
as determined by fitting an exponential function to the profile outside of 
$r_{\rm ref}$.  This choice of coordinates helps to reveal possible deviations 
of the disks from a canonical exponential profile ($n = 1$), which appears as 
a straight line.  It is apparent that the light distributions of the disk very 
rarely follow a pure exponential function (Figure~\ref{figure:compprofb}), 
especially in their outer regions.  Most show a downward turn compared to a
single exponential, but not an insignificant number show an upward turn.  This 
phenomenon is well-known \citep[e.g.,][]{phil91, erwi05, erbp05, potr06}.  We 
code the three profile types with different colors, with Type~I profiles (no 
break) in red, Type~II profiles (downward break) in purple, and Type~III 
profiles (upturn) in blue.  The scatter among the normalized profiles is 
smaller in the redder bands, indicating that dust extinction and young stars 
have a greater effect on the profile shapes in the blue.  
Figure~\ref{figure:compprofi} shows the equivalent montage for the $I$ band.

Table~1 tabulates the frequency of the different 
profiles for each bin of morphological type for the disk galaxies. 
Occasionally, there are some galaxies with complicated surface brightness 
profiles, which cannot be classified as any of the three standard types listed 
above. The profiles for such objects are listed as ``Other'' in 
Table~1\footnote{
Since the brightness profile in the outer regions of the galaxy depends 
sensitively on the accuracy of the sky subtraction, we exclude objects 
with unreliable sky determination.  We also omit galaxies whose light 
distribution is severely adversely affected by very bright foreground stars, 
by excessively crowded field stars, or by an interacting neighbor.  The 
excluded objects are flagged in Table~2. Note that a star that is bright 
and excluded in one filter may not be equally bright or rejected in another; 
therefore, the number of objects in a morphological bin is not the same 
among all the filters.  We further omit the 11 extra galaxies that do not 
formally meet the CGS selection criteria.}.
Type~II and Type~III profiles are common in our sample, and their fractions 
depend on the galaxy morphology.  Type~II profiles occur more frequently in 
late-type disk galaxies ($\sim 70\%$ among Sc--Sd spirals), whereas Type~III 
profiles are preferentially found in more bulge-dominated, earlier-type 
systems, especially 
among the S0--S0/a class ($\sim 50\%-60\%$).  The fraction of galaxies with 
Type~I profiles, on the other hand, seems to be roughly constant, at 
$\sim 20\%$, across all morphological types; the fraction increases 
systematically toward redder bandpasses, except for the Sdm--Sm galaxies, 
where the fraction seems to be roughly constant, although the number of 
objects is small. A detailed analysis of the different profile types and their 
dependence on other physical parameters will be presented in a separate paper.

\section{Color Information}

Table~2 presents integrated colors and color gradients for 
the sample, corrected for foreground Galactic extinction using values from 
\citet{schl98}.  We list $B-I$, $V-I$, and $R-I$, from which other color 
combinations can be readily derived; we use total magnitudes within the last 
reliable isophote (1 $\sigma$ above the sky), as given in Table~4 of Paper~I. 
For each of 

\begin{figure*}[t]
\centerline{\psfig{file=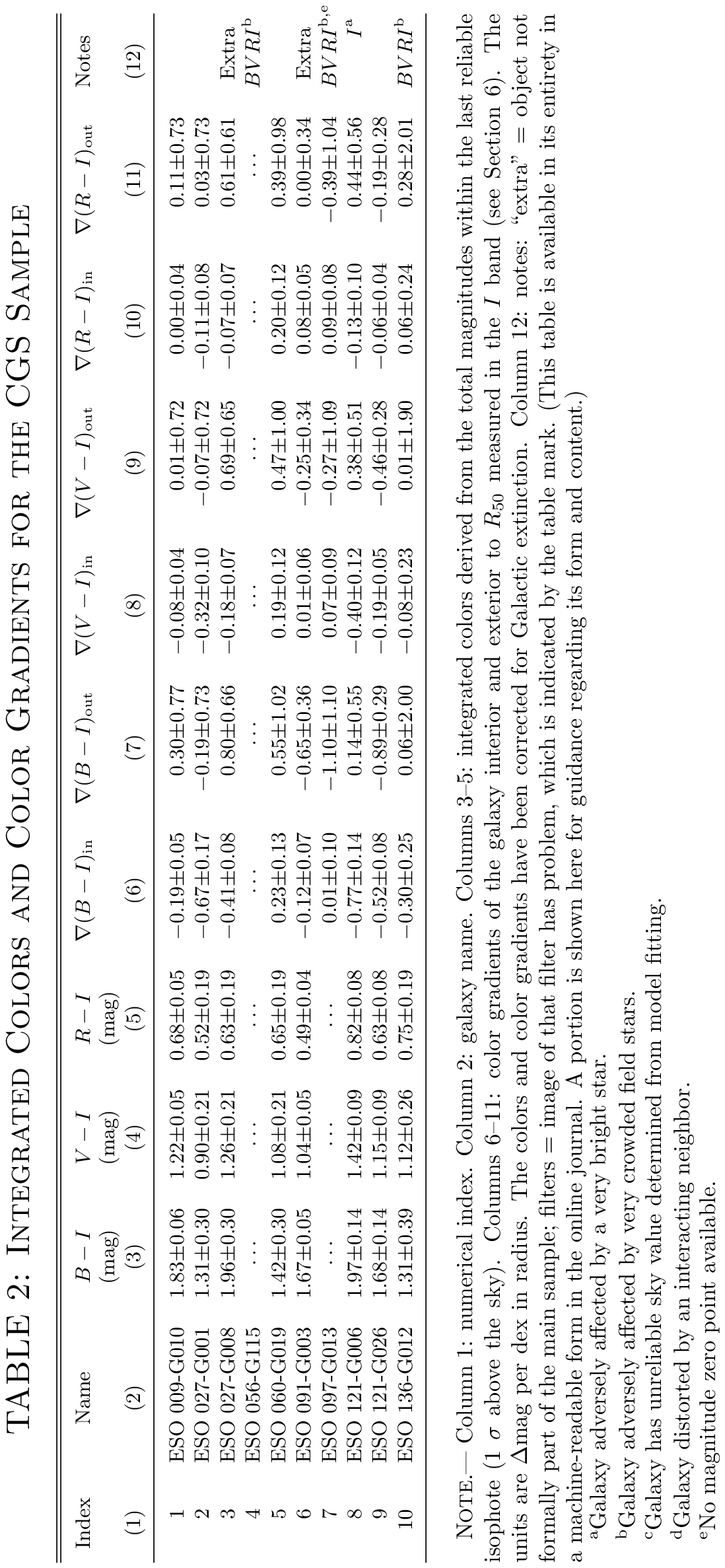,height=10.5in,angle=180}}
\end{figure*}

\begin{figure*}[t]
\centerline{\psfig{file=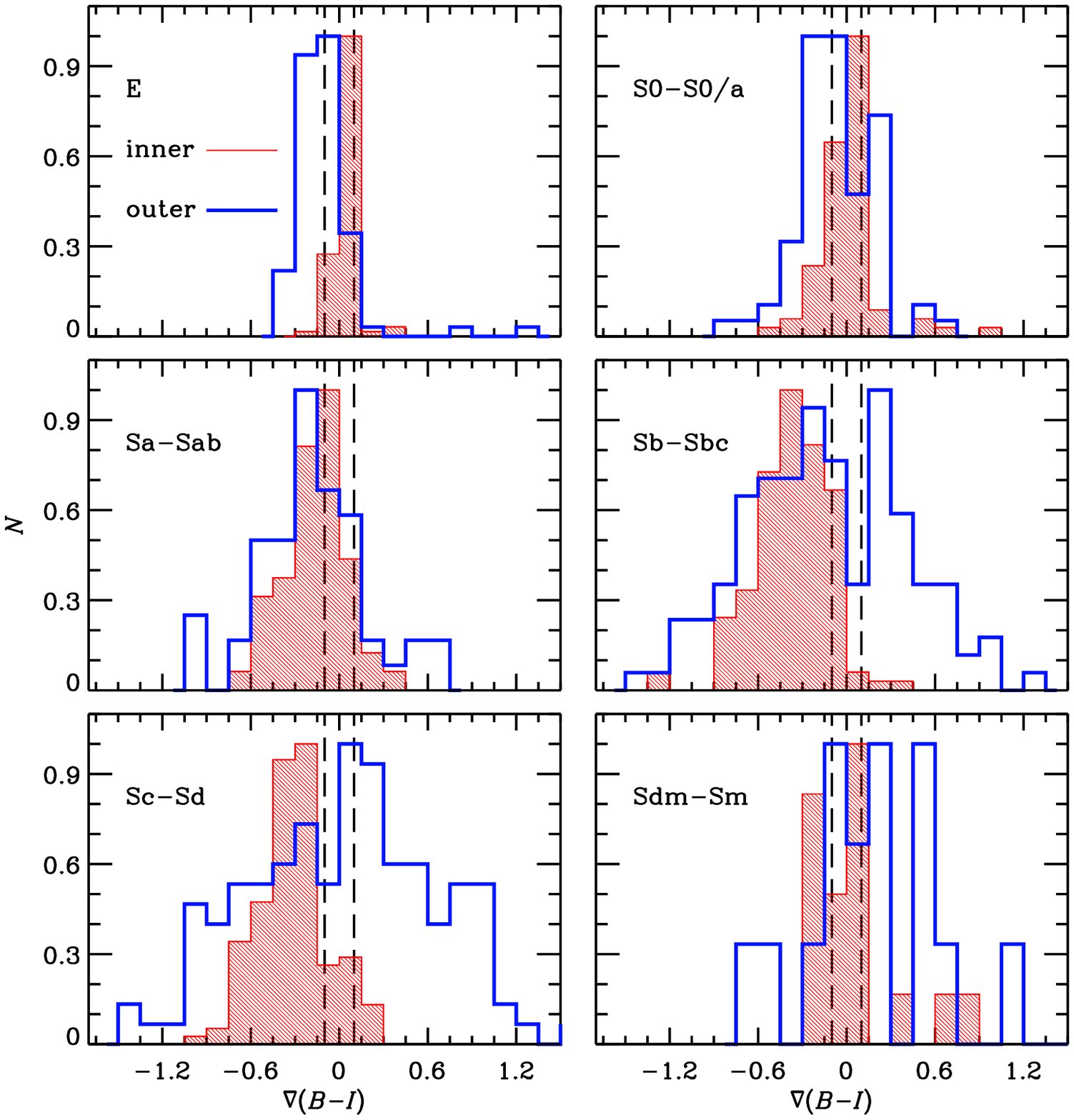,width=17.0cm,angle=0}}
\figcaption[fig6.eps]{
Normalized histograms of the inner and outer $B-I$ color gradients,
divided by morphological type. The inner and outer color gradients are
represented in red and blue solid histograms, respectively. The
vertical dashed lines in each panel mark the adopted boundaries for negative
($\nabla(B-I) < -0.1$), flat ($-0.1 \leq \nabla(B-I) \leq 0.1$), and positive
($\nabla(B-I) > 0.1$) color gradients. A positive color gradient means that the
color becomes redder outward, while a negative value indicates that the color
gets bluer outward.
(A color version of this figure is available in the online journal.)
\label{figure:cghistbi}}
\end{figure*}

\begin{figure*}[t]
\centerline{\psfig{file=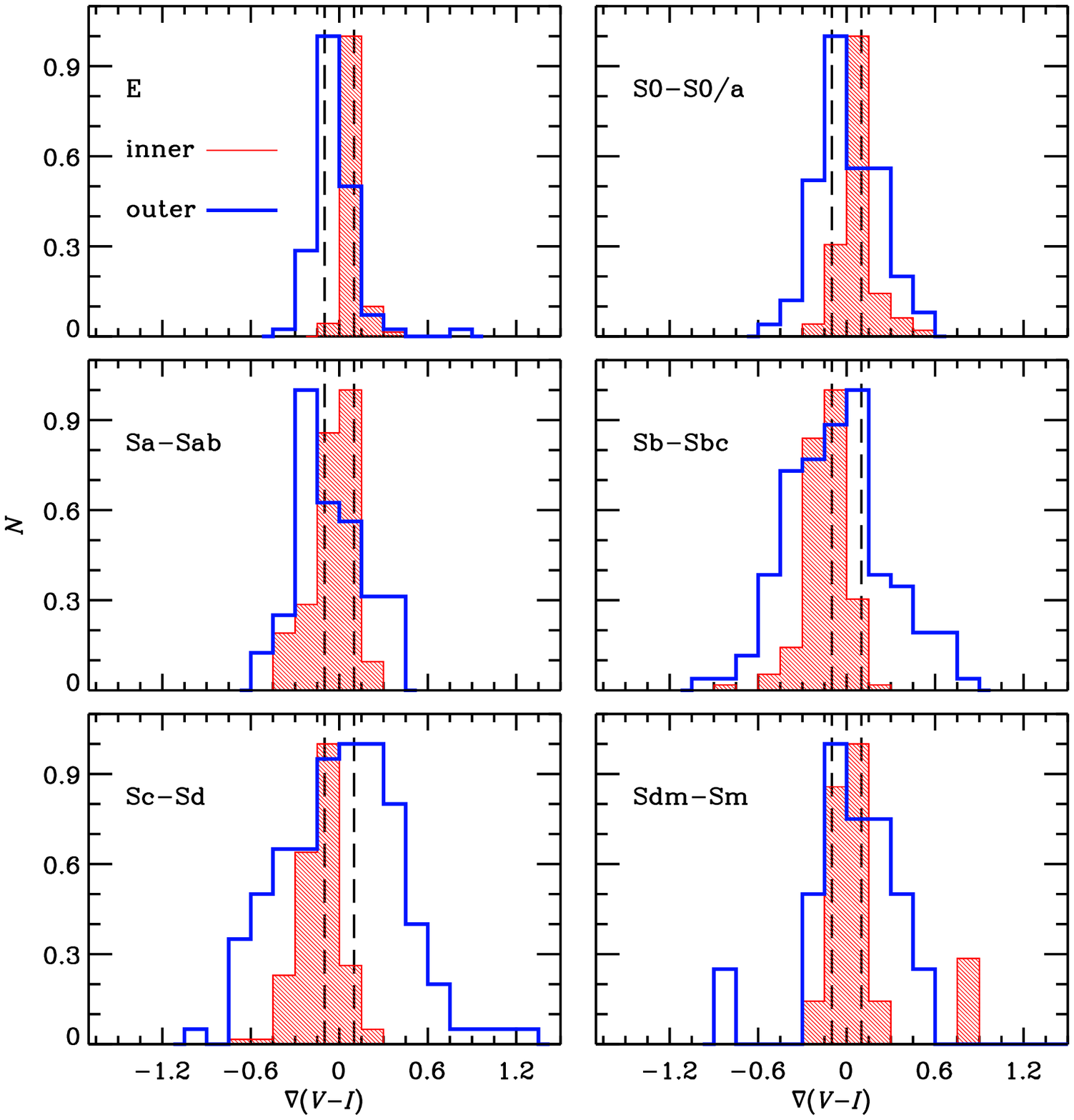,width=17.0cm,angle=0}}
\figcaption[fig7.eps]{
Normalized histograms of the inner and outer $V-I$ color gradients.
See Figure~\ref{figure:cghistbi} for details.
(A color version of this figure is available in the online journal.)
\label{figure:cghistvi}}
\end{figure*}

\begin{figure*}[t]
\centerline{\psfig{file=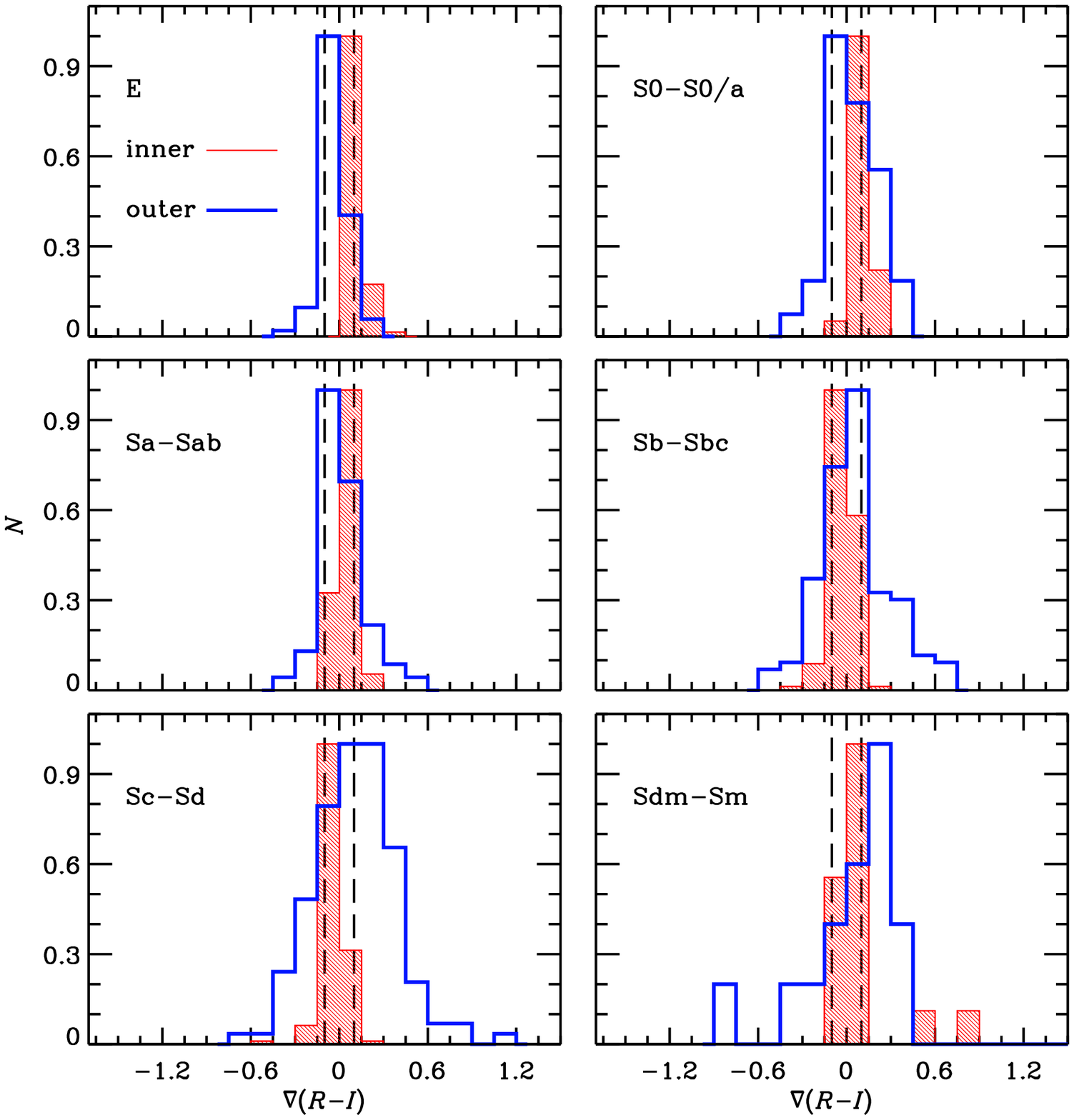,width=17.0cm,angle=0}}
\figcaption[fig8.eps]{
Normalized histograms of the inner and outer $R-I$ color gradients.
See Figure~\ref{figure:cghistbi} for details.
(A color version of this figure is available in the online journal.)
}
\label{figure:cghistri}
\end{figure*}
\clearpage
\begin{figure*}[t]
\centerline{\psfig{file=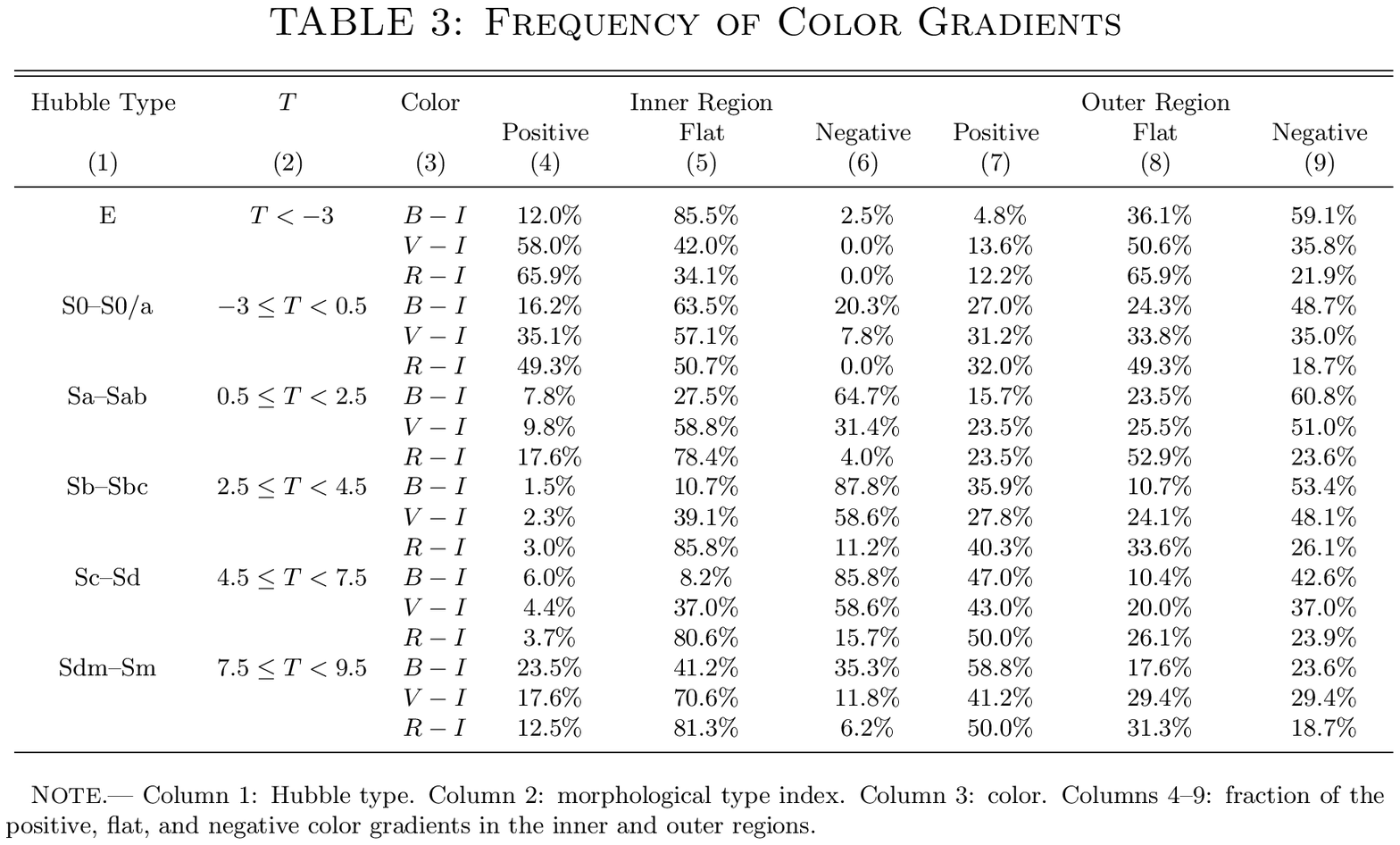,width=16.5cm,angle=0}}
\end{figure*}

\noindent
these color combinations, we also calculate two simple measures
of the color gradient, after resampling the color profiles with 300 equally
spaced data points in linear space, to overcome the heavily sampled points in
the central region.
Similar to \citet{tayl05}, the color gradient is derived 
both inside and outside the half-light radius, $R_{50}$, as determined in the 
$I$ band.  The inner region ranges from 3 times the seeing FWHM (to avoid 
spurious effects from seeing mismatches) to $R_{50}$, while the outer region 
extends between $R_{50}$ and $2.5\ R_{50}$.  Tests show that most of the color 
profiles of the CGS galaxies can reach further than $2.5\ R_{50}$. We confine 
the color gradient into specific radial ranges marked by the three anchor 
points for each galaxy\footnote{In the future, we will also derive the color 
gradient in physically interesting regions, such as spiral arms, bars, or the 
break points of the surface brightness profiles.}.
The resulting color profile slope represents the change in color ($\Delta$ mag) 
per dex in radius, where a positive slope indicates that the galaxy is getting redder 
with increasing radius from its center.  

We implement a Monte Carlo method to compute the 
color gradients and their uncertainties.  The essence of our approach is to 
iteratively sample the two observables, the radius and the color at that 
radius, while incorporating uncertainties from Poisson noise, sky subtraction, 
radius measurement, and stochastic fluctuations in the color profile (due to,
for instance, substructure from dust lanes, star clusters, or spiral arms). 
Except for Poisson noise, which is random, the other uncertainties generally 
contribute to errors in a systematic way. Our Monte Carlo approach makes it 
possible to derive an effective random uncertainty from a series of systematic 
errors.  We start with the premise that the effective radius is uncertain by a 
Gaussian error with $\sigma \approx 0.1 R_{\rm 50}$, which is frequently true 
when the data are not in the noise-dominated regime. The systematic 
uncertainty on $R_{\rm 50}$ arises from the fact that there is a range of 
plausible, acceptable models that scatter around the best-fitting solution.
We draw a radius from that distribution, and the color at that corresponding 
radius. The color value is also sampled from a Gaussian distribution given by 
the Poisson noise, centered on the original color at the sampling radius. The 
net random uncertainty of the color includes uncertainty in the sky value. 
The same sampling process is applied at $2.5 R_{\rm 50}$, now with an 
effective scatter of $2.5 \, \sigma (R_{\rm 50})$, and in the central 
region with a scatter of the seeing FWHM. Having sampled 
the radii and colors around $R_0$, $R_{\rm 50}$, and 
$2.5 R_{\rm 50}$, we calculate the color gradient following

\begin{eqnarray}
\nabla({\rm Color})_{\rm in/out} = \frac{({\rm Color})_{r_1} - 
({\rm Color})_{r_0}}{\log{r_1} - \log{r_0}}.
\end{eqnarray} 

\noindent
This procedure is repeated 10$^4$ times to generate a 
distribution of color, the median and width of which are the gradient and the
uncertainty in the gradient, respectively.

Figure~\ref{figure:cghistbi} shows the normalized histograms of the inner and 
outer gradients of the $B-I$ color, sorted into bins of different 
morphological types; the corresponding gradients for the $V-I$ and $R-I$ 
colors are shown in Figures~\ref{figure:cghistvi} and 
\ref{figure:cghistri}, respectively.  We define positive gradients 
as those larger than 0.1, flat gradients as those between $-$0.1 and 0.1, and 
negative ones as those smaller than $-$0.1. The vertical dashed lines in each 
panel mark the boundaries between these three categories.  

Table~3 summarizes the statistics for CGS\footnote{We 
excluded the galaxies flagged in Table~2 as adversely affected by field stars.}.
The color 
gradients depend strongly on galaxy morphology and color. Elliptical galaxies 
generally have very little, if any, measurable color gradients.  Interestingly,
their central regions show a slight tendency to exhibit {\it positive}\  
gradients in all three colors, whereas beyond their effective radii the trend 
reverses and there is a mild preference for negative gradients.  In either 
case, the distribution of gradients is narrowly peaked, with a dispersion 
of $\sim 0.11$.  S0 and S0/a galaxies largely follow the same pattern as the 
ellipticals.  By contrast, spirals of types Sa through Sd behave quite 
differently.  The inner regions of these galaxies show a wide dispersion 
in gradients ($\sim 0.23$), and they are predominantly negative: the colors 
get redder toward the center.  The gradients in the outer regions, on the 
other hand, are predominantly flat (peak near 0), and there is roughly 
an equal number of positive and negative values, although the scatter is
large.  Galaxies belonging to the latest types (Sdm and Sm) 
display no preference for gradients of either sign, neither in their interior 
nor in their exterior regions.  The above trends stand out most clearly in 
$B-I$, the color combination with the greatest wavelength separation, and they 
become less pronounced in $V-I$, and even more so in $R-I$, although they are 
still noticeable.

\section{Bars}

Two fundamental quantities that characterize a bar are its length and strength.
There are many ways to estimate the characteristic size of a bar.  Apart from
simple visual inspection \citep{korm79}, the most commonly used approaches 
involve measurement of the maximum value of the bar ellipticity 
\citep[e.g.,][]{lain02, mene07}, the radial variation of the position angle 
\citep[e.g.,][]{erwi05, mene07}, 
the radial variation of the phase angle of the second Fourier mode 
\citep{ague03}, and the bar-interbar contrast \citep{ague03}, as well as 
direct decomposition of the image into different components \citep{prie97}. 
The strength of the bar can be ascertained by quantifying the maximum 
ellipticity in the bar region \citep{mart95, mafr97, mene07}, the torques generated by 
the bar \citep{bubl01}, the amplitude of the even Fourier modes of the 
isophotal intensity distribution \citep{ohta90, atmi02}, and direct 
decomposition of the galaxy into its constituent light fractions 
\citep{laurikainen05, gadotti08, peng10}.  Here we present a preliminary 
appraisal of the bar properties of the CGS sample based on information that 
can be readily extracted from our 1-D isophotal data.  We describe analyses 
based on the geometric parameters and Fourier components.

\subsection{Geometric Analysis}

In the absence of confusion from dust, star-forming regions, and projection 
effects, bars usually leave a distinctive imprint on the $e$ and PA profile of 
a galaxy.  The bar is marked by a region wherein the ellipticity rises 
steadily until it reaches a peak and drops, and, unless the bar semi-major 
axis is closely aligned with the major axis of the 
disk, the constant position angle in the bar region abruptly changes value 
as it transitions into the disk region \citep{gadotti07}.

We begin with the $e$ and PA profiles of the $I$-band image, as extracted from 
the second step of running the task {\em ellipse} (Section 4.1), during 
which only the galaxy center was held fixed. The $I$ band is preferred over 
the other bands at shorter wavelengths because it mitigates contamination by 
dust and young stars.  We consider the profiles from an inner radius 
corresponding to 3 times the seeing disk to the radius where the isophotal 
intensity is $1 \, \sigma$ above the sky background, beyond which we truncate 
the surface brightness profile.  Similar to \citet{mene07} and \citet{ague09}, 
bars are required to have a maximum projected ellipticity ($e_{\rm max}$) 
greater than 0.2, and within the bar region the position angle should be 
constant to within $\Delta\rm PA<20^\circ$.  If none of the data points in the 
$e$ profile exceeds 0.2, or if $\Delta e \leq 0.1$ throughout the entire $e$ 
profile, we classify the galaxy as unbarred.  If $\Delta e > 0.1$ somewhere 
along the $e$ profile but the associated $\Delta$PA $\leq 10^\degree$, it is 
possible that a bar exists but happens to align fortuitously with the major 
axis of the outer disk. We flag these cases as ``possibly'' barred and 
carefully inspect the galaxy image visually to see if we can confirm their 
reality.  If the galaxy is barred, we set the inner boundary of the bar region 
to be the first data point in the $e$ profile that exceeds 0.2. \citet{mene07} 
find that near the end of the bar the ellipticity and position angle usually 
begin to show large deviations, typically at the level of $\Delta e \geq 0.1$ 
and $\Delta$PA $\geq 10^\degree$.  We adopt these criteria to define the 
radius of the outer boundary of the bar.

The projected bar size is set to be the semi-major axis of the isophote where 
$e$ peaks \citep{mene07}, with the associated error as the semi-major axis 
range containing the tip of the $e$ profile (i.e. $e \geq e_{\rm max}-0.01$)
in the bar region. The reason we do not simply equate the bar size with the 
outer boundary of the bar is because the change in $e$ and PA in the
transition zone between the bar and spiral arms can be influenced by the 
latter.  The outer boundary of the bar can be overestimated if the bar is 
aligned with the spiral arms. This effect can be mitigated by using the 
position where $e$ peaks as the bar size, since spiral features cannot 
produce ellipticity values as high as those of a bar.  Assuming that the 
intrinsic shape of the galaxy disk is purely circular, we correct the 
observed, projected bar length ($R^{o}_{\rm bar}$) to its intrinsic value 
($R^{i}_{\rm bar}$) following 

\begin{eqnarray}
R^{i}_{\rm bar} = R^{o}_{\rm bar}\sqrt{(\cos{\Delta{\rm PA)}}^2 
+ \left(\frac{\sin{\Delta{\rm PA}}}{1 - e_{\rm gal}}\right)^2},
\end{eqnarray}

\bigskip
\noindent
where $\Delta{\rm PA} = {\rm PA}_{\rm gal} - {\rm PA}_{\rm bar}$, and
${\rm PA}_{\rm gal}$ and $e_{\rm gal}$ are the position angle and ellipticity 
of the outer disk of the galaxy, respectively. Although a full 2-D analytical deprojection 
of the bar is more accurate \citep[see Appendix A in][]{gadotti07}, in 
practice the size estimates from the two methods agree very well.  The typical 
difference in bar radii measured by 1-D fitting compared to 2-D deprojection
is about $0\farcs3$. Since the difference is very small, we use the 1-D method
for simplicity. The projected $e_{\rm max}$ then represents 
the strength of the bar \citep{mene07}, with the fitted error of $e_{\rm max}$ 
as its uncertainty. The position angle of the bar, ${\rm PA}_{\rm bar}$, is 
given by the average PA over the bar region, with the rms as its error. 

\begin{figure*}[t]
\centerline{\psfig{file=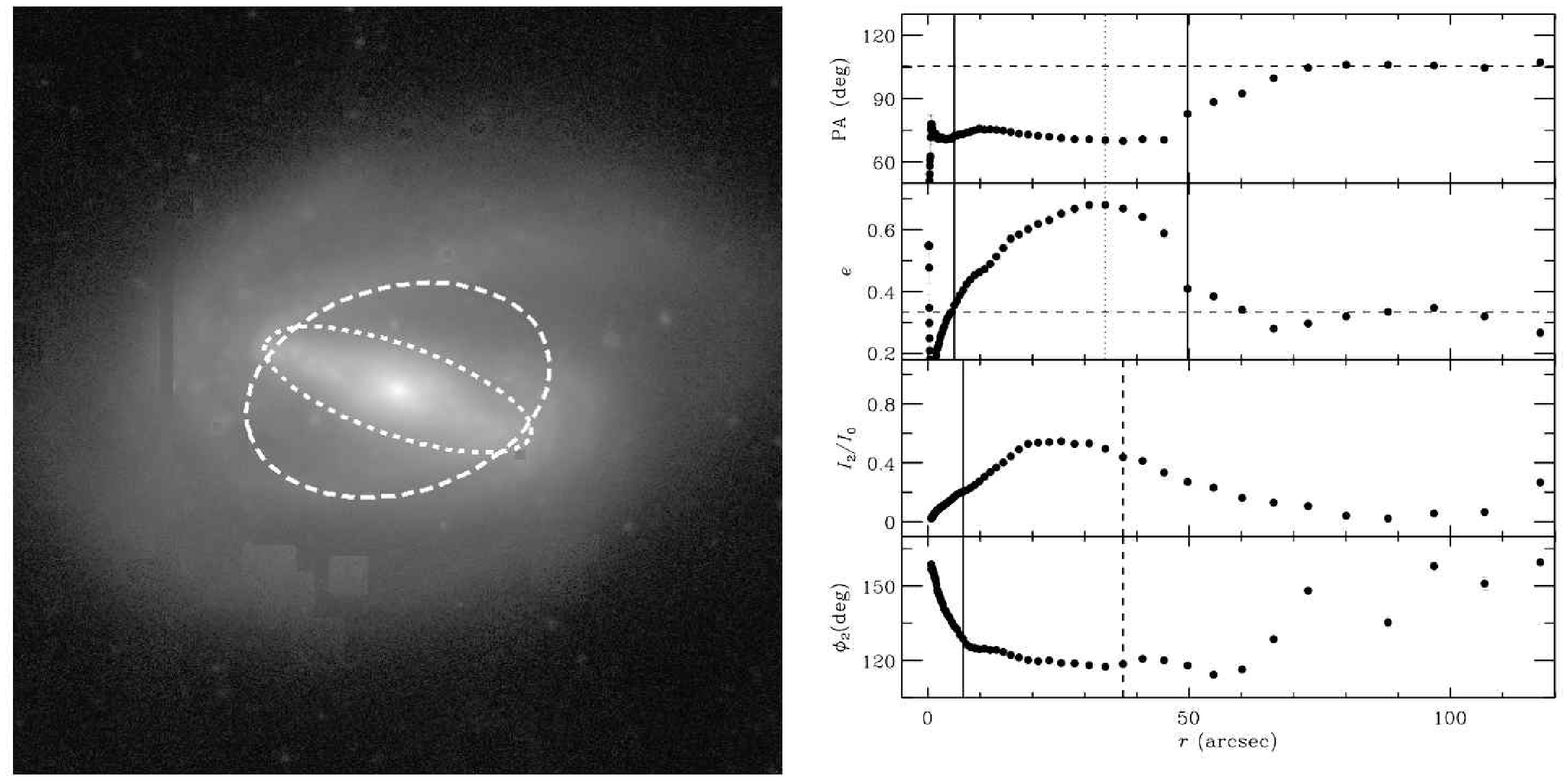,width=17.5cm,angle=0}}
\figcaption[fig9.ps]{
Illustration of how we determine the size and strength of a bar.
Left: star-cleaned $I$-band image of NGC~7513; the size of the image
is $\sim$3\amin$\times$3\amin.  Right: radial profiles of PA, $e$,
$I_2/I_0$, and $\phi_2$.  The horizontal dashed lines in the PA and $e$ panels
denote the characteristic values of the galaxy.  The solid vertical lines
mark the inner and outer boundaries of the bar-dominated region, and the
dotted vertical line represents the projected size of the bar determined
using the geometric method, where the bar is required to have $e_{\rm max} \geq
0.2$ and $\Delta {\rm PA} \leq 20$\deg. In the $I_2/I_0$ and $\phi_2$ panels,
the solid vertical line marks the inner boundary of the bar.  The vertical
dashed line represents both the outer boundary and size of the bar,
determined using the Fourier method, where the bar criteria are $I_2/I_0 \geq
0.2$ and $\Delta \phi_2 \leq 20$\deg. The corresponding isophotal ellipses for
the two methods are overplotted on the left-hand image with the same type
of lines as in the right-hand panel.
\label{figure:barexp}}
\end{figure*}

Figure~\ref{figure:barexp} illustrates how we identify and measure the 
bar size and strength, as applied to the star-cleaned $I$-band image of the 
SBb galaxy NGC~7513.  The radial profiles of $e$ and PA clearly show the 
hallmark features of a bar: a distinctly broad peak in $e$ above our minimum 
threshold of 0.2, reaching $e_{\rm max}$ = 0.68, and an extended plateau of 
near-constant PA $\approx$ 75\deg\ ($\Delta {\rm PA} \leq 20$\deg).  The outer 
disk of the galaxy has a clearly different $e$ (0.33) and PA (105\deg).  The 
two vertical solid lines mark the inner and outer boundaries of the 
bar-dominated region; the vertical dotted line gives the projected bar radius. 
Although widely used in the literature, the measured $e_{\rm max}$ of the bar 
is actually $\sim$ 20\% lower than that derived from 2-D image 
decomposition when the bulge and disk components are also included 
\citep{gadotti08}.  Indeed, reliable bar parameters can only be determined 

\begin{figure*}[t]
\hbox{
\hskip -0.1in
\psfig{file=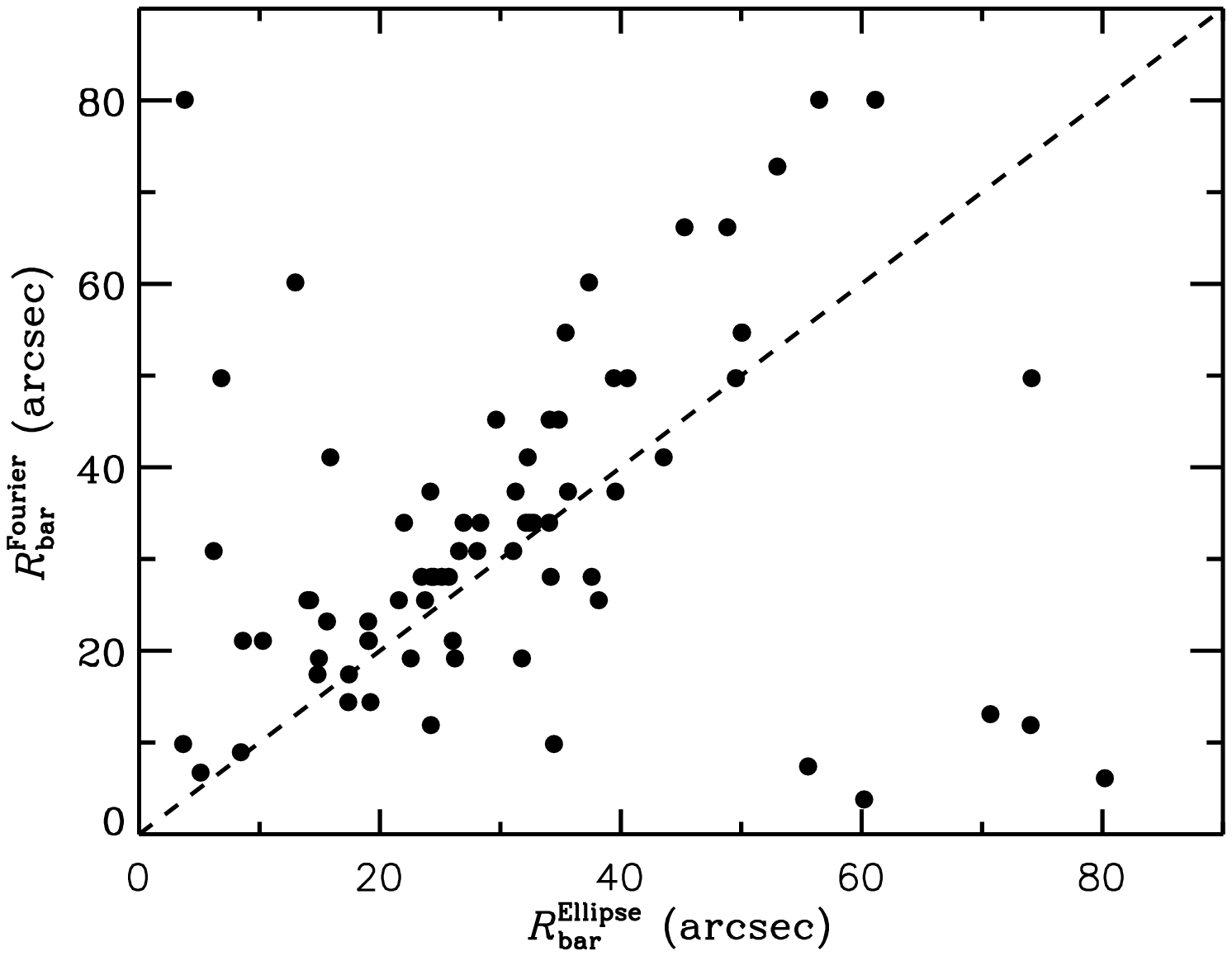,width=9.8cm,angle=0}
\hskip -0.3in
\psfig{file=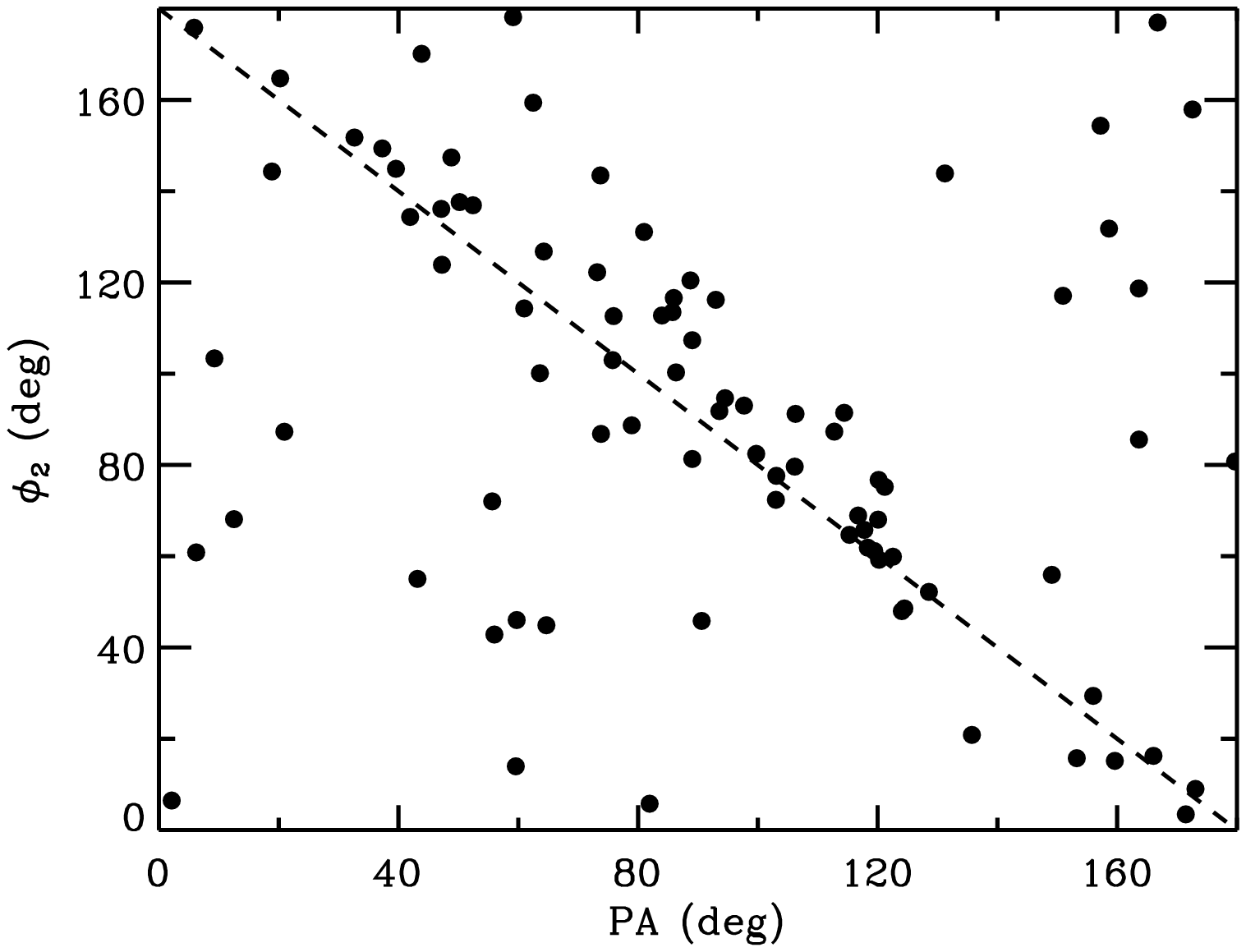,width=9.8cm,angle=0}
}
\figcaption[fig10.ps]{
Left: correlation between $R_{\rm bar}^{\rm Ellipse}$, the
deprojected bar size measured from the geometric, ellipse method and
$R_{\rm bar}^{\rm Fourier}$, that measured from the Fourier decomposition.
The dashed line represents $y = x$. Right: correlation between
$\phi_2$ and PA of the bar. The dashed line represents $y = 180^\degree - x$.
\label{figure:barcmp}}
\end{figure*}

\noindent
by 
2-D decomposition of the images with all of the other components included. 
Such analysis is outside the scope of this work, but future papers will 
present results from 2-D decompositions of the CGS images.

\subsection{Fourier Analysis}

In addition to the geometric method described above, we also derive bar 
properties using the radial profiles of the relative amplitude of the $m$ = 2 
Fourier mode ($I_2/I_0$) and its associated phase angle ($\phi_2$).  As 
before, we work with the $I$-band images.  Bars are usually associated with 
the first local maximum in the $I_2/I_0$ profile, where the bar/interbar 
contrast is the largest, over an extended region where $\phi_2$ keeps 
approximately constant.  Subsequent maxima in the $I_2/I_0$ profile, if 
present, trace spiral arms or ring structures, but in these instances 
$\phi_2$ varies with radius.  Spiral arms always produce varying phase angles, 
and thus the region where they dominate can be easily excluded from the bar 
size measurement.

Adopting a procedure similar to that used by \citet{ague09}, we define the bar 
to be the region wherein the maximum relative $m = 2$ Fourier amplitude 
$({I_2/I_0})_{\rm max} > 0.2$ and the phase angle remains constant to 
$\Delta{\phi_2} < 20^{\degree}$.  We set the inner boundary of the bar to be 
the first data point outside of 3 times the seeing radius in which $I_2/I_0 > 
0.2$.  Past the peak, we 

\begin{figure*}[t]
\centerline{\psfig{file=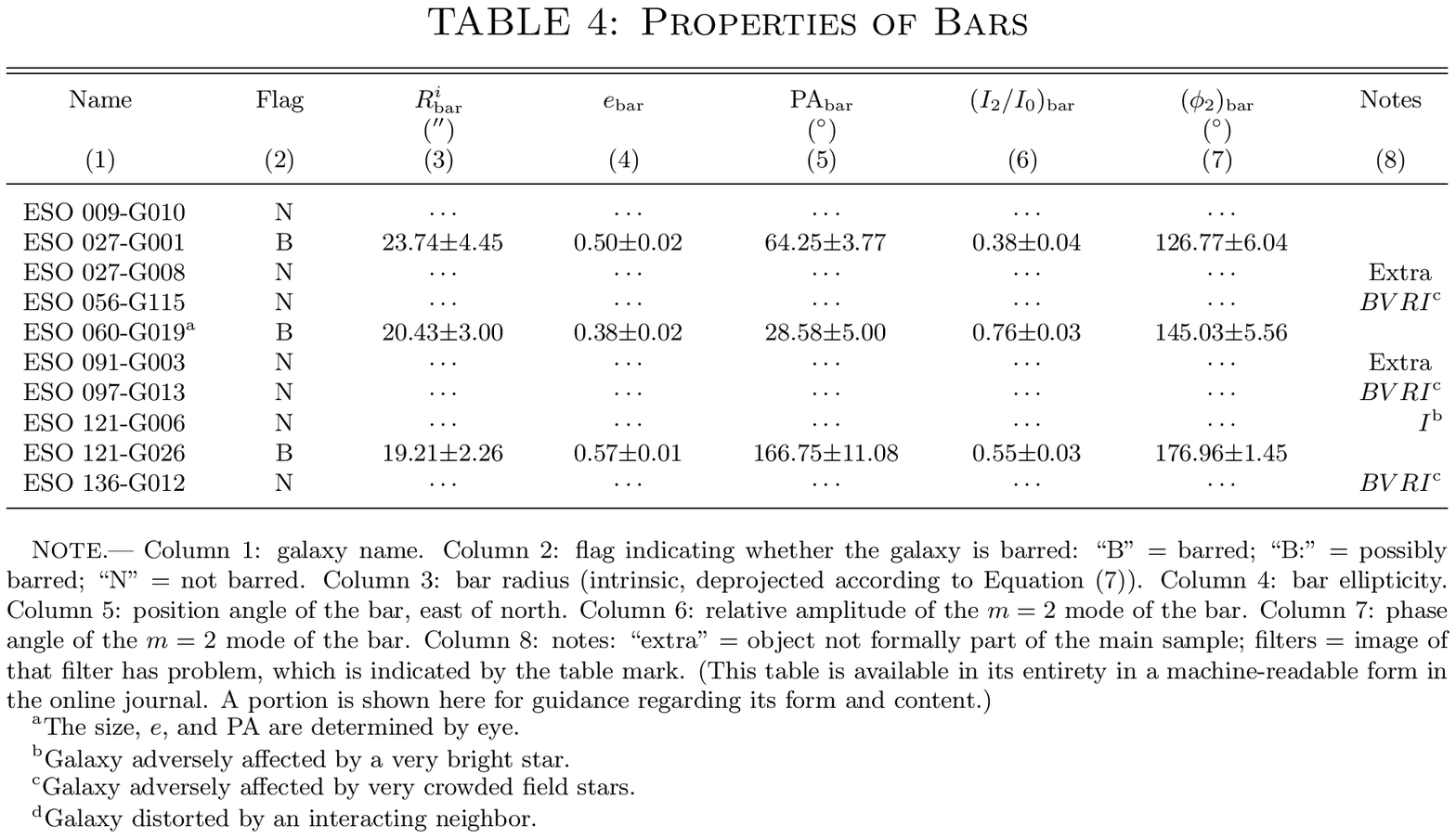,width=16.5cm,angle=0}}
\end{figure*}

\begin{figure*}[t]
\centerline{\psfig{file=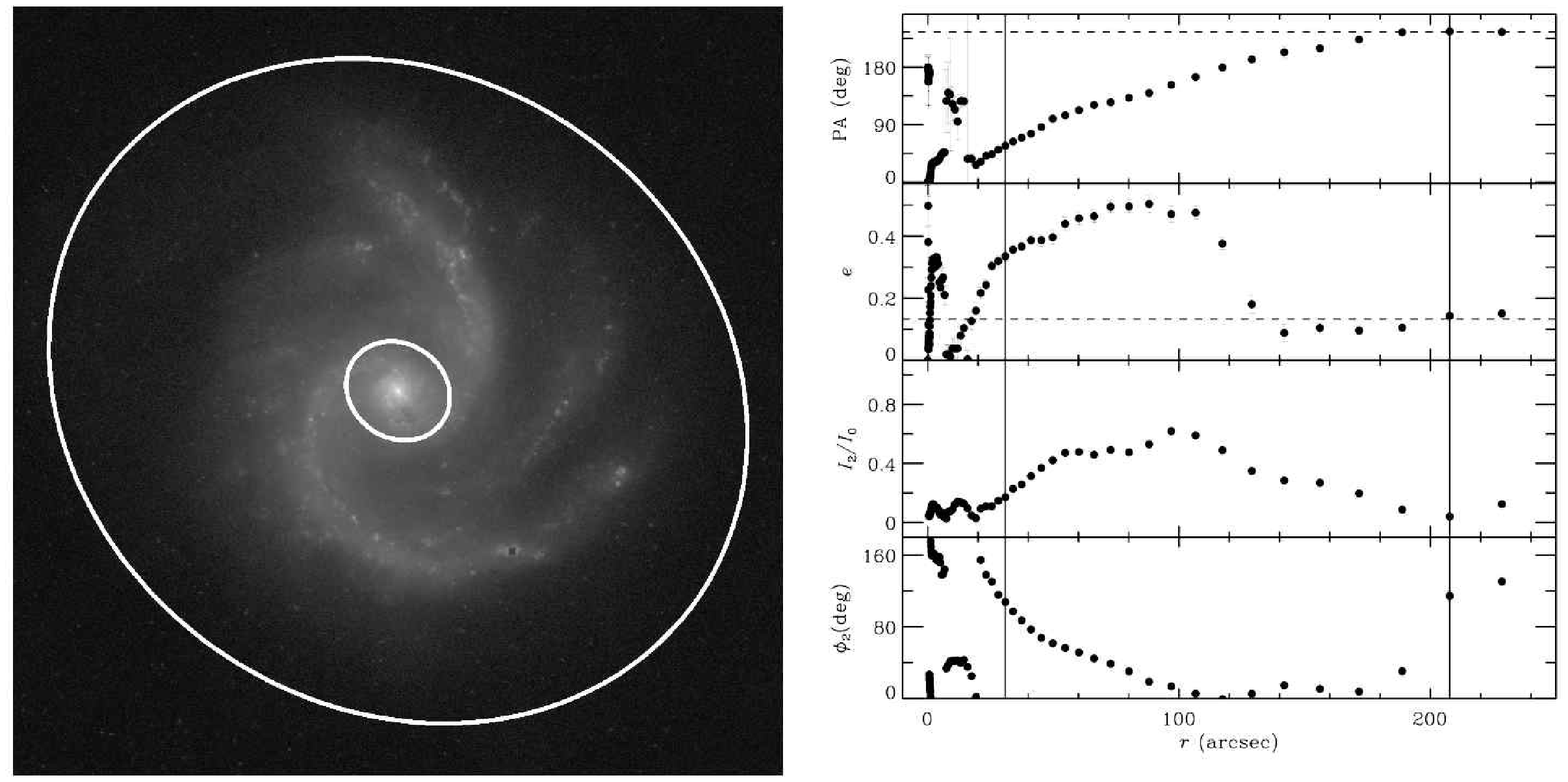,width=17.5cm,angle=0}}
\figcaption[fig11.ps]{
Illustration of how we measure the strength of the spiral arms using
the Fourier method.  Left: star-cleaned $I$-band image of NGC~5247; the
size of the image is $\sim$7\farcm3$\times$7\farcm3. Right: radial
profiles of PA, $e$, $I_2/I_0$, and $\phi_2$.  The solid vertical lines
mark the inner and outer boundaries of the disk-dominated region, which
is defined to be outside the central bulge or bar component, but inside the
radius where the isophotal intensity is 3 $\sigma$ above the sky background.
The corresponding isophotal ellipses are overplotted on the left-hand image.
The horizontal dashed lines in the PA and $e$ panels denote the characteristic
values of the galaxy.
\label{figure:armexamp}}
\end{figure*}
\vskip 0.3cm

\noindent
designate the radius where $I_2/I_0 = 
({I_2/I_0})_{\rm max}-0.1$ as the outer boundary of the bar region.  In the 
event that there is a secondary maximum and the local minimum between the two 
peaks exceeds $({I_2/I_0})_{\rm max} - 0.1$, we set the position of the local 
minimum to be the outer boundary of the bar region.

We assign $({I_2/I_0})_{\rm max}$ to be the bar strength, with the uncertainty 
set by the statistical error derived from the Fourier decomposition process.  
As the Fourier method is minimally affected by spiral arms, we simply set the 
bar size to be equal to the radius of its outer boundary; its associated 
uncertainty is the semi-major axis range between the outer boundary and the 
radius where $I_2/I_0 = ({I_2/I_0})_{\rm max} - 0.05$.  The Fourier analysis
is performed on the isophotes extracted in the third step of the ellipse
fitting, where the geometric parameters were all fixed to those of the 
outermost isophote (see Section 4.1). For simplicity, we just assume 
that the disk is purely circular in its face-on view, so the semi-major 
axes of the isophotal ellipses are the radii of the circles. The size of the 
bar is denoted by the semi-major 

\begin{figure*}[t]
\centerline{\psfig{file=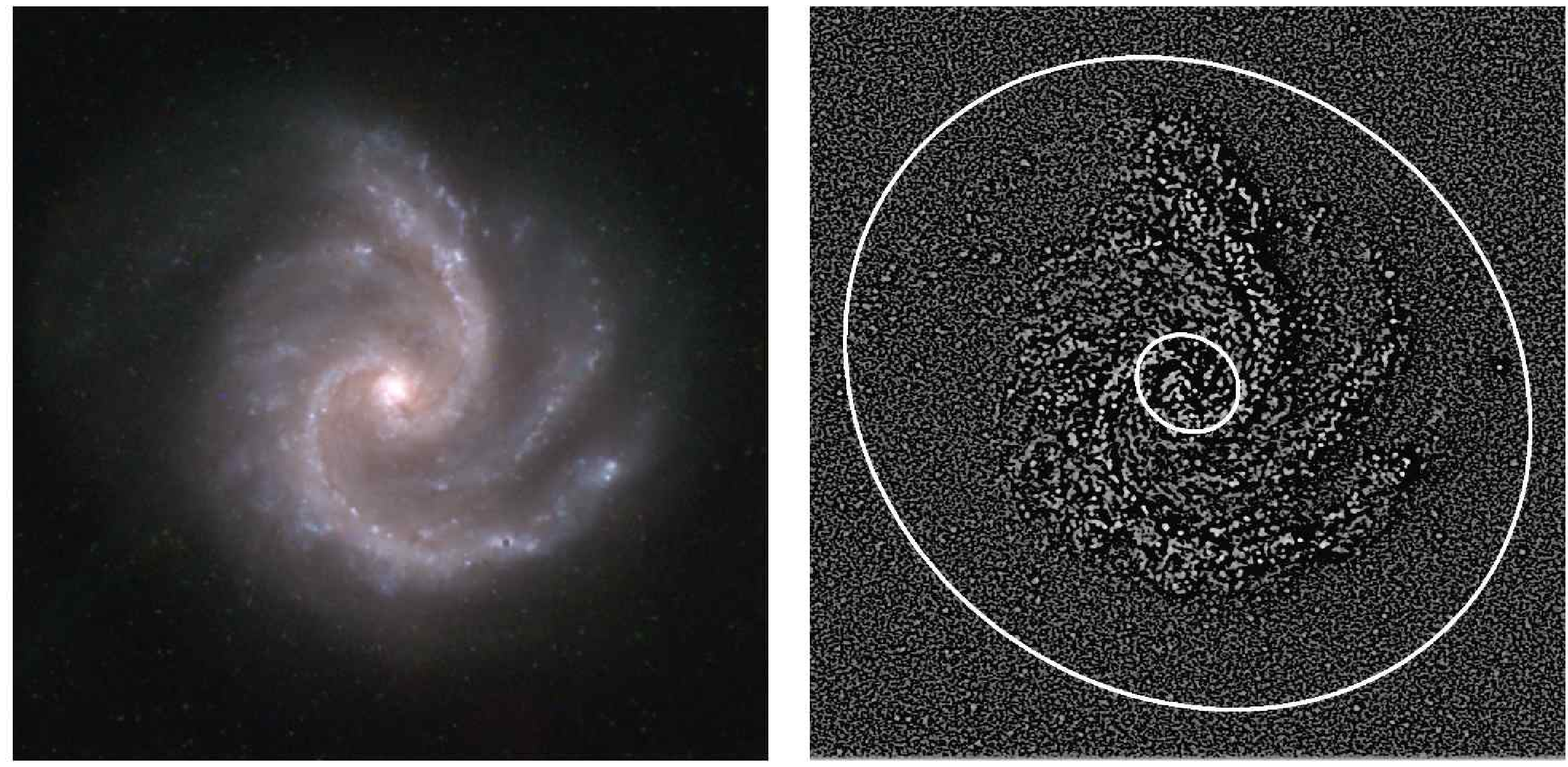,width=17.5cm,angle=0}}
\figcaption[fig12.ps]{
Illustration of how we measure the strength of the spiral arms using
the structure map.  Left: star-cleaned composite color image of
NGC~5247; the size of the image is $\sim$7\farcm3$\times$7\farcm3.
Right: structure map of the star-cleaned $B$-band image.  The two
overplotted ellipses mark the inner and outer boundaries where the spiral
arms populate. After rejecting the masked objects, we use the rms of the
pixels within these two ellipses to estimate the strength of the spiral arms.
(A color version of this figure is available in the online journal.)
\label{figure:struexamp}}
\end{figure*}
\vskip 0.3cm

\noindent
axis of the particular isophote that encloses 
the bar region, which is actually the radius of the circle that passes right 
through the end of the bar in a face-on view.  Under this simplified 
assumption, the bar size is already deprojected.  The characteristic phase 
angle of the bar is the average $\phi_2$ within the bar region, with 
$\Delta \phi_2$ as the uncertainty.  The two bottom-right panels of 
Figure~\ref{figure:barexp} show our Fourier technique applied to NGC~7513.

Fourier analysis offers a very useful approach to studying bars. It not only 
provides another independent method to identify and quantify bars, but also 
recovers bars missed by the geometric method in galaxies with internal 
structure too complex to yield an unambiguous bar signature in their $e$ and 
PA profiles.  In fact, the bar parameters derived from these two methods 
usually agree quite well.  Figure~\ref{figure:barcmp} (left panel) 
compares the bar sizes measured from the ellipse method ($R_{\rm bar}^{\rm 
Ellipse}$; deprojected) versus those measured from the Fourier method 
($R_{\rm bar}^{\rm Fourier}$).  Overall, there is a good correlation, but some 
glaring outliers stand out.  Although our analysis is done in the $I$ band, 
dust extinction can still be significant in some galaxies. The dust extinction 
features near the central region of a galaxy tends to trick the Fourier method 
by producing a false peak in $I_2/I_0$ with roughly constant $\phi_2$; this 
yields an unusually compact, incorrect bar size.  The Fourier method can also 
be unreliable for weak bars embedded in disks with strong spiral arms.  Under 
these circumstances, the relatively weak local peak of the bar in the 
$I_2/I_0$ profile can be overshadowed by a stronger peak generated by the 
spiral arms, leading to an overestimate of the bar size.  The position angle 
of the bar correlates strongly with $\phi_2$ (Figure~\ref{figure:barcmp}, 
right panel). Ideally, $\phi_2 = 180^\degree - {\rm PA}$. However, when the PA 
of the bar approaches $0^\degree$ or $180^\degree$, the corresponding phase 
angle can have values similar to the PA; this is responsible for the points 
lying on the lower-left and upper-right portions of the figure. Dust 
in the central regions of the galaxy further contributes to the scatter.

\subsection{Final Bar Classification}

Since neither of the methods discussed above is foolproof, we use both to 
assign the final bar classification to the galaxies in CGS.  If a galaxy is 
not classified as barred in both the geometric and Fourier analysis, it is 
labeled unbarred. If the galaxy is classified as barred by only one method, we 
call it ``possibly'' barred. If both methods consider a galaxy barred, we 
check whether the derived bar sizes are consistent between them. We only 
classify it as definitely barred if the bar sizes from the two methods differ
by less than 10\asec; if they disagree by more than 10\asec, we call it 
``possibly'' barred.

For the ``possibly'' barred galaxies, we visually examine their $I$-band images 
and further scrutinize their geometric and Fourier profiles to check whether 
we are being misled by internal structural complexities such as spiral arms or 
dust features. If any of the measurements from either of the two methods is 
suspect, we manually set the inner and outer boundaries of the bar region and 
redo the measurements.  Not all ambiguous cases can be resolved, and in our 
final classification we continue to flag their bar status as uncertain.  For 
our final measurement of the bar size, we usually give higher priority to the 
results derived from the geometric method, which generally gives more accurate 
sizes than those derived from the Fourier method.  The definition of the bar 
size in the Fourier method is of limited utility because it is largely 
arbitrary and more prone to being influenced by the presence of spiral arms.
Table~4 summarizes the final bar classification and some 
basic bar parameters derived from our analysis.  Among the 501 disk galaxies 
($T \geq -3$) in the final catalog, 44 (9\%) are deemed definitely barred, 136 
as possibly barred (27\%), and 321 (64\%) as unbarred.  As bar identification 
is uncertain in highly inclined galaxies, we reexamine the statistics in a 
subsample restricted to have ellipticities smaller than $e_{\rm gal} = 0.6$.  
As expected, the bar fraction increases.  Out of 387 disk galaxies, 173 
($45\%$) are barred and 214 ($55\%$) are unbarred.  A more detailed comparison 
between our results and those in the literature will be deferred to a future 
paper.

\vskip 1cm

\begin{figure*}[t]
\centerline{\psfig{file=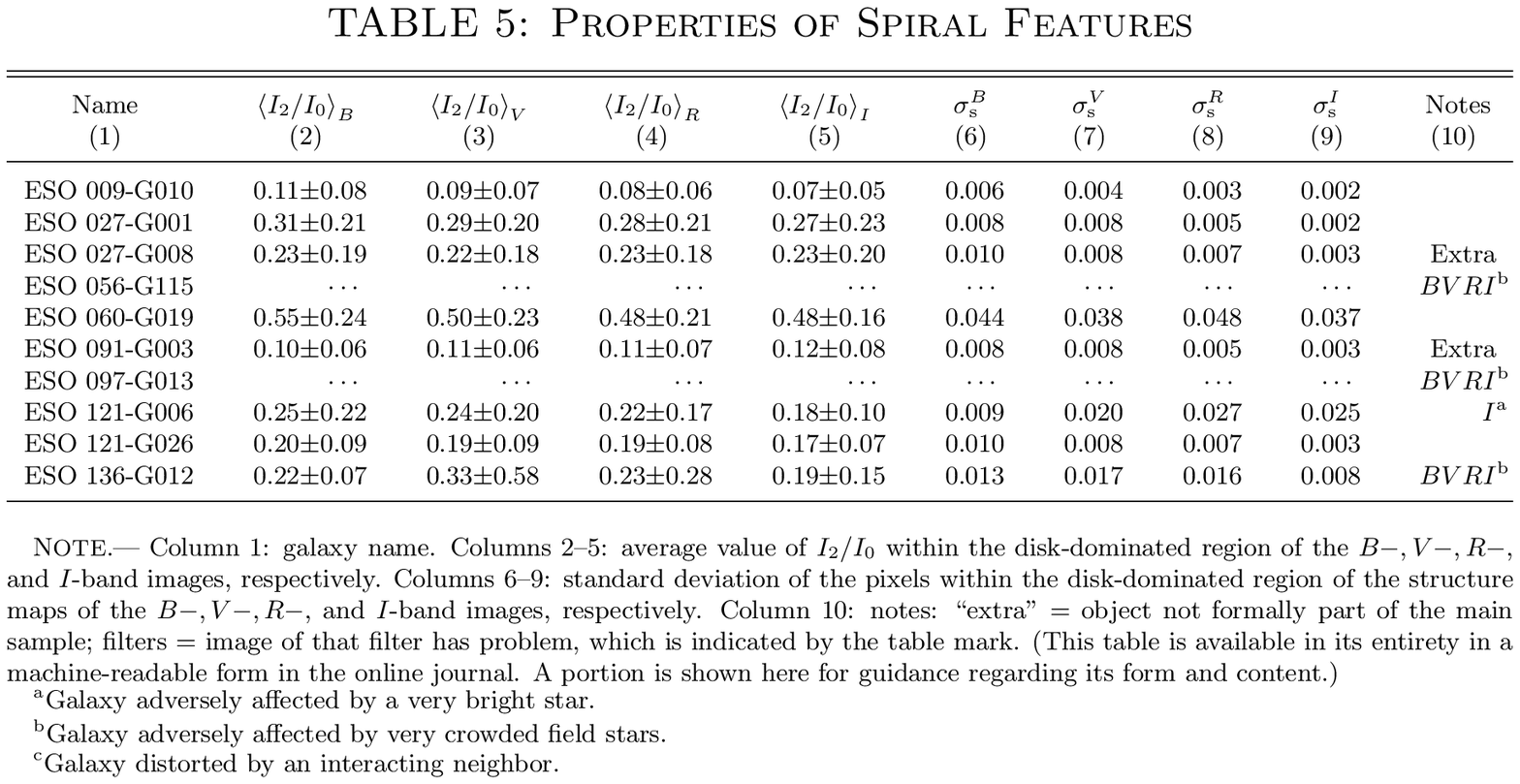,width=16.5cm,angle=0}}
\end{figure*}

\section{Spiral Arms}

We provide two quantitative measurements (Table~5) that 
can be used to access the presence and strength of spiral arms in galaxies.  
The analysis is applied uniformly to all non-elliptical galaxies in the 
sample, including disk galaxies traditionally deemed to lack spiral arms, such 
as S0s.  

We perform a simple measurement of the average strength of $I_2/I_0$ in the 
disk region outside the central bulge and the bar. Spiral arms are the main 
contributor to any significant $m=2$ mode in this region\footnote{The $m=2$ 
mode is most sensitive to systems with grand design, two-arm spirals.  However,
flocculent or multiple-arm spirals still exhibit significant amplitude in the 
$m=2$ mode.  We defer a full treatment of spiral arms, including exploration of 
high-order modes, to a separate paper.}. If neither a featureless, classical 
bulge nor a bar is present, the minimum inner boundary for our calculation is 
set to 3 times the seeing radius.  For barred galaxies, the inner boundary is 
naturally set to the bar radius, which almost always lies exterior to the 
bulge.  For unbarred galaxies with classical bulges, we define the inner 
boundary to be the radius where $e > 0.2$, beyond which the disk usually 
dominates over the bulge.  This criterion fails for face-on galaxies with very 
weak spiral arms and classical bulges, because the disk becomes 
indistinguishable from the bulge on the basis of its ellipticity alone.
Fortunately, under these circumstances both the bulge and the disk contribute 
little to $I_2/I_0$ anyway, and it makes little difference whether the bulge 
is excluded or not.  The outer boundary is the radius where the isophotal 
intensity reaches $3 \, \sigma$ above the background in the $I$-band image; we 
apply the same boundary for the other filters. We then calculate a 
characteristic value of $I_2/I_0$ by averaging its profile between the inner 
and outer boundaries.  We illustrate our methodology in 
Figure~\ref{figure:armexamp}, applied to the Sbc galaxy NGC~5247.  

\begin{figure*}[t]
\centerline{\psfig{file=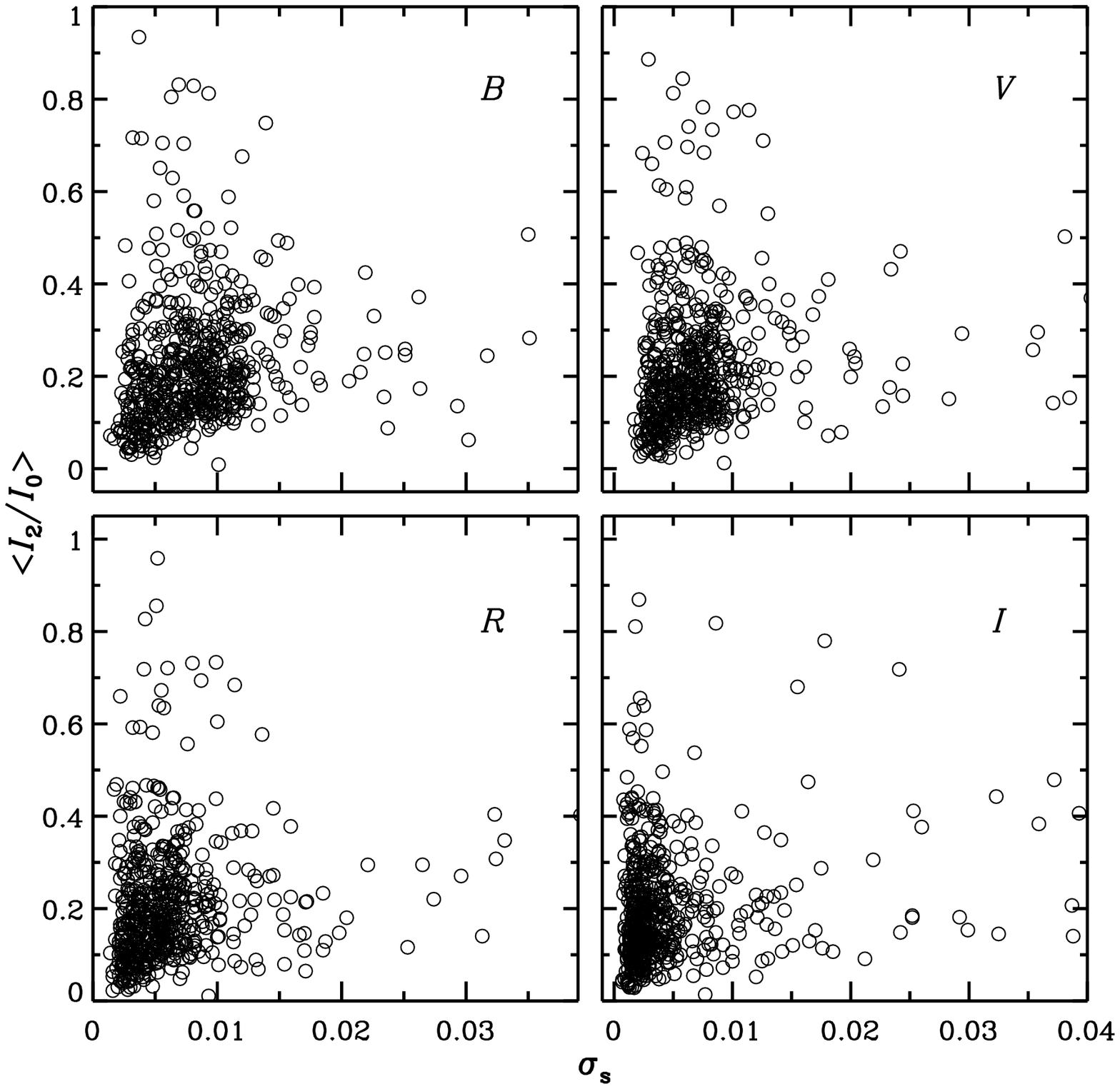,width=17.5cm,angle=0}}
\figcaption[fig13.ps]{
Comparison between the two measures of the spiral arms
($\left<I_2/I_0\right>$ and $\sigma_{\rm s}$) for \emph{BVRI} filters. The 
correlation is very weak or absent, because the two measures essentially probe
structures on quite different scales. $\sigma_{\rm s}$ is mainly determined by
structures with typical scales of the point-spread function, while
$\left<I_2/I_0\right>$ is more sensitive to more extended, large-scale
features coherent over significant portions of the spiral arms.
\label{figure:armcmp}}
\end{figure*}

Our second method makes use of the structure maps (Paper~I) to estimate the 
strength of the spiral features (Figure~\ref{figure:struexamp}). After masking 
out the field stars and background galaxies, we compute the standard deviation 
of all the remaining pixels within the inner and outer boundaries of the 
disk-dominated region, as determined above.  The agreement between the two 
different measurements is not good, as can be seen in 
Figure~\ref{figure:armcmp}, where we plot $\left<I_2/I_0\right>$ against 
$\sigma_{\rm s}$ for all the filters.  The two parameters trace structures on
different scales.  The structure map optimally filters spatial features on the 
scale of the point-spread function. It effectively highlights features such as 
dust lanes and thin arms, but it is not very sensitive to smooth and wide 
spiral arms, which can be better probed via $\left<I_2/I_0\right>$. 

\section{Lopsidedness}

The relative amplitude of the $m$ = 1 Fourier mode is widely used to study the 
lopsidedness of galactic stellar \citep{rixzar95, zari97, bour05, joco08, 
reic08} and gaseous \citep[e.g.,][]{vaneymeren11} disks.  This approach is 
well-defined, quantitative, and relatively straightforward to implement, and 
hence can be applied to study large samples of galaxies.  A lopsided disk 
stands out as a region of enhanced $I_1/I_0$ and roughly constant phase angle 
$\phi_1$.  A one-arm spiral also exhibits a large $I_1/I_0$, but $\phi_1$ 
increases monotonically as a function of radius.

Our method to measure lopsidedness differs somewhat from that used in previous 
works.  \citet{rixzar95} perform a bulge-to-disk decomposition of the surface 
brightness profile to determine the scale length of the disk, and then compute 
the average relative Fourier amplitude between 1.5 and 2.5 scale lengths.  As 
we do not yet have robust structural decompositions for the entire CGS sample 
(this work is in progress), we resort to a simpler strategy, one based on the 
expectation that the lopsided portion of the disk should be characterized by a 
roughly flat $\phi_1$ radial profile.  Through careful experimentation 

\begin{figure*}[t]
\centerline{\psfig{file=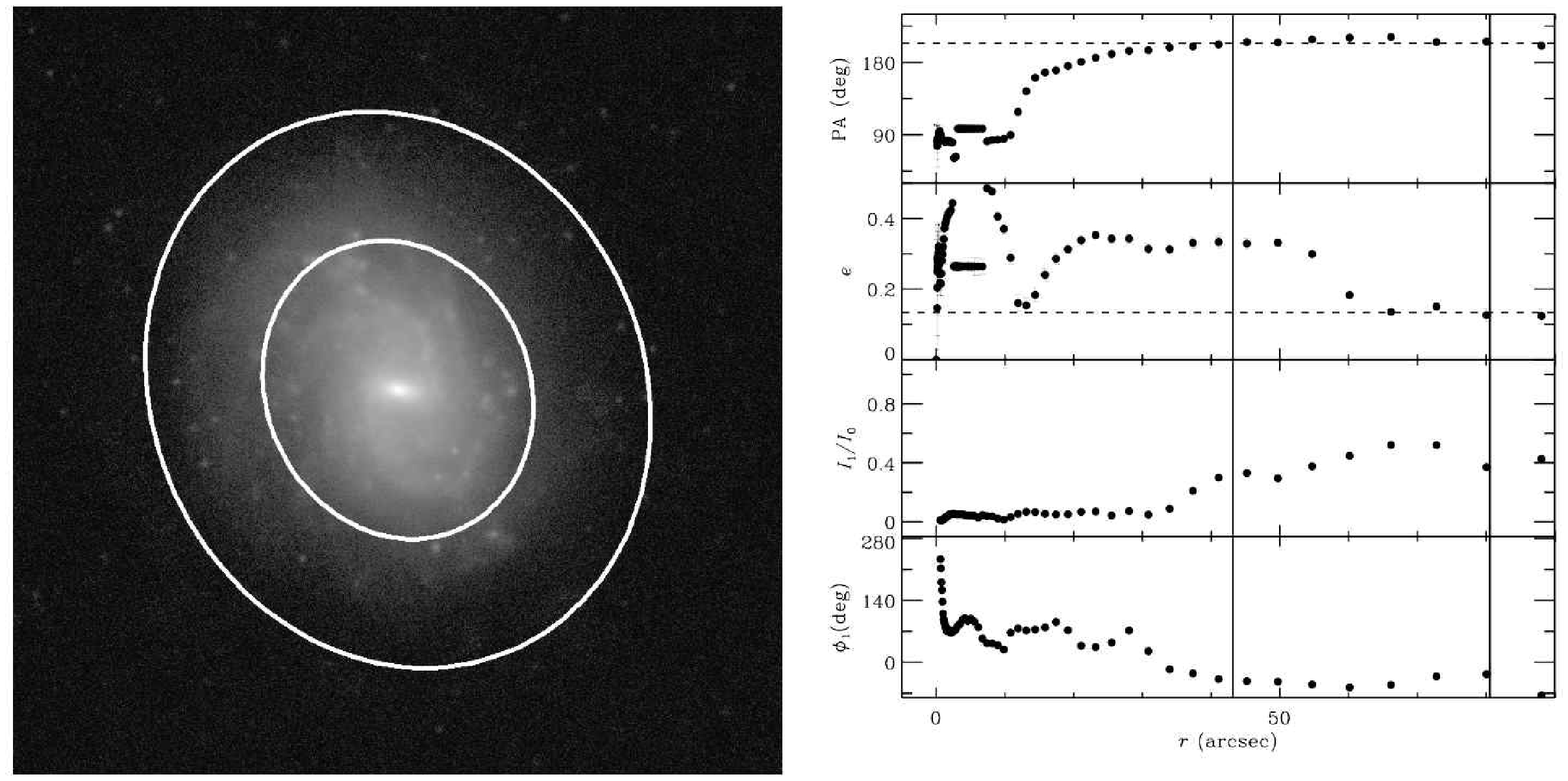,width=17.5cm,angle=0}}
\figcaption[fig14.ps]{
Illustration of how we measure lopsidedness.  Left:
star-cleaned $I$-band image of NGC~7070; the size of the image is
$\sim$3\farcm6$\times$3\farcm6.  Right: radial profiles of PA, $e$,
$I_1/I_0$, and $\phi_1$.  The horizontal dashed lines in the PA and $e$ panels
denote the characteristic values of the galaxy.  The solid vertical lines, and
the corresponding isophotal ellipses overplotted in the left-hand image, mark
the inner and outer boundaries of the region used to compute the lopsidedness,
which we define to be that in which the radial variation of $\phi_1$ is
smaller than $70^\circ$.  The lopsidedness is measured simply by averaging
$I_1/I_0$ within this region.
\label{figure:lopexamp}}
\end{figure*}

\noindent
with a
number of galaxies with prominent lopsided disks, we find that the lopsided
region can be effectively isolated by requiring that the phase angle be
constant to within $\Delta \phi_1 \leq 70^\degree$.   We
apply this criterion to the phase angle profile of the $I$-band image to
define the inner and outer radii of the lopsided region, and then adopt these 
values for the other filters.  For each filter, the lopsidedness is the 
average value of $I_1/I_0$ within that region, with the standard deviation as 
its associated uncertainty. The characteristic value of $\phi_1$ and its error 
are calculated similarly.   Figure~\ref{figure:lopexamp} illustrates our 
method on the Scd galaxy NGC~7070.  

Lopsidedness measurements for the entire sample are given in 
Table~6\footnote{We have flagged the galaxies whose 
lopsidedness measurements may be unreliable because of contamination 
by bright stars or excessive crowding by field stars.}. 
Table~7 summarizes the 
frequency of galaxies with significant lopsidedness, defined as 
$\left<I_1/I_0\right>_I \ge 0.2$ \citep[e.g.,][]{zari97}, for the
subsample of 350 disk galaxies with $e_{\rm gal} \le 0.6$ for which robust 
measurements could be made.  Consistent with previous studies, the fraction 
of galaxies with significant lopsidedness is high, and the frequency increases 
substantially in late-type galaxies.


\section{Data Verification}

\subsection{Internal Comparison}

\vskip 0.3cm

\begin{figure*}[t]
\centerline{\psfig{file=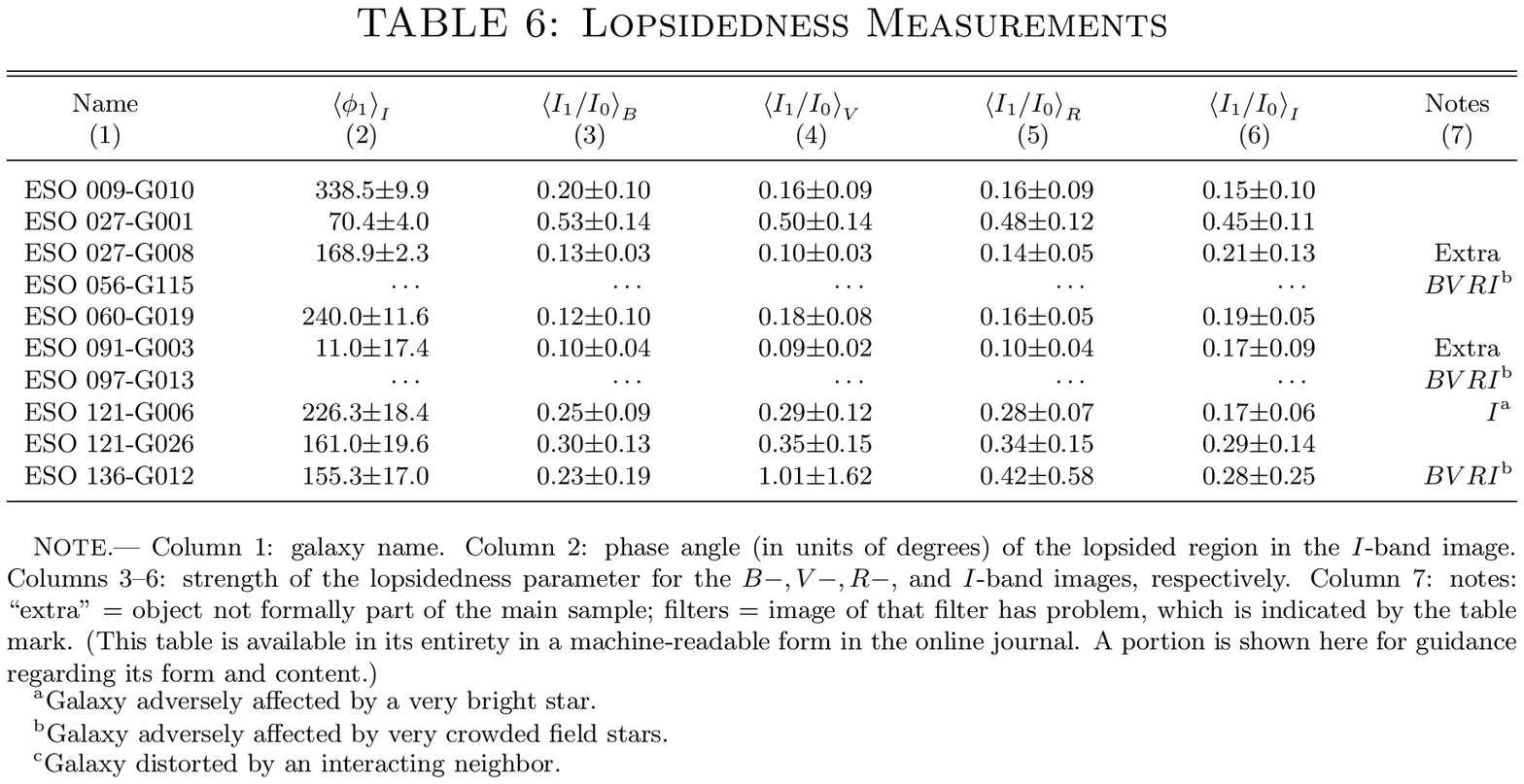,width=16.5cm,angle=0}}
\end{figure*}

Several galaxies were observed more than once during different nights 
throughout the survey.  Although only the best images are included in the 
final CGS catalog, the repeat observations, which were reduced and analyzed in 
the same manner as the rest, afford an opportunity to access the accuracy of 
our calibration methods and the reliability of the parameter measurements.  

Table~8 lists the galaxies that have pairs of repeat 
observations useful for internal comparison.  Note that for this exercise we 
only select objects that have reasonably good data.  Many of the duplicate 
observations were taken precisely because the original observation was deemed 
to be of exceptionally low quality, either because of weather conditions 
(bad seeing, excessive cloud cover) or technical problems (poor telescope 
focus, tracking errors).  Because one of the observations in the comparison
pair is---by definition---suboptimal, the following assessment, in some sense,
gives an overly conservative estimate of the magnitude of internal errors.

Figure~\ref{figure:prof_diff} compares the surface brightness profiles for 

\begin{figure*}[b]
\centerline{\psfig{file=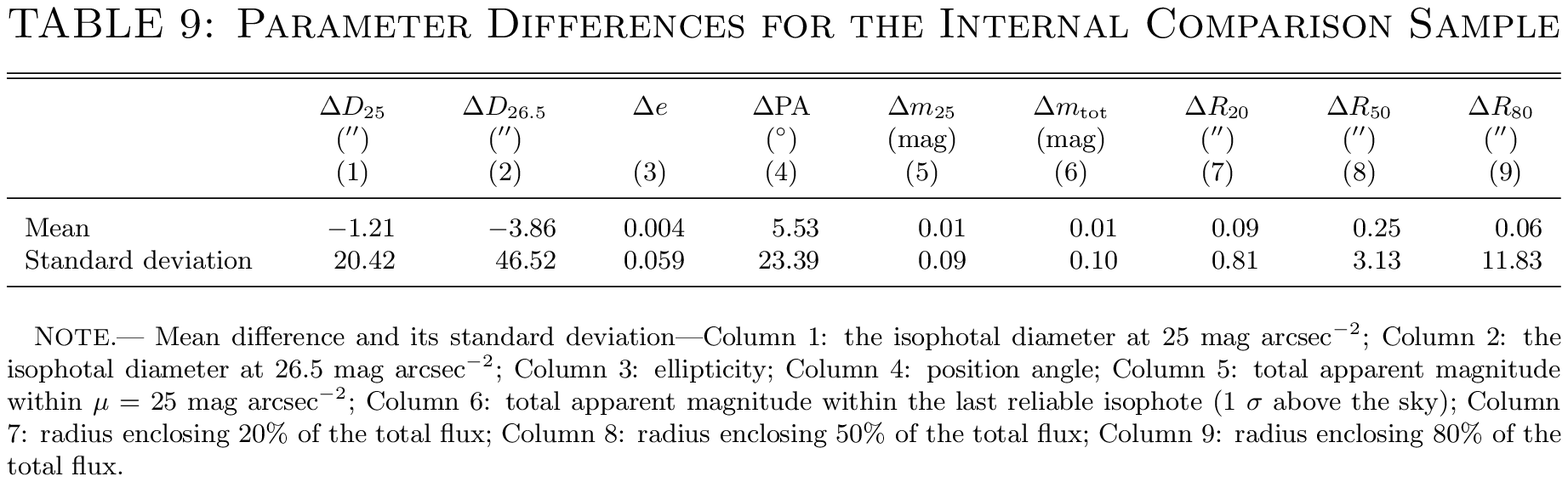,width=17.5cm,angle=0}}
\end{figure*}

\vskip 0.3cm
\psfig{file=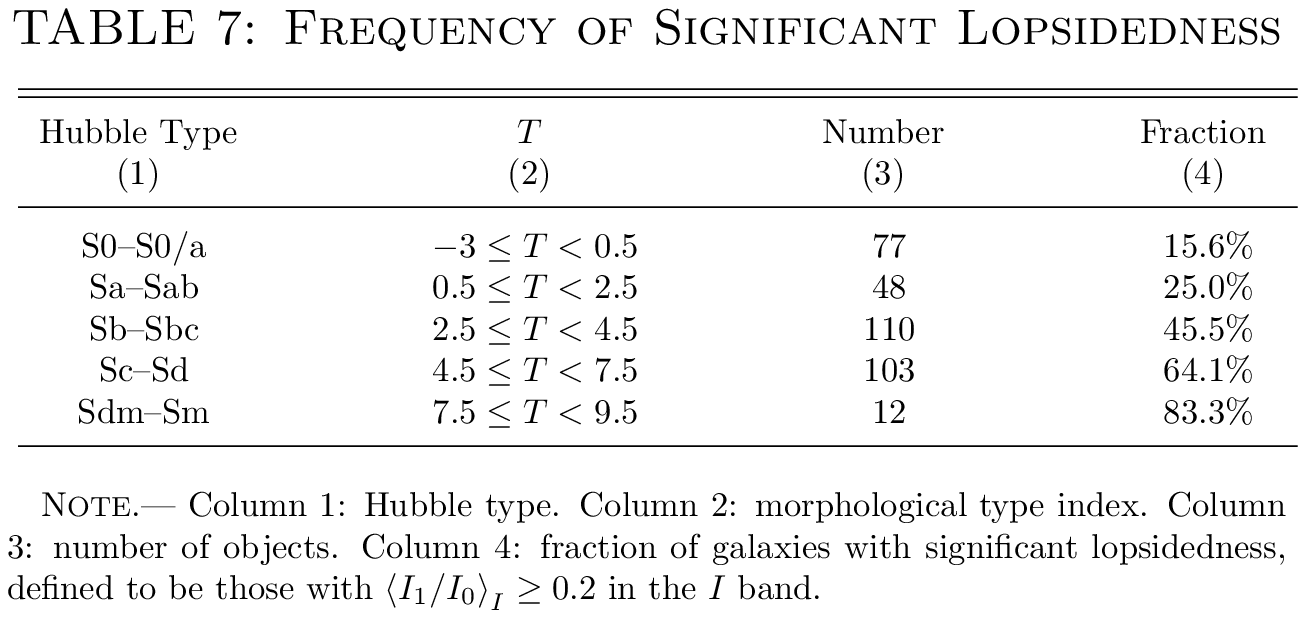,width=8.75cm,angle=0}

\noindent
galaxies observed on different nights.  In most cases, they agree quite well. 
The weighted average of the profile differences (dashed line, calculated by
 averaging the profile differences weighted by the corresponding error at 
each data point) resides well inside the formal $1 \, \sigma$ photometric 
uncertainty \citep{noovan07}. This suggests that 
our photometric errors are robust, both for the photometric and non-photometric 
observations.   A few objects (NGC 1374, 2196, 6810, 6861, 7590) show slightly 
larger, but by no means alarming, discrepancies.  Two types of differences in 
profile {\it shape}\ can be seen.  The innermost portions of the profiles 
often show systematic deviations, sometimes as large as $\sim 0.5$ mag~$\rm 
arcsec^{-2}$. This effect can be entirely attributed to mismatches in seeing, 
but is well confined within $\sim$3 times the radius of the seeing disk.  This 
is the reason we restrict all of our scientific analysis to radii beyond this.
Additionally, many of the profiles show some level of systematic deviation at 
large radii.  This most likely arises from errors in sky subtraction.  For most 
objects, the deviations occur at the level of $\sim 0.2$ mag~$\rm arcsec^{-2}$,
but they lie well within the error bars of the individual isophotal 
intensities, which again indicates that our error budget is realistic.  The 
most extreme deviations occur in galaxies that are too extended 

\vskip 0.3cm
\psfig{file=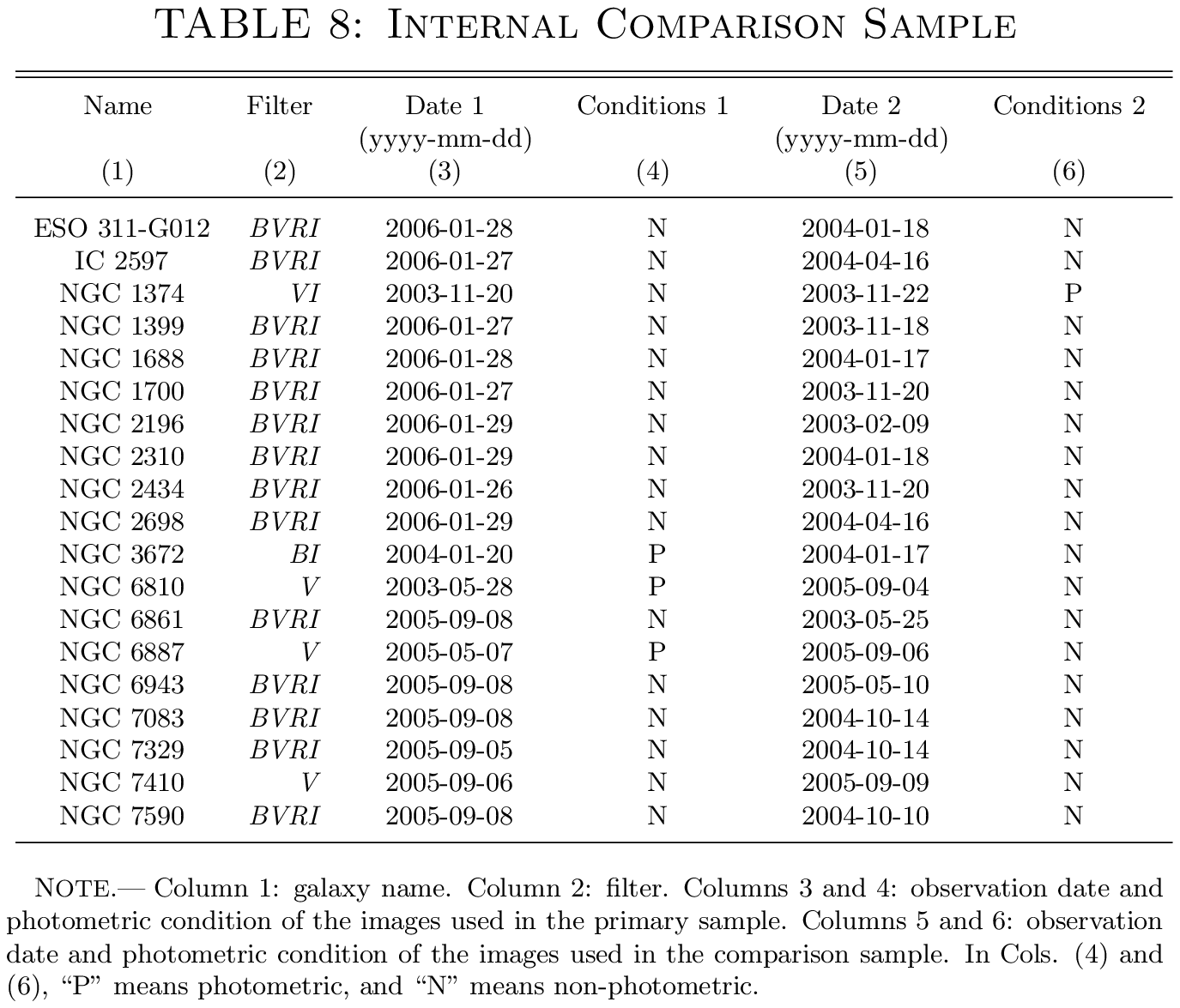,width=8.75cm,angle=0}
\vskip 0.5cm

\noindent
for standard 
sky subtraction, for which we had to resort to an indirect estimate based on 
profile fitting (Section~3). In these cases, the deviations in the outer 
profiles may be as large as $\sim 0.5$ mag~$\rm arcsec^{-2}$.  NGC~2434 is 
such an example.  However, our formal error bars appear to be realistic even in 
these extreme situations.

Figure~\ref{figure:interpara} compares nine measured parameters derived from 
the set of repeat observations.  Observation 1 denotes the measurement with 
better quality that has been adopted in the final database of the survey,
and Observation 2 gives the comparison measurement.  We can see that overall 
the agreement is quite good.  Notable exceptions can be identified with 
galaxies that have especially unreliable sky values, such as NGC~2434, which 
is the most deviant outlier in the $R_{80}$ plot (this quantity is

\vskip 1.5cm

\begin{figure*}[t]
\centerline{\psfig{file=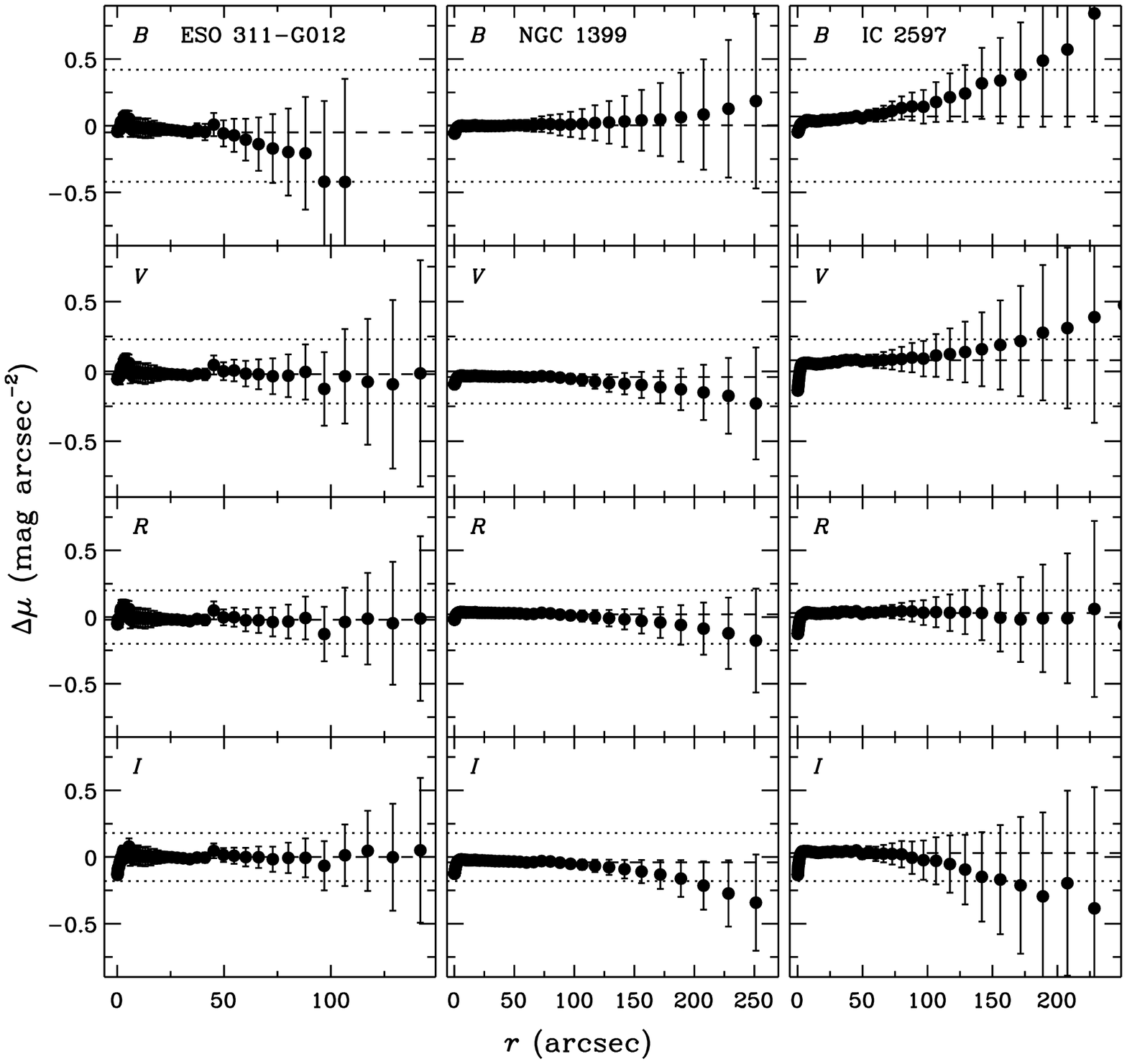,width=17.0cm,angle=0}}
\figcaption[fig15.eps]{
Internal comparison of the surface brightness profiles for the
objects for which we have repeat observations. The data points are calculated
by subtracting one profile from another. The horizontal dashed line in each
panel is the weighted average of the residual data, and the two horizontal
dotted lines mark the 1 $\sigma$ photometric uncertainty of the two
observations.
\label{figure:prof_diff}}
\end{figure*}

\addtocounter{figure}{-1}
\begin{figure*}[t]
\centerline{\psfig{file=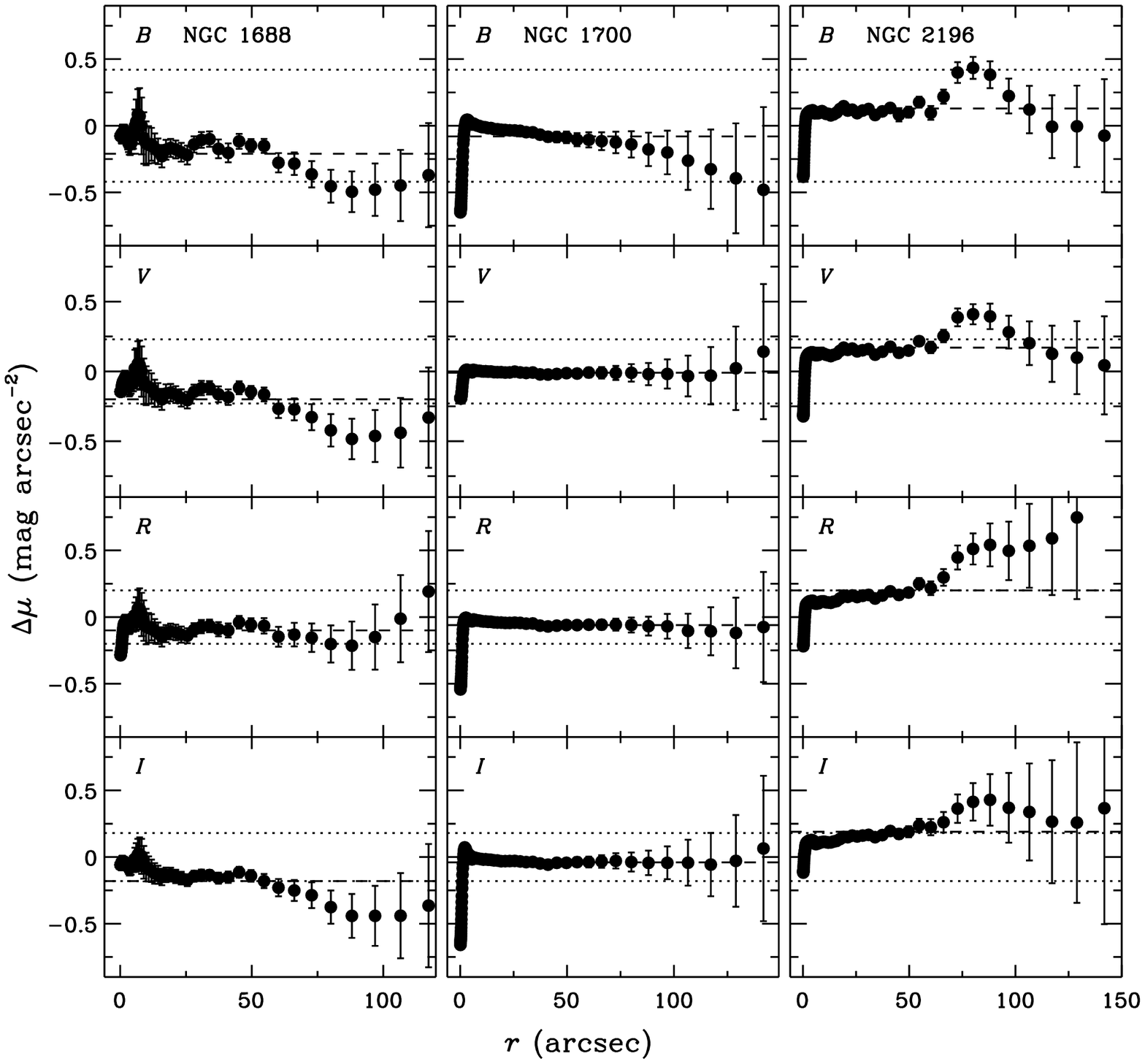,width=17.0cm,angle=0}}
\figcaption[fig15.eps]{
continued
\label{figure:prof_diff}}
\end{figure*}

\addtocounter{figure}{-1}
\begin{figure*}[t] 
\centerline{\psfig{file=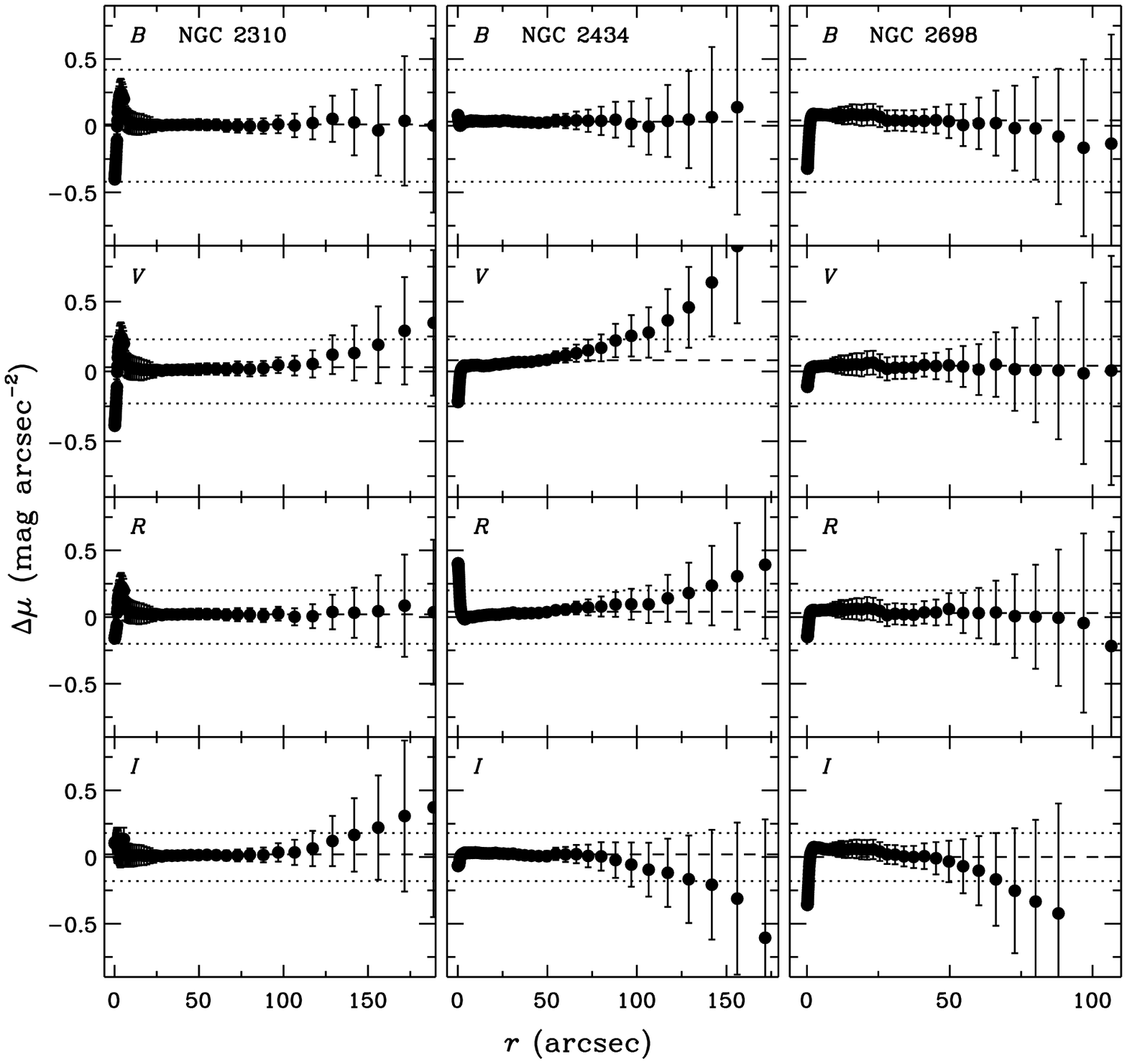,width=17.0cm,angle=0}}
\figcaption[fig15.eps]{
continued
\label{figure:prof_diff}}
\end{figure*}

\addtocounter{figure}{-1}
\begin{figure*}[t] 
\centerline{\psfig{file=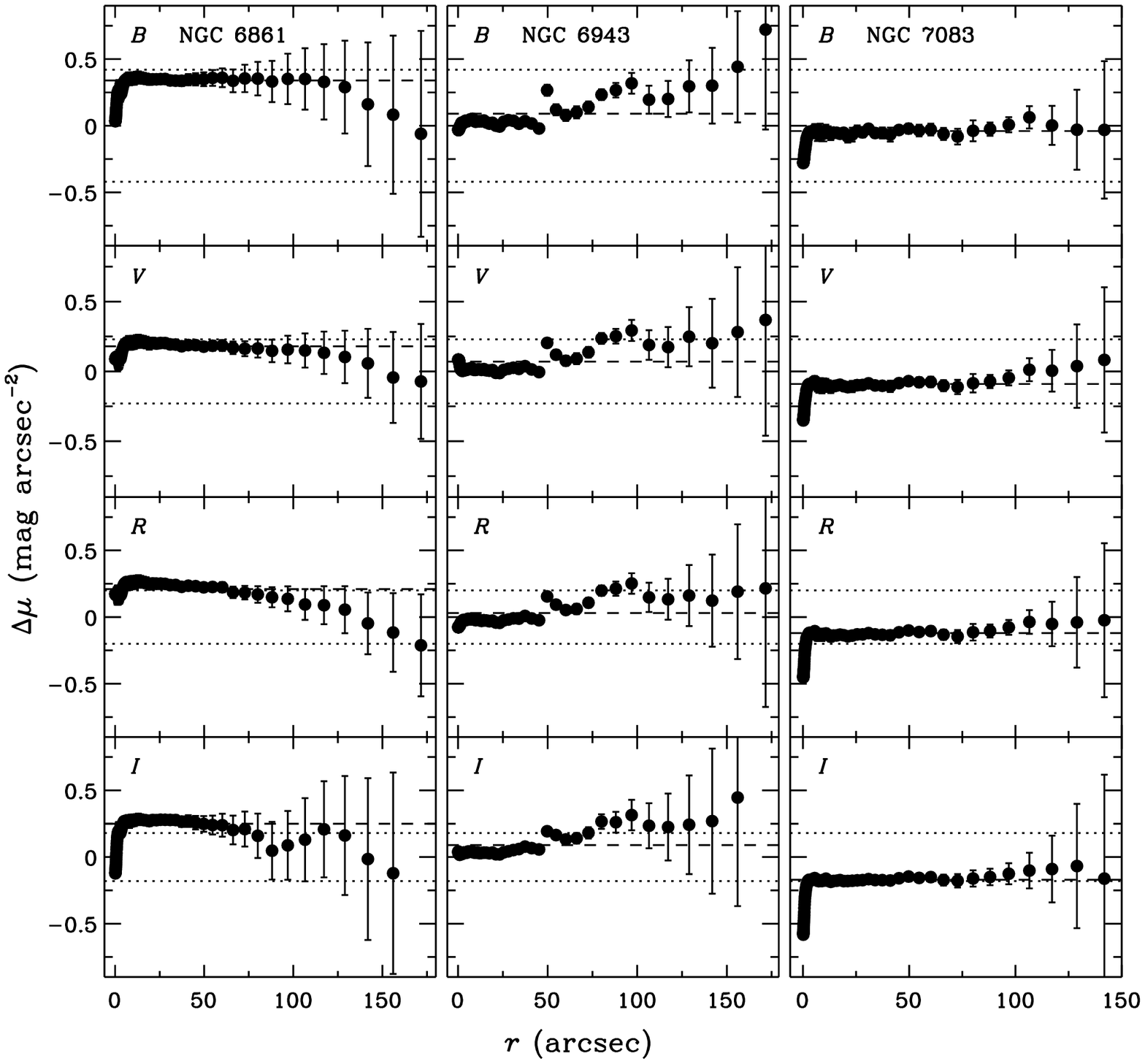,width=17.0cm,angle=0}}
\figcaption[fig15.eps]{
continued
\label{figure:prof_diff}}
\end{figure*}

\addtocounter{figure}{-1}
\begin{figure*}[t] 
\vbox{
\psfig{file=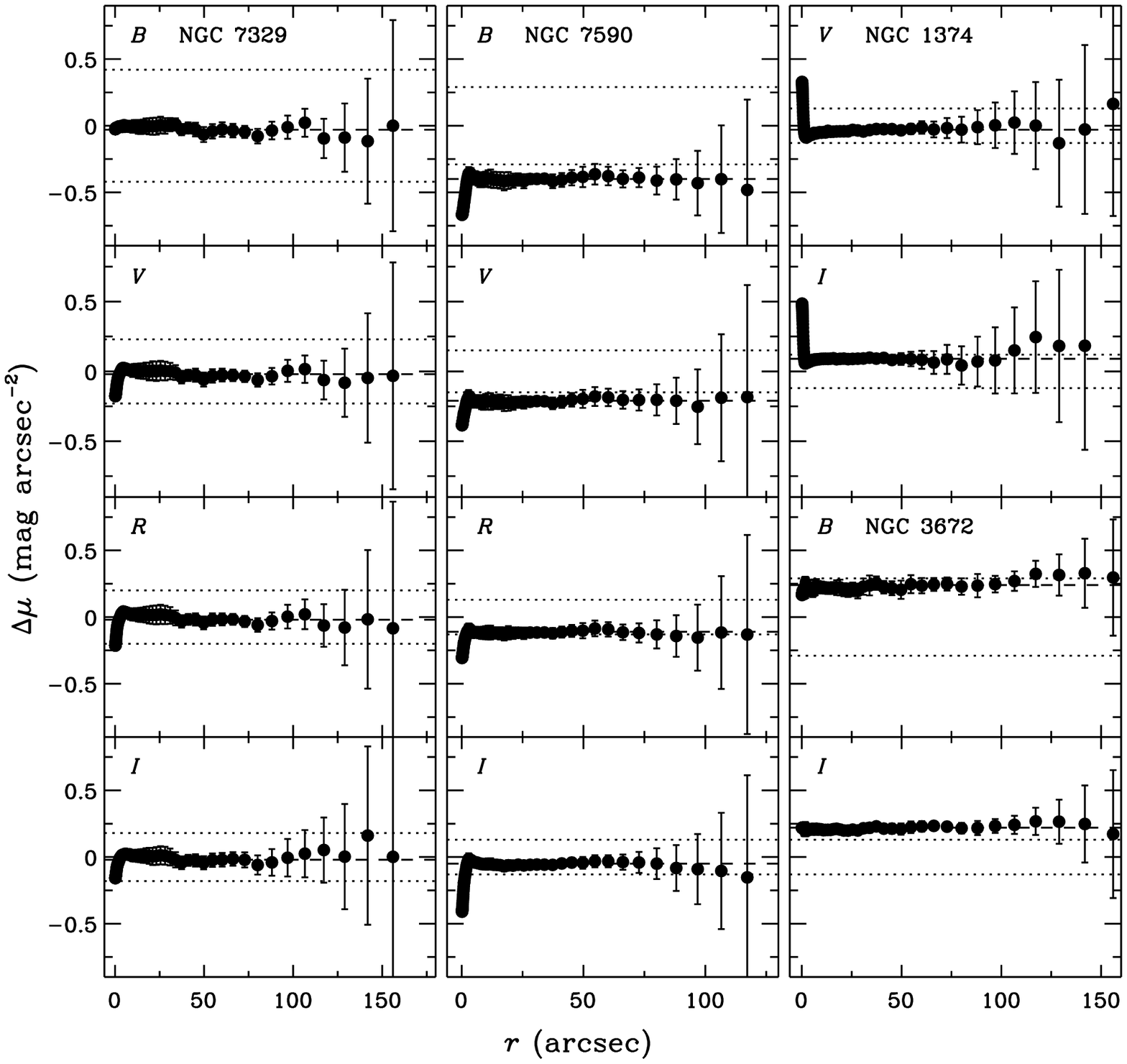,width=17.0cm,angle=0}
\psfig{file=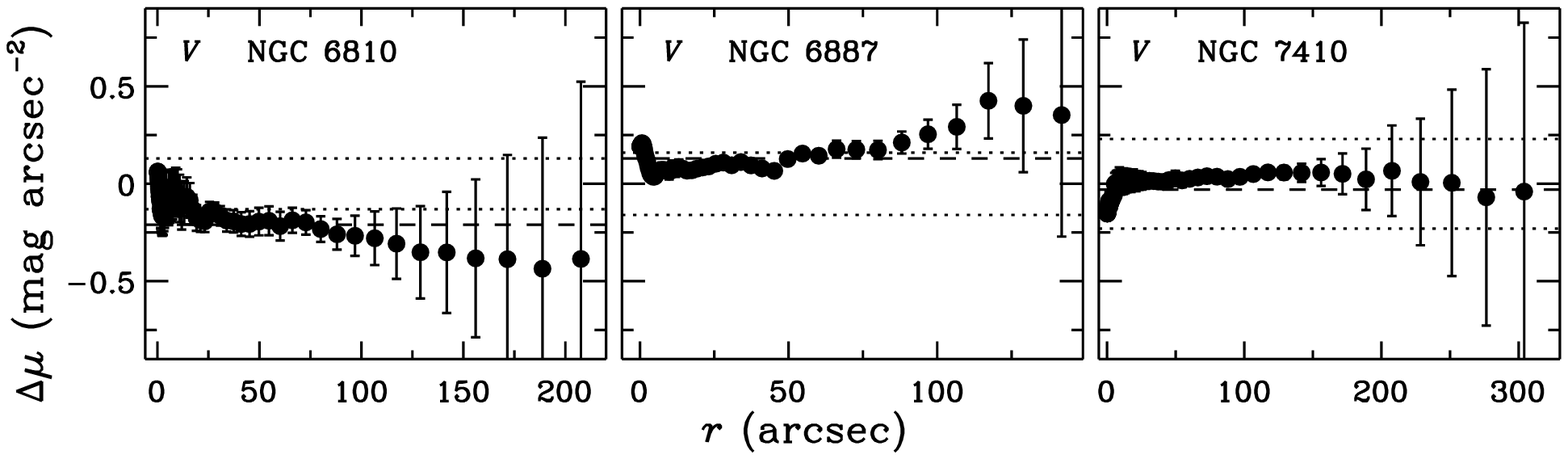,width=17.0cm,angle=0}
}
\figcaption[fig15.eps]{
continued
\label{figure:prof_diff}}
\end{figure*}

\begin{figure*}[t]
\centerline{\psfig{file=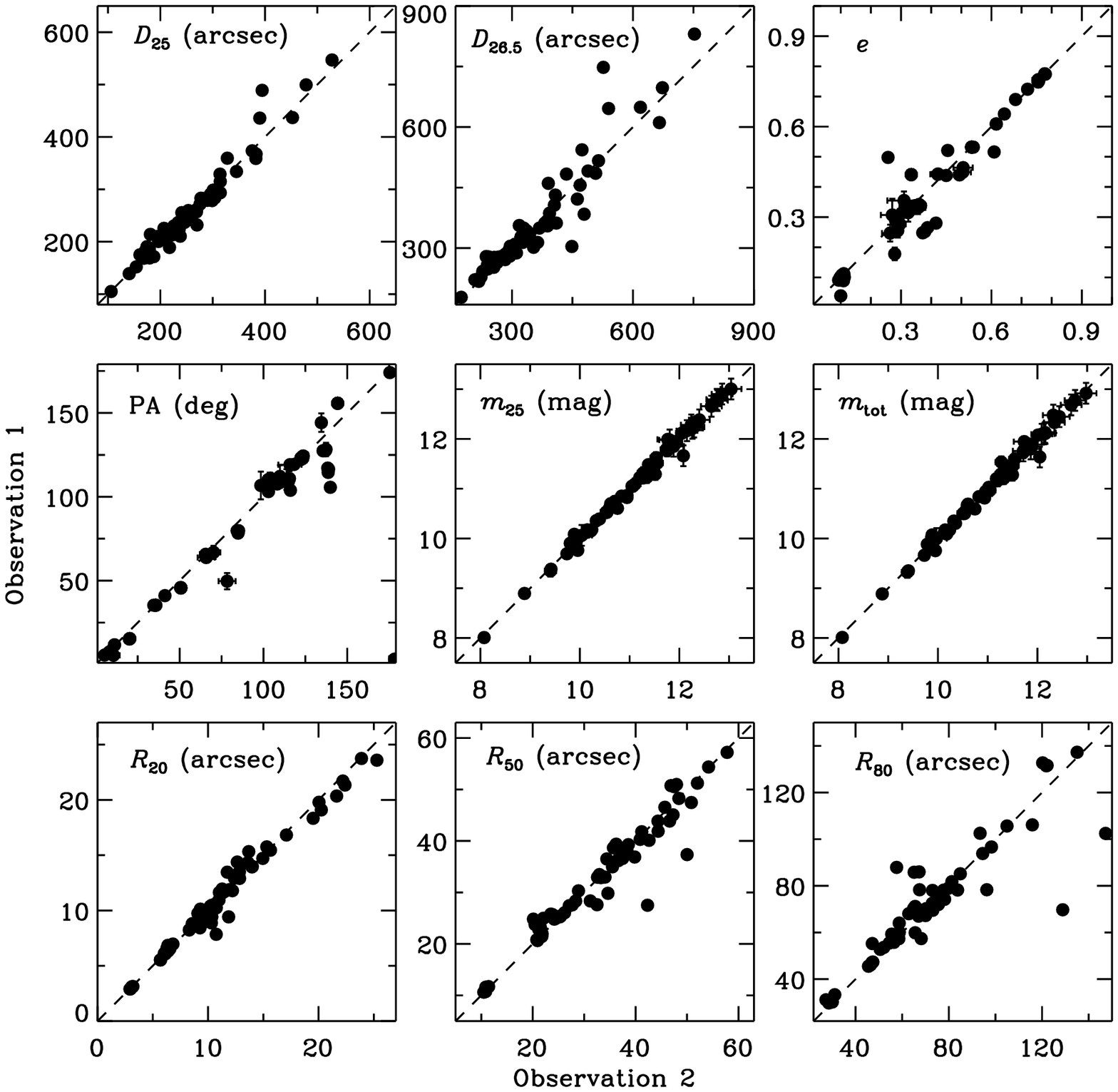,width=17.5cm,angle=0}}
\figcaption[fig16.eps]{
Internal comparison of derived parameters for objects with repeat
observations.  Observation 1 denotes values we adopt in the survey, while
Observation 2 gives the reference values for internal comparison.
The dashed line denotes $y = x$.
\label{figure:interpara}}
\end{figure*}
\vskip 0.3cm

\begin{figure*}[t]
\centerline{\psfig{file=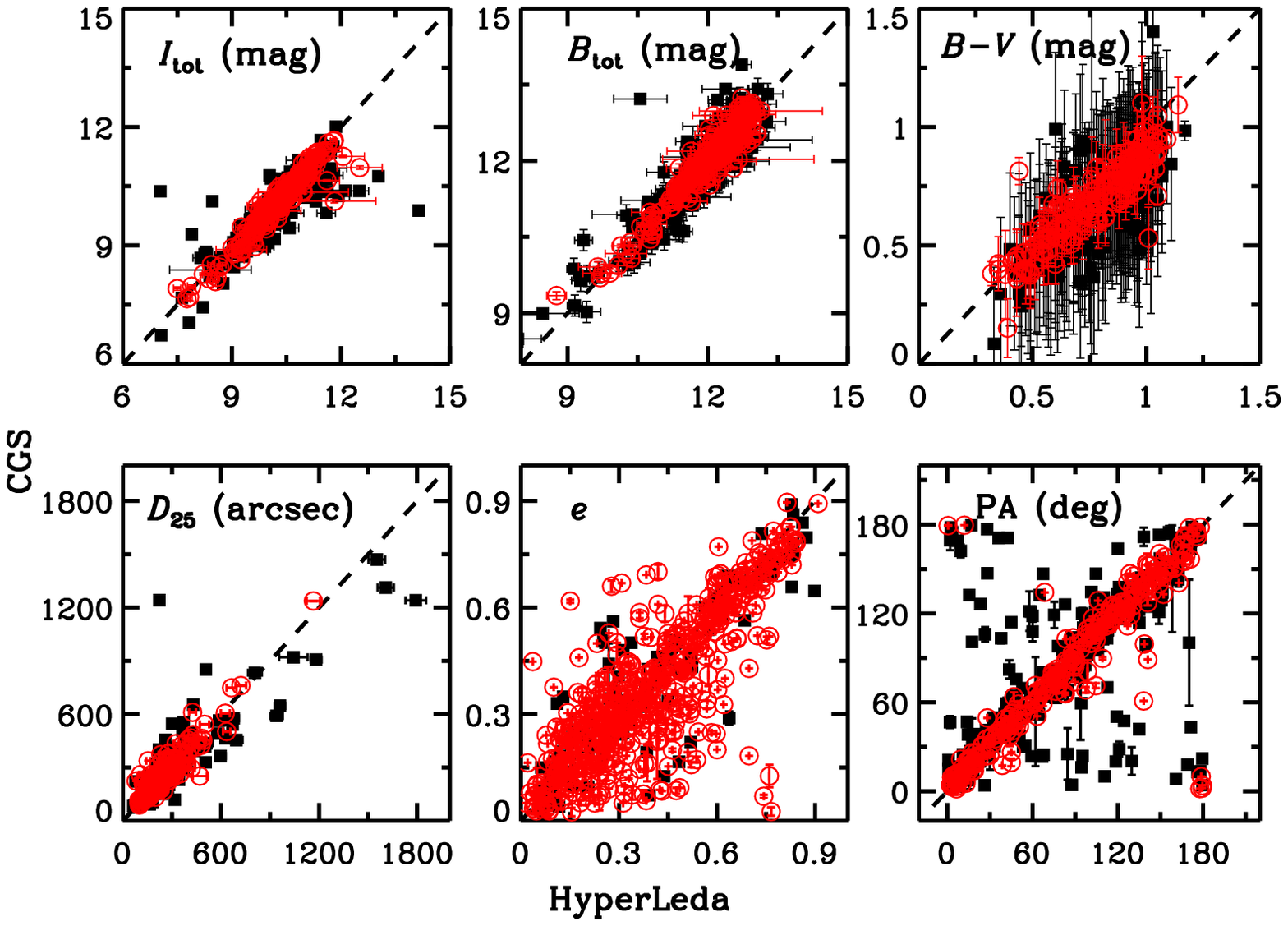,width=17.5cm,angle=0}}
\figcaption[fig17.eps]{
External comparison of derived parameters between CGS and HyperLeda.
The full sample available for comparison is shown as black squares, while the
red open points mark the subset with the smallest measurement uncertainty.  For
$I_{\rm tot}$, $B_{\rm tot}$, $B-V$, and $D_{25}$, the most reliable points are
those that were observed under photometric conditions, that have well-measured
sky values, and that do not suffer from contamination by nearby bright field
stars. For $e$ the red points highlight objects with minimal bright star
contamination, and for the PA we additionally require that $e \geq 0.3$.
The dashed line denotes $y = x$.
(A color version of this figure is available in the online journal.)
\label{figure:cingshype}}
\end{figure*}
\vskip 0.3cm

\begin{figure*}[t]
\centerline{\psfig{file=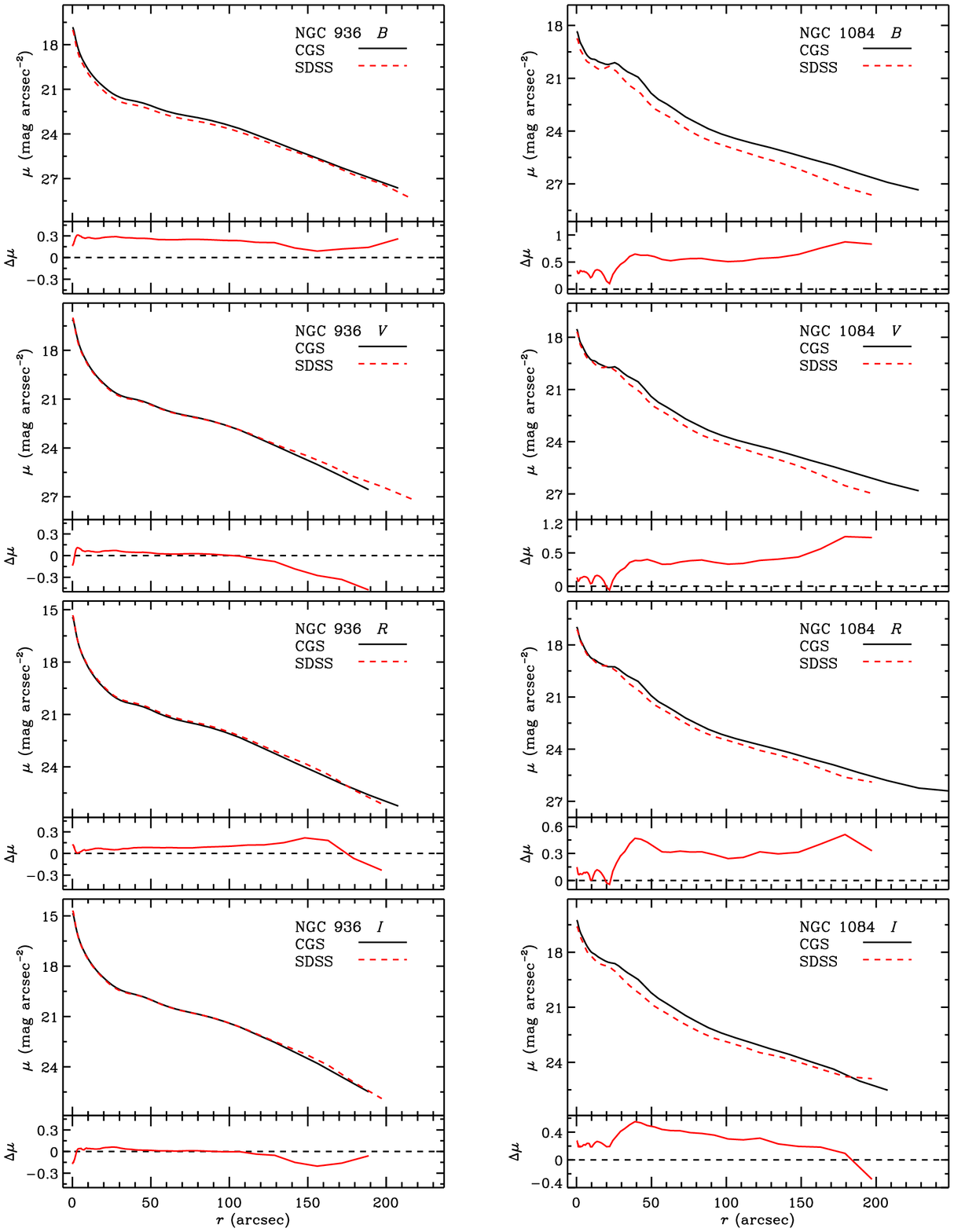,width=17.0cm,angle=0}}
\figcaption[fig18.ps]{
Comparison between the surface brightness profiles derived from CGS
and SDSS.  The SDSS data were analyzed in the same way as the CGS data,
and their photometric system was transformed to ours as described in
Section~10.3.   Within each panel, the upper subpanel shows the two profiles,
truncated at the radius where the isophotal intensity is $1 \, \sigma$ above
the local sky background; the bottom subpanel shows the difference between the
CGS and SDSS profiles.
(A color version of this figure is available in the online journal.)
\label{figure:cingssdss}}
\end{figure*}

\addtocounter{figure}{-1}
\begin{figure*}[t]
\centerline{\psfig{file=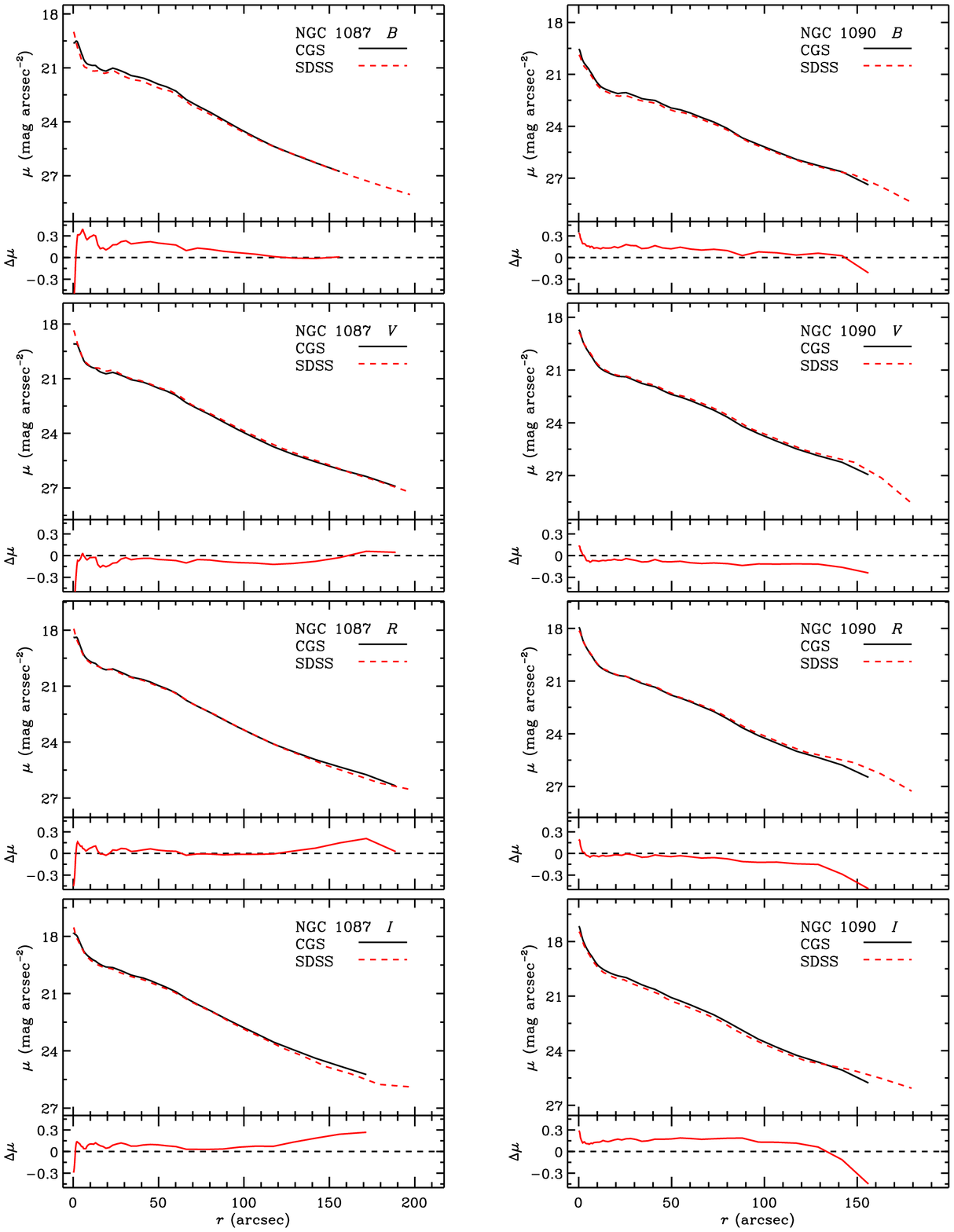,width=17.0cm,angle=0}}
\figcaption[fig18.ps]{
continued
\label{figure:cingssdss}}
\end{figure*}
\vskip 0.3cm

\noindent
sensitive
to data at large radii).  The average differences and standard deviations of
the parameters plotted in Figure~\ref{figure:interpara} are listed in
Table~9. The average differences are quite close to 0,
and the scatter is small.

\subsection{External Comparison with HyperLeda}

Several of the global parameters measured in our sample have independent data 
listed in HyperLeda\footnote{{\tt http://leda.univ-lyon1.fr}} \citep{patu03}, 
which we can use to perform an external comparison of our errors.  
Figure~\ref{figure:cingshype} compares the following six parameters 
between CGS and HyperLeda: total $I$-band magnitude ($I_{\rm tot}$), total 
$B$-band magnitude ($B_{\rm tot}$), integrated $B-V$ color, isophotal 
diameter at 25 $B$ mag~$\rm arcsec^{-2}$ ($D_{25}$), $e$, and PA.  
The overall agreement is quite good for the integrated magnitudes, $B-V$ 
color, and $D_{25}$ diameters, especially for the red open points, which 
represent galaxies that were observed under photometric conditions, that have 
no contamination from the nearby bright stars, and that have reliable sky 
values.  The PA comparison improves dramatically after isolating the subset 
with $e \geq 0.3$.  When the galaxy is round, the PA is hard to determine 
because the semi-major axis for any given isophote is ill-defined.  In 
addition, extreme outliers lying on the lower-right and upper-left corners of 
the distribution can be attributed to the 180\deg\ ambiguity for PAs close 
to 0\deg\ or 180\deg.  The ellipticities show by far the worst agreement.
The large scatter may be due in part to the fact that the HyperLeda values 
pertain to measurements made at $\mu_B = 25$ mag~$\rm arcsec^{-2}$, whereas 
ours are made at a significantly lower surface brightness threshold, at $\mu_B 
\approx 27$ mag~$\rm arcsec^{-2}$.  A more serious problem may be related to 
inherent biases and errors that are known to plague the axial ratios 
(ellipticities) contained in the HyperLeda database (see the Appendix of Ho 2007). 

\subsection{External Comparison with SDSS}

Roughly 9\% of the CGS galaxies overlap with the Sloan Digital Sky Survey 
\citep[SDSS;][]{york00, stou02}, which provides well-calibrated, uniform optical
images with a photometric accuracy of $2\%-3\%$.  The SDSS images are 
available in the \emph{ugriz} filters, but the $u$ and $z$ images have very low 
S/N, and we concentrate our attention on the $g, r,$ and $i$ bands.  We 
analyze the SDSS data following exactly the same procedures applied to the 
CGS.  After registering the $g$ and $r$ images to the $i$ image, we extract 
isophotal intensity profiles as described in Section~4.1.  As the SDSS images 
were observed in a drift-scan mode, they have both a very large field-of-view 
and an exceptionally uniform background \citep{potr06}. This allows us to 
determine accurate sky values using the method of \citet{noovan07}, as 
described in Section~3.  To convert the SDSS \emph{gri} photometry into our 
standard \emph{BVRI} system, we use the transformation equations given in 
\citet{jest05}:

\begin{eqnarray}
B & = & g + 0.39(g - r) + 0.21 \\
V & = & g - 0.59(g - r) - 0.01 \\
R & = & V - 1.09(r - i) - 0.22 \\
I & = & R - 1.00(r - i) - 0.21.
\end{eqnarray}

\bigskip
Figure~\ref{figure:cingssdss} compares the surface brightness profiles from 
CGS with those derived from SDSS, for a subset of four relatively small 
galaxies (NGC~936, 1084, 1087 and 1090) that have well-measured sky values.  
The profiles are truncated at the radius where the isophotal intensity is 
$1 \, \sigma$ above the local sky value.  It is clear that most of the profiles 
agree well with each other.  The absolute value of the average profile 
differences is 0.24, 0.14, 0.08, and 0.14 mag~$\rm arcsec^{-2}$ for $B, V, 
R,$ and $I$ bands, respectively.  This level of discrepancy is not unexpected, 
given that all four of these galaxies were observed under non-photometric 
conditions in CGS, not to mention of additional uncertainties introduced by 
the photometric transformation from the SDSS to the CGS system.  This 
comparison confirms that the basic reduction and calibration of the CGS 
data are sound.

Despite the short exposures of the SDSS images (54~s), they have superior 
background uniformity and better sky determination than CGS.  These advantages 
translate to better sensitivity in terms of surface brightness, by $\sim 0.4$, 
0.2, 0.6, and 0.9 mag~$\rm arcsec^{-2}$ in the $B$, $V$, $R$, and $I$ bands, 
respectively.  However, the longer integration times, better seeing, and finer 
pixel scale of the CGS images imply that they have much higher S/N and 
sensitivity to compact structures compared to SDSS, typically by a factor of 
$\sim4-5$.

\section{Summary}

We present a comprehensive isophotal analysis of optical (\emph{BVRI}) images 
for a statistically complete, magnitude-limited sample of 605 bright, southern 
galaxies, observed as part of the Carnegie-Irvine Galaxy Survey (CGS).  We 
discuss our strategy for determining the sky level and its error.  It is 
challenging to achieve very accurate sky subtraction with our images because 
the CGS galaxies are relatively large and the background suffers from 
low-level non-uniformities due to residual flat-fielding errors.  Nevertheless, 
crosschecks with internal and external data indicate that our calibration 
and sky subtraction strategies are robust, and that our quoted measurement 
uncertainties are sound.

This paper focuses on the derivation of radial profiles of surface brightness,
color, and various geometric parameters that characterize the shape and 
orientation of the isophotes.  We construct composite brightness profiles as 
a function of Hubble type to highlight statistical trends.  Non-exponential 
disks are seen in many S0 and spiral galaxies.  We perform a Fourier analysis 
of the isophotes to characterize their non-axisymmetric deviations from pure 
ellipses.  The relative amplitude of the $m = 1$ mode effectively identifies 
lopsided structures in the light distribution, which we find to be common in 
our sample, especially among late-type galaxies.  Bars and spiral arms, by 
contrast, are best revealed by the relative amplitude of the $m = 2$ Fourier 
mode.  We present a uniform set of quantitative measurements of bar size and 
bar strength, spiral arm strength, and lopsidedness amplitudes.

Forthcoming papers will utilize the databases assembled here and in Paper~I 
to explore a number of scientific issues, including the following.

\begin{enumerate}

\item{Statistics of bars, bar properties, and their possible connection to
spiral arms.}

\item{Incidence of lopsidedness and its dependence on global galaxy and 
environmental parameters.}

\item{Disk profiles, truncations, and correlations with color gradients.}

\end{enumerate}

\acknowledgements
We thank the referee for a prompt and helpful review of this manuscript.
This work was supported by the Carnegie Institution for Science (L.C.H.), 
the UC Irvine School of Physical Sciences (A.J.B.), the China Scholarship 
Council (Z.-Y.L.), and the Plaskett Fellowship of the Herzberg Institute of 
Astrophysics, National Research Council of Canada (C.Y.P.).  Z.-Y.L. is 
grateful to Professor X.-B. Wu of the Department of Astronomy in Peking University 
for his support and helpful suggestions on this project. We made use of 
HyperLeda and the NASA/IPAC Extragalactic Database (NED), which is operated by 
the Jet Propulsion Laboratory, California Institute of Technology, under 
contract with the National Aeronautics and Space Administration.
Funding for the SDSS and SDSS-II has been provided by the Alfred P. Sloan 
Foundation, the Participating Institutions, the National Science Foundation, 
the U.S. Department of Energy, the National Aeronautics and Space 
Administration, the Japanese Monbukagakusho, the Max Planck Society, and the 
Higher Education Funding Council for England. The SDSS Web site is 
{\tt http://www.sdss.org}.

\clearpage


\clearpage
\appendix

\section{Database of Isophotal Parameters}
Figures 19.1-19.616 present the database of isophotal parameters for the 605
galaxies in CGS; we also include the 11 extra galaxies that are not part of
the formal sample.  One full-page figure is devoted to each galaxy, ordered
sequentially following the numerical indices listed in Column~1 of 
Table~2.  In the lower-left panel, we give the $B$, $V$, 
$R$, and $I$ surface brightness profiles, followed by the $B-I$, $V-I$, and 
$R-I$ color profiles.  The lower-right panel shows the radial profiles of $e$, 
PA, $A_3$ (blue triangles) and $B_3$ (red circles), $A_4$ (blue triangles) and 
$B_4$ (red circles), $I_1/I_0$ (black triangles) and $\phi_1$ (red circles), 
and $I_2/I_0$ (black triangles) and $\phi_2$ (red circles).  The horizontal 
dashed lines in the $e$ and PA subpanels denote the characteristic values of 
the galaxy.  The top two panels show the \emph{BVI} color composite image, 
displayed using an arcsinh stretch, and the star-cleaned $I$-band structure 
map, displayed using a linear stretch.  These are helpful to show side-by-side 
with the isophotal plots to aid in the identification and interpretation of 
specific features in the 1-D radial profiles.  We only display three sample 
pages for illustration (Figures 19(a)--19(c); the full set of figures is 
available in the electronic version of the paper, as well as on the project Web 
site {\tt http://cgs.obs.carnegiescience.edu}.

\clearpage

\figurenum{19(a)}
\begin{figure}
\epsscale{0.95}
\plotone{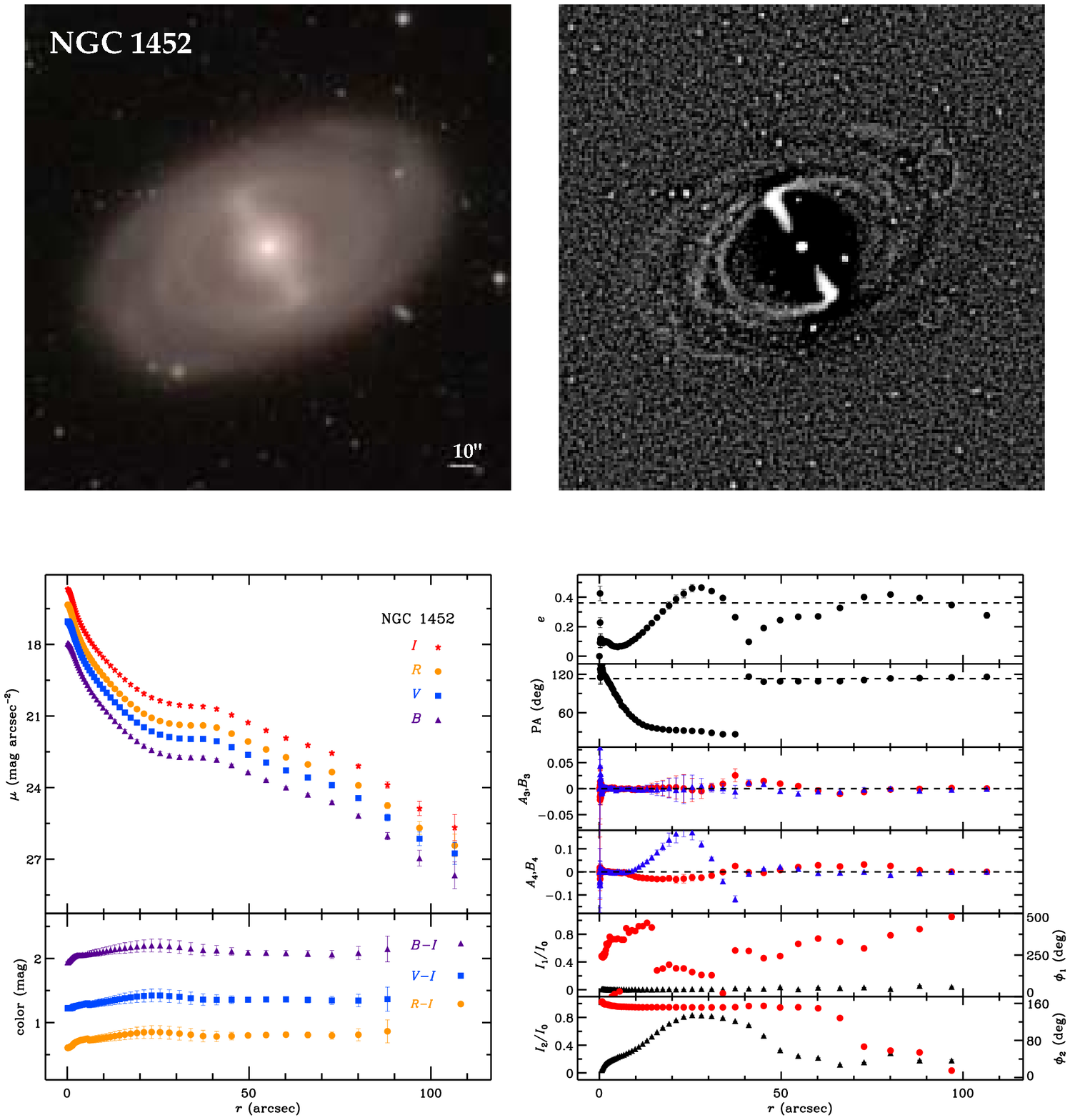}
\caption{Brightness profiles and isophotal parameters for the barred SB0/a 
galaxy NGC~1452. The color composite image is shown in the upper left, 
displayed using an arcsinh stretch, and the star-cleaned $I$-band structure 
map is shown in the upper right, displayed using a linear stretch.  The 
lower-left panel gives the $B$, $V$, $R$, and $I$ surface brightness profiles, 
followed by the $B-I$, $V-I$, and $R-I$ color profiles.  The lower-right panel 
shows the radial profiles of $e$, PA, $A_3$ (blue triangles) and $B_3$ (red 
circles), $A_4$ (blue triangles) and $B_4$ (red circles), $I_1/I_0$ (black 
triangles) and $\phi_1$ (red circles), and $I_2/I_0$ (black triangles) and 
$\phi_2$ (red circles). The horizontal dashed lines in the $e$ and PA 
subpanels denote the characteristic values of the galaxy.
(A color version and the complete figure set (616 images) are available in 
the online journal.)
}
\label{figure:barprof}
\end{figure}

\figurenum{19(b)}
\begin{figure}
\epsscale{0.95}
\plotone{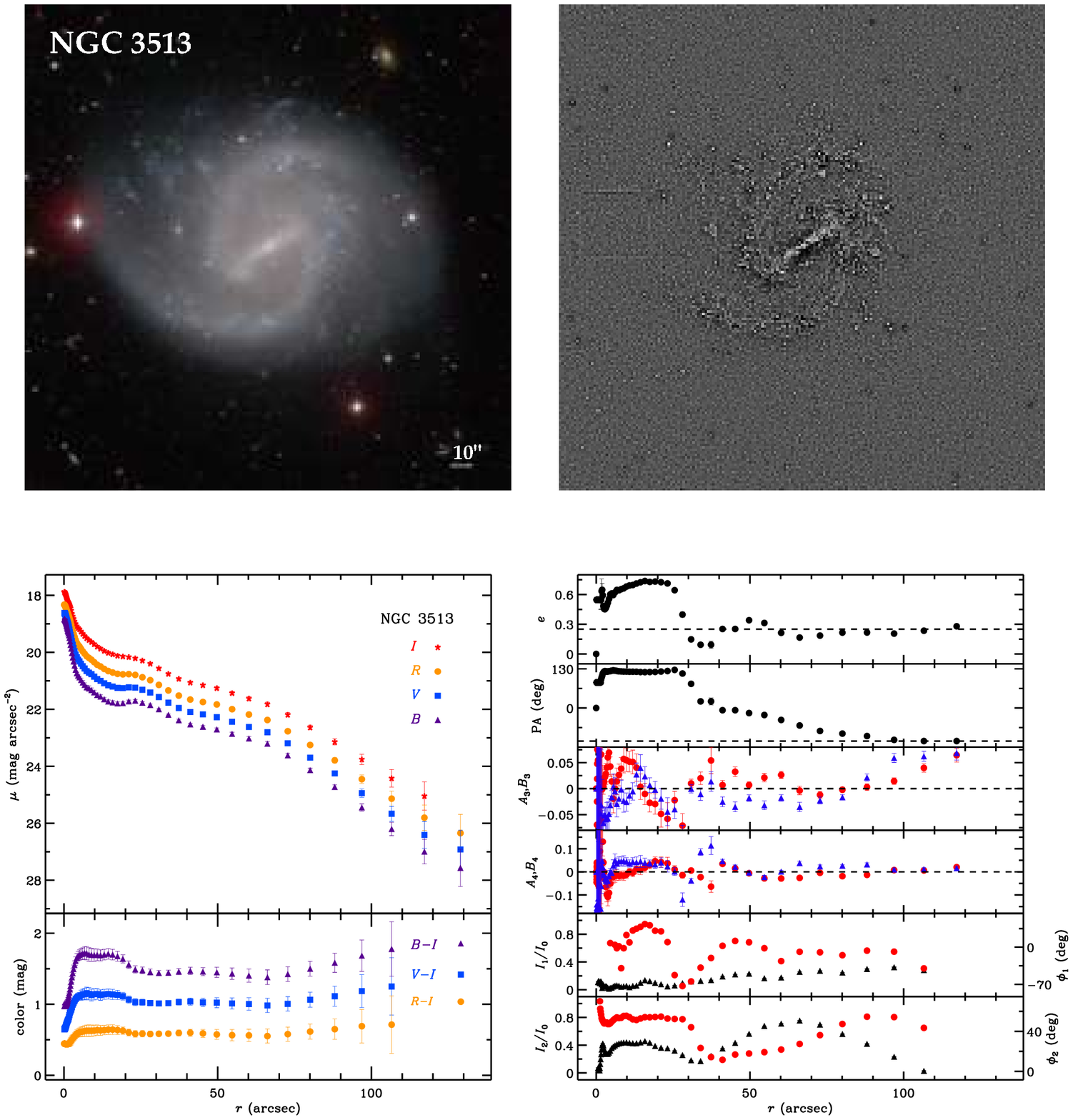}
\caption{Brightness profiles and isophotal parameters for the lopsided SBc
galaxy NGC~3513. The color composite image is shown in the upper left,
displayed using an arcsinh stretch, and the star-cleaned $I$-band structure
map is shown in the upper right, displayed using a linear stretch.  The
lower-left panel gives the $B$, $V$, $R$, and $I$ surface brightness profiles,
followed by the $B-I$, $V-I$, and $R-I$ color profiles.  The lower-right panel
shows the radial profiles of $e$, PA, $A_3$ (blue triangles) and $B_3$ (red
circles), $A_4$ (blue triangles) and $B_4$ (red circles), $I_1/I_0$ (black
triangles) and $\phi_1$ (red circles), and $I_2/I_0$ (black triangles) and
$\phi_2$ (red circles). The horizontal dashed lines in the $e$ and PA 
subpanels denote the characteristic values of the galaxy.
(A color version of this figure is available in the online journal.)
}
\label{figure:lopprof}
\end{figure}

\figurenum{19(c)}
\begin{figure}
\epsscale{0.95}
\plotone{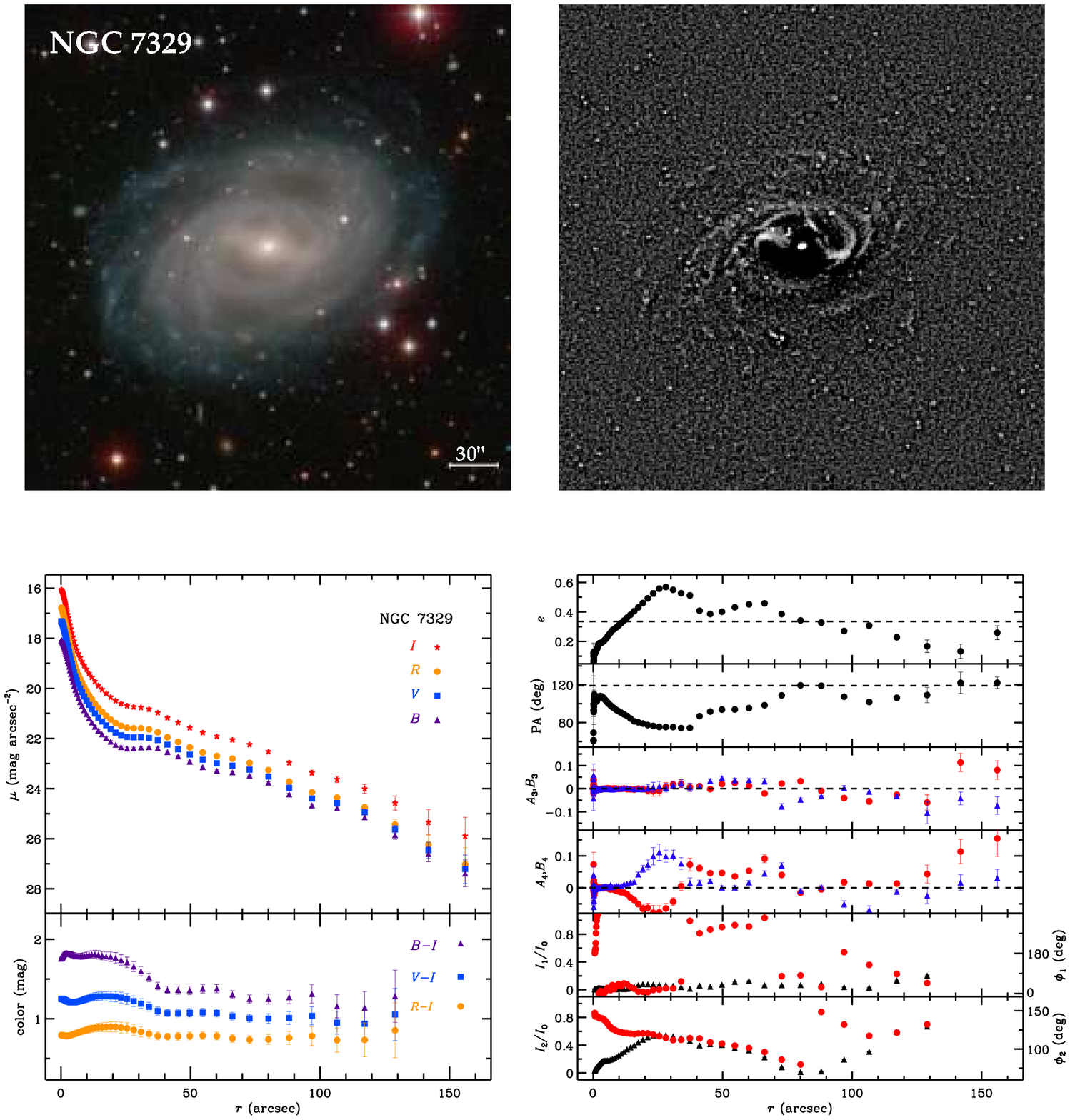}
\caption{Brightness profiles and isophotal parameters for the barred SBb galaxy
NGC~7329. The color composite image is shown in the upper left,
displayed using an arcsinh stretch, and the star-cleaned $I$-band structure
map is shown in the upper right, displayed using a linear stretch.  The
lower-left panel gives the $B$, $V$, $R$, and $I$ surface brightness profiles,
followed by the $B-I$, $V-I$, and $R-I$ color profiles.  The lower-right panel
shows the radial profiles of $e$, PA, $A_3$ (blue triangles) and $B_3$ (red
circles), $A_4$ (blue triangles) and $B_4$ (red circles), $I_1/I_0$ (black
triangles) and $\phi_1$ (red circles), and $I_2/I_0$ (black triangles) and
$\phi_2$ (red circles). The horizontal dashed lines in the $e$ and PA 
subpanels denote the characteristic values of the galaxy.
(A color version of this figure is available in the online journal.)
}
\label{figure:spiralprof}
\end{figure}


\begin{thebibliography}{}

\bibitem[Aguerri et al.(2003)]{ague03} Aguerri, J. A. L., Debattista, V. P., \& Corsini, E. M. 2003, \mnras, 338, 465

\bibitem[Aguerri et al.(2005)]{agu05} Aguerri, J. A. L., Elias-Rosa, N., Corsini, E. M., \& Mu\~{n}oz-Tu\~{n}\'{o}n, C. 2005, \aap, 434, 109

\bibitem[Aguerri et al.(2009)]{ague09} Aguerri, J. A. L., M\'{e}ndez-Abreu, J., \& Corsini, E. M. 2009, \aap, 495, 491

\bibitem[Andredakis et al.(1995)]{andr95} Andredakis, Y. C., Peletier, R. F., \& Balcells, M. 1995, \mnras, 275, 874

\bibitem[Andredakis \& Sanders(1994)]{ansa94} Andredakis, Y. C., \& Sanders, R. H. 1994, \mnras, 267, 283

\bibitem[Athanassoula \& Misiriotis(2002)]{atmi02} Athanassoula, E., \& Misiriotis, A. 2002, \mnras, 330, 35

\bibitem[Barazza et al.(2008)]{bara08} Barazza, F. D., Jogee, S., \& Marinova, I. 2008, \apj, 675, 1194


\bibitem[Bournaud et al.(2005)]{bour05} Bournaud, F., Combes, F., Jog, C. J., \& Puerari, I. 2005, \aap, 438, 507

\bibitem[Buta(1986)]{buta86} Buta, R. 1986, \apjs, 61, 631

\bibitem[Buta \& Block(2001)]{bubl01} Buta, R., \& Block, D. L. 2001, \apj, 550, 243

\bibitem[Caon et al.(1993)]{caon93} Caon, N., Capaccioli, M., \& D'Onofrio, M. 1993, \mnras, 265, 1013


\bibitem[Carter(1978)]{cart78} Carter, D. 1978, \mnras, 182, 797

\bibitem[Courteau et al.(1996)]{cour96} Courteau, S., de Jong, R. S., \& Broeils, A. H. 1996, \apj, 457, 73


\bibitem[de~Grijs(1998)]{degr98} de Grijs, R. 1998, \mnras, 299, 595

\bibitem[de~Grijs et al.(2001)]{degr01} de Grijs, R., Kregel, M., \& Wesson, K. H. 2001, \mnras, 324, 1074

\bibitem[de~Jong(1996)]{dejo96} de Jong, R. S. 1996, \aaps, 118, 557

\bibitem[de~Vaucouleurs(1948)]{deva48} de Vaucouleurs, G. 1948, Ann. Astrophys, 11, 247

\bibitem[de~Vaucouleurs(1959)]{deva59} de Vaucouleurs, G. 1959, in Handbuch der Physik, Bd. LIII, ed. S. Flugge (Berlin: Springer-Verlag), 275

\bibitem[de~Vaucouleurs et al.(1991)]{deva91} de Vaucouleurs, G., de Vaucouleurs, A., Corwin Jr., H. G., et al. 1991, Third Reference Catalogue of Bright Galaxies (New York: Springer-Verlag)

\bibitem[D'Onofrio et al.(1994)]{dono94} D'Onofrio, M., Capaccioli, M., \& Caon, N. 1994, \mnras, 271, 523

\bibitem[Elmegreen \& Elmegreen(1985)]{elmegreen85} Elmegreen, B.~G., \& Elmegreen, D.~M. 1985, \apj, 288, 438


\bibitem[Elmegreen et al.(1989)]{elmegreen89} Elmegreen, B.~G., Seiden, P. E., \& Elmegreen, D.~M. 1989, \apj, 343, 602


\bibitem[Erwin(2005)]{erwi05} Erwin, P. 2005, \mnras, 364, 283

\bibitem[Erwin et al.(2005)]{erbp05} Erwin, P., Beckman, J. E., \& Pohlen, M. 2005, \apj, 626, 81

\bibitem[Erwin et al.(2008)]{erwi08} Erwin, P., Pohlen, M., \& Beckman, J. E. 2008, \aj, 135, 20

\bibitem[Ferguson \& Clarke(2001)]{fecl01} Ferguson, A. M. N., \& Clarke, C. J. 2001, \mnras, 325, 781

\bibitem[Ferrarese et al.(2006)]{ferr06} Ferrarese, L., et al. 2006, \apjs, 164, 334

\bibitem[Fisher \& Drory(2008)]{fidr08} Fisher, D. B., \& Drory, N. 2008, \aj, 136, 773

\bibitem[Freeman(1970)]{free70} Freeman, K. C. 1970, \apj, 160, 811

\bibitem[Fry et al.(1999)]{fry99} Fry, A. M., Morrison, H. L., Harding, P., \& Boroson, T. A. 1999, \aj, 118, 1209


\bibitem[Gadotti(2008)]{gadotti08} Gadotti, D.~A. 2008, \mnras, 384, 420

\bibitem[Gadotti et al.(2007)]{gadotti07} Gadotti, D.~A., Athanassoula, E., Carrasco, L., et al. 2007, \mnras, 381, 943

\bibitem[Governato et al.(2007)]{gove07} Governato, F., Willman, B., Mayer, L., et al. 2007, \mnras, 374, 1479

\bibitem[Graham(2001)]{grah01} Graham, A. W. 2001, \aj, 121, 820

\bibitem[Graham \& Colless(1997)]{grco97} Graham, A., \& Colless, M. 1997, \mnras, 287, 221

\bibitem[Graham et al.(1996)]{grah96} Graham, A., Lauer, T. R., Colless, M., \& Postman, M. 1996, \apj, 465, 534

\bibitem[Gunn et al.(1998)]{gunn98} Gunn, J. E., Carr, M., Rockosi, C., et al. 1998, \aj, 116, 3040

\bibitem[Ho(2007)]{Ho2007} Ho, L.~C. 2007, \apj, 668, 94

\bibitem[Ho et al.(2011)]{Ho2011} Ho, L. C., Li, Z.-Y., Barth, A. J., Seigar, M., \& Peng, C. Y. 2011, \apjs, in press (Paper~I)


\bibitem[Hubble(1926)]{hubb26} Hubble, E. 1926, \apj, 64, 321


\bibitem[Jedrzejewski(1987)]{jedr87} Jedrzejewski R. I. 1987, \mnras, 226, 747

\bibitem[Jester et al.(2005)]{jest05} Jester, S., Schneider, D. P., Richards, G. T., et al. 2005, \aj, 130, 873

\bibitem[Jog \& Combes(2008)]{joco08} Jog, C. J., \& Combes, F. 2009, Phys. Rep., 471, 75

\bibitem[Jogee et al.(2004)]{jogee04} Jogee, S., Barazza, F. D., Rix, H.-W., et al. 2004, \apj, 615, 105


\bibitem[Katz(1991)]{katz91} Katz, N. 1991, \apj, 368, 325



\bibitem[King(1978)]{king78} King, I. R. 1978, \apj, 222, 1

\bibitem[Kormendy(1979)]{korm79} Kormendy, J. 1979, \apj, 227, 714


\bibitem[Kormendy \& Djorgovski(1989)]{kodj89} Kormendy, J., \& Djorgovski, S. 1989, ARA\&A, 27, 235

\bibitem[Kormendy et al.(2009)]{korm09} Kormendy, J., Fisher, D. B., Cornell, M. E., \& Bender, R. 2009, \apjs, 182, 216

\bibitem[Kormendy \& Kennicutt(2004)]{koke04} Kormendy, J., \& Kennicutt, R. C. 2004, ARA\&A, 42, 603

\bibitem[Laine et al.(2002)]{lain02} Laine, S., Shlosman, I., Knapen, J. H., \& Peletier, R. F. 2002, \apj, 567, 97


\bibitem[Lauer et al.(1995)]{lauer95} Lauer, T.~R., Ajhar, E. A., Byun, Y.-I., et al.  1995, \aj, 110, 2622

\bibitem[Laurikainen et al.(2005)]{laurikainen05} Laurikainen, E., Salo, H., \& Buta, R. 2005, \mnras, 362, 1319


\bibitem[MacArthur et al.(2003)]{maca03} MacArthur, L. A., Courteau, S., \& Holtzman, J. A. 2003, \apj, 582, 689

\bibitem[Marinova \& Jogee(2007)]{marjog07} Marinova, I., \& Jogee, S. 2007, \apj, 659, 1176

\bibitem[Martin(1995)]{mart95} Martin, P. 1995, \aj, 109, 2428

\bibitem[Martinet \& Friedli(1997)]{mafr97} Martinet, L., \& Friedli, D. 1997, \aap, 323, 363

\bibitem[Matthews \& Gallagher(1997)]{maga97} Matthews, L. D., \& Gallagher III, J. S. 1997, \aj, 114, 1899

\bibitem[Men\'{e}ndez-Delmestre et al.(2007)]{mene07} Men\'{e}ndez-Delmestre, K., Sheth, K., Schinnerer, E., Jarrett, T. H., \& Scoville, N. Z. 2007, \apj, 657, 790

\bibitem[Milvang-Jensen \& J\o rgensen(1999)]{miljog99} Milvang-Jensen, B., \& J\o rgensen, I. 1999, Baltic Astron., 8, 535

\bibitem[Noordermeer \& van~der~Hulst(2007)]{noovan07} Noordermeer, E., \& van der Hulst, J. M. 2007, \mnras, 376, 1480

\bibitem[Odewahn et al.(2002)]{odewahn02} Odewahn, S.~C., Cohen, S.~H., Windhorst, R.~A., \& Philip, N.~S. 2002, \apj, 568, 539

\bibitem[Ohta et al.(1990)]{ohta90} Ohta, K., Hamabe, M., \& Wakamatsu, K. 1990, \apj, 357, 71

\bibitem[Paturel et al.(2003)]{patu03} Paturel G., Petit C., Prugniel Ph., et al. 2003, A\&A, 412, 45

\bibitem[Peletier et al.(1990)]{pele90} Peletier, R. F., Davies, R. L., Illingworth, G. D., Davis, L. E., \& Cawson, M. 1990, \aj, 100, 1091

\bibitem[Peng et al.(2010)]{peng10} Peng, C. Y., Ho, L. C., Impey, C. D., \& Rix, H.-W. 2010, AJ, 139, 2097

\bibitem[Phillipps et al.(1991)]{phil91} Phillipps, S., Evans, R., Davies, J. I., \& Disney, M. J. 1991, \mnras, 253, 496


\bibitem[Pohlen et al.(2000)]{pohl00} Pohlen, M., Dettmar, R.-J., \& L\"{u}tticke, R. 2000, \aap, 357, L1

\bibitem[Pohlen et al.(2002)]{pohl02} Pohlen, M., Dettmar, R.-J., L\"{u}tticke, R., \& Aronica, G. 2002, \aap, 392, 807

\bibitem[Pohlen \& Trujillo(2006)]{potr06} Pohlen, M., \& Trujillo, I. 2006, \aap, 454, 759

\bibitem[Prieto et al.(1997)]{prie97} Prieto, M., Gottesman, S. T., Aguerri, J. A. L., \& Varela, A. M. 1997, \aj, 114, 1413

\bibitem[Reichard et al.(2008)]{reic08} Reichard, T. A., Heckman, T. M., Rudnick, G., et al. 2008, \apj, 677, 186

\bibitem[Rix \& Zaritsky(1995)]{rixzar95} Rix, H.-W., \& Zaritsky, D. 1995, \apj, 447, 82


\bibitem[Robertson et al.(2004)]{robe04} Robertson, B., Yoshida, N., Springel, V., \& Hernquist, L. 2004, \apj, 606, 32


\bibitem[Sandage(1961)]{sand61} Sandage, A. 1961, The Hubble Atlas of Galaxies (Washington, DC: Carnegie Inst. Washington)


\bibitem[Schlegel et al.(1998)]{schl98} Schlegel, D. J., Finkbeiner, D. P., \& Davis, M. 1998, \apj, 500, 525

\bibitem[Schombert(1986)]{schombert86} Schombert, J.~M. 1986, \apjs, 60, 603


\bibitem[S\'{e}rsic(1968)]{sers68}  S\'ersic, J.~L. 1968, Atlas de Galaxias Australes (C\'ordoba: Obs. Astron., Univ. Nac. C\'ordoba)

\bibitem[Silva \& Elston(1994)]{silels94} Silva, D. R., \& Elston, R. 1994, \apj, 428, 511

\bibitem[Slyz et al.(2002)]{slyz02} Slyz, A. D., Devriendt, J. E. G., Silk, J., \& Burkert, A. 2002, \mnras, 333, 894



\bibitem[Stoughton et al.(2002)]{stou02} Stoughton, C., Lupton, R. H., Bernardi, M., et al. 2002, \aj, 123, 485

\bibitem[Taylor et al.(2005)]{tayl05} Taylor, V. A., Jansen, R. A., Windhorst, R. A., Odewahn, S. C., \& Hibbard, J. E. 2005, \apj, 630, 784

\bibitem[Trujillo et al.(2001)]{truj01} Trujillo, I., Graham, A. W., \& Caon, N. 2001, \mnras, 326, 869

\bibitem[Trujillo et al.(2002)]{truj02} Trujillo, I., Ramos, A. A., Rubi\~{n}o-Mart\'{i}n, J. A., et al. 2002, \mnras, 333, 510

\bibitem[van~Albada(1982)]{vanalbada82} van Albada, T.~S. 1982, \mnras, 201, 939

\bibitem[van~der~Kruit(1979)]{vand79} van der Kruit, P. C. 1979, \aaps, 38, 15


\bibitem[van~der~Kruit \& Freeman(2011)]{vanderkruitfreeman11} van der Kruit, P.~C., \& Freeman, K.~C. 2011, \annrev, 49, 301

\bibitem[van~der~Kruit \& Searle(1981)]{vase81} van der Kruit, P. C., \& Searle, L. 1981, \aap, 95, 116

\bibitem[van~Eymeren et al.(2011)]{vaneymeren11} van Eymeren, J., Juette, E., Jog, C. J., Stein, Y., \& Dettmar, R.-J. 2011, \aa, 530, A30


\bibitem[Wu et al.(2002)]{wu02} Wu, H., Burstein, D., Deng, Z., et al. 2002, \aj, 123, 1364

\bibitem[York et al.(2000)]{york00} York, D.~G., Adelman, J., Anderson Jr., J. E., et al. 2000, \aj, 120, 1579



\bibitem[Yoshii \& Sommer-Larsen(1989)]{yoso89} Yoshii, Y., \& Sommer-Larsen, J. 1989, \mnras, 236, 779

\bibitem[Zaritsky \& Rix(1997)]{zari97} Zaritsky, D., \& Rix, H.-W. 1997, \apj, 477, 118

\bibitem[Zhang \& Wyse(2000)]{zhwy00} Zhang, B., \& Wyse, R. F. G. 2000, \mnras, 313, 310

\end{thebibliography}
\end{document}